\documentclass[11pt,a4paper]{amsart}
\usepackage[scale=0.805,marginratio={1:1, 1:1},ignoreall]{geometry}

\usepackage{amsmath,amssymb}

\usepackage[pagewise]{lineno}

\usepackage{setspace}

\usepackage{proof-at-the-end}
\usepackage{hyperref}
\usepackage{enumitem}
\usepackage[round]{natbib}

\theoremstyle{plain}
\newtheorem{thm}{Theorem}[section]

\theoremstyle{definition}
\newtheorem{example}[thm]{Example}
\newtheorem{Def}[thm]{Definition}

\theoremstyle{remark}
\newtheorem{Rem}{Remark}[thm]

\numberwithin{equation}{section}

\DeclareMathOperator{\supp}{supp}
\DeclareMathOperator*{\argmax}{arg\;\!max}

\newcommand{\witilde}{\widetilde}
\newcommand{\wihat}{\widehat}

\newcommand{\cplm}[1]{{#1}^{\mathsf{c}}}

\begin{document}


\title{Evolution of Preferences in Multiple Populations}
\author{Yu-Sung Tu}
\address{Institute of Economics\\
         Academia Sinica\\
         Taipei 11529, Taiwan}
\email[Yu-Sung Tu]{yusungtu@gate.sinica.edu.tw}
\author{Wei-Torng Juang}
\email[Wei-Torng Juang]{wjuang@econ.sinica.edu.tw}


\begin{abstract}
We study the evolution of preferences in multi-population settings that allow matches across distinct populations.
Each individual has subjective preferences over potential outcomes,
and chooses a best response based on his preferences and the information about the opponents' preferences.
Individuals' realized fitnesses are given by material payoff functions.
Following \citet{eDek-jEly-oYil:ep}, we assume that individuals observe their opponents' preferences with probability $p$.
We first derive necessary and sufficient conditions for stability for $p=1$ and $p=0$,
and then check the robustness of our results against small perturbations on observability for the case of pure-strategy outcomes.
\end{abstract}
\maketitle


\section{Introduction}


Evolution of preferences can be studied through the \emph{indirect evolutionary approach},
which is often used to explain how behavior appears inconsistent with material self-interest,
such as altruism, vengeance, punishment, fairness, and reciprocity.\footnote{%
See, for example, \citet{wGut-mYaa:erbssg}, \citet{wGut:eaecbri}, \citet{hBes-wGut:iaes}, \citet{sHuc-jOec:ieaefa}, \citet{eOst:caesn}, and \citet{rSet-eSom:per}.}
In this setting, players choose strategies to maximize their own subjective preferences instead of playing pre-programmed strategies,
but their actual fitnesses are determined by material payoff functions which may be distinct from their preferences.
Eventually, evolutionary selection is driven by differences in average fitness values.

In the indirect evolutionary approach literature,
almost every concept of static stability is built on symmetric two-player games played by a single population of players without identifying their player positions.
However, on many occasions, players' roles are fixed and their interactions are asymmetric.
For example, they may be males and females, buyers and sellers, employers and employees, or parents and their children.\footnote{%
An example, the Battle of the Sexes, put forward by \citet{rDaw:sg} is described in Example~\ref{exam:20171114}.}
If individuals always have different roles, then, following the method described by \citet{rSel:nessaac},
asymmetric interactions can be analyzed by
embedding them in a larger symmetric game in which each player is randomly assigned one of the roles,
and then chooses his strategy according to the assigned role;
see \citet{iAlg-jWei:hmpeiiam} and \citet{yHel-eMoh:cdp}.
In this paper, we adopt another approach based on multi-population matching settings:
players are drawn from distinct populations; they may have different action sets and different material payoff functions;
each player knows his player position and has personal preferences over potential outcomes.\footnote{%
Although \citet[p.~61]{svW:eneup} gives a similar definition for two-population models,
he only shows some illustrative examples rather than a general study.}

Recalling the standard evolutionary game theoretic model where individuals are programmed to play some strategies,
there are two quite different ways of extending the definition of an \emph{evolutionarily stable strategy} (ESS) 
from a single-population setting to a multi-population setting.\footnote{%
\citet{jMay-gPri:lac} introduced the concept of an ESS for a symmetric two-player game.}
The multi-population stability criterion suggested by \citet{pTay:essttp} is based on average fitness values aggregated over all player positions.
Such a criterion may be particularly appropriate for coevolutionary games.
\citet{rCre:scegt} introduced a seemingly weaker criterion for multi-population evolutionary stability:
the stability is ensured if entrants earn less in at least one population.
Indeed, it can be shown that both criteria with no intraspecific interactions
are equivalent to the following one:
for any $n$-player game, a strategy profile is \emph{evolutionarily stable} if and only if it is a strict Nash equilibrium.\footnote{%
In \citet{rSel:nessaac}, the definition of an ESS is extended to general two-player games with asymmetric contests.
It turns out that a strategy in the symmetrized game is an ESS if and only if the associated strategy pair is a strict Nash equilibrium of the underlying game.
Thus, those two-species definitions followed from \citet{pTay:essttp} and \citet{rCre:scegt} with no intraspecific interactions
are all equivalent to this role-conditioned single-population definition;
see \citet{jSwi:esee} and \citet[p.~167]{jWei:egt}.}
However, unlike those in standard evolutionary game theory, we will show that under complete information,
the properties of multi-population stability of preferences are strictly dependent on the stability criterion adopted and on the number of populations considered.

\citet{eDek-jEly-oYil:ep} use the indirect evolutionary approach to study endogenous preferences in single-population settings.
They offer two methodological contributions to the work on the evolution of preferences,
namely that they allow for all possible preferences, and they consider various degrees of observability.\footnote{%
\citet{lSam:iep} regards the indirect evolutionary approach as incomplete, 
since only a few possible preferences are considered for applications in some special games,
and those new results always rely on the assumption that preferences are perfectly observable
(see also \citet{aRob-lSam}).}
In this paper,
we extend the static stability criterion of \citet{eDek-jEly-oYil:ep} to the multi-population model
by applying the concept of a two-species ESS introduced in \citet{rCre:scegt}.
This allows us to study endogenous preferences over outcomes in any $n$-player game,
and to analyze the stability characteristics in multi-population matching environments.
It will also be seen that our stability criterion is weaker than the one derived from \citet{pTay:essttp} (or \citet{rSel:nessaac}).

The model we propose has the following features.
There are $n$ large populations, each of which may be \emph{polymorphic}, meaning that not all individuals in a population have the same preferences.
In every match, individuals drawn from distinct populations play an \emph{equilibrium},
which is determined by their subjective preferences and the information about the opponents.
The objective game, whose entries represent the actual fitnesses, may be either symmetric or asymmetric.
The outcome of a matched tuple may not be a Nash equilibrium of the objective game.
In a population, only the preference types which can earn the highest average fitness can survive, that is, the population evolves.
The criteria of evolutionary stability are developed for a \emph{configuration},
which consists of a distribution of types in all $n$ populations and
a particular equilibrium specifying what strategies will be chosen.
We say that a configuration is stable if it simultaneously satisfies two characteristics after an entry of a mutant sub-profile:
every incumbent will be able to survive,
and the post-entry equilibrium behavior can get arbitrarily close to the pre-entry one as long as the mutants are rare enough.
In other words,
the amount of change caused by rare mutants either in the type distribution or in the aggregate behavior can be ignored.

In this paper, all possible preferences are considered as potential mutations.
Thus, under complete information,
there may exist mutants' strategic interactions which look \emph{as if} the mutants play by forming coalitions.
It should be stressed that such a coalition formation mechanism does not exist in our model.
Besides,
for a given strategy, there exist mutants whose preferences make it a dominant strategy.
For a given incumbent type,
there exist mutants who behave like the incumbents, and hence earn the same average fitness.
It follows that we allow the coexistence of incumbents and mutants in a stable configuration,
and that we are interested in the evolutionarily viable outcomes of an objective game,
rather than the emergence of a particular preference type.

As in \citet{eDek-jEly-oYil:ep}, our multi-population stability is defined for every degree of observability.
We begin by studying two extreme cases: in one each player observes the opponents' preferences,
and in the other each player knows only the distribution of opponents' preferences.
We then study intermediate cases,
which allow as to investigate the robustness of the preceding results against small perturbations in the degrees of observability.

Under the assumption that preferences are observable,
the key features of the indirect evolutionary approach are that
players can effectively make commitments by their observed preferences,
and that players can adjust their actions depending on the observed types of their opponents.
Since we allow for all possible preferences in every population,
an inefficient outcome would be destabilized by entrants who have some kind of secret handshake.\footnote{%
\citet{aRob:eegdnsh} demonstrates that any inefficient ESS can be destabilized by the ``secret handshake'' mutant,
which refers to the mutants playing the inefficient outcome when matched against the incumbents and attaining a more efficient outcome when matched against themselves.}
Therefore, under perfect observability,
a configuration is stable only if the outcome of each matched tuple is a Pareto-efficient strategy profile,
rather than a Nash equilibrium.
This result also indicates that an individual endowed with \emph{materialist preferences},
which coincide with fitness maximization, may have no fitness advantage.

In the existing literature on the indirect evolutionary approach,
the ``stable only if efficient'' result is a general property of the single-population models with complete information.\footnote{%
See, for example, \citet{aPos:tsespsg}, \citet{eDek-jEly-oYil:ep}, and \citet{svW:eneup}.}
The value of \emph{efficiency} in a symmetric game is undoubtedly unique.\footnote{%
Efficiency of a strategy in a symmetric two-player game means that no other strategy yields a strictly higher fitness when played against itself.}
In contrast, there may be many Pareto-efficient allocations in an arbitrary game.
Even so, in our polymorphic multi-population model,
the uniqueness of the fitness vector is still ensured for a stable configuration,
in the sense that the action profiles played for all matches always yield the same Pareto-efficient fitness vector.

We will see that it is not easy to draw conclusions about the stability of a Pareto-efficient outcome for general $n$-player games.
When the number of populations is equal to two, we obtain a simple sufficient condition for stability:
under complete information, Pareto-efficient strict Nash equilibria are stable.
However, if the number of populations increases,
mutants may have opportunities to gain evolutionary advantages by applying various multilateral deviations,
regardless of the incumbents' responses,
and so the stability may be difficult to attain.
We present several examples of three-player games which are useful
in helping us understand how mutants interact with one another to destabilize Pareto-efficient strict Nash equilibria,
even though every multilateral deviation will hurt at least one deviator. 
To obtain a sufficient condition for stability in the $n$-population case,
we define a \emph{strict union Nash equilibrium} to be a strategy profile in which
no subgroup of individuals can maintain the sum of their payoffs whenever a deviation occurs in the subgroup.
It implies that not all losses caused by deviations can be compensated from other matches.
In addition to guaranteeing stability in multi-population settings,
it also reveals a connection between the core of a normal-form game and the stable outcome of preference evolution.

We then study the case of unobservable preferences,
where players cannot observe one another's preferences,
and know only the distribution of opponents' preferences in every population.
The interactions among players can be described as an $n$-player Bayesian game.
Since players with no observability cannot adjust their actions depending on specific opponents,
a non-Nash outcome can be destabilized by entrants adopting material payoff-maximizing behavior.
It is also easy to see that under incomplete information, individuals endowed with materialist preferences have fitness advantages.\footnote{%
\citet{eOk-fVeg:eipiis} introduce an evolutionary model to study general preference evolution with no observability.
They find that if the subgroups are relatively small (and the effective matching uncertainty is therefore large),
materialist preferences are stable in a vast set of environments.
For the issue of preference evolution with unobservable preferences, see also \citet{jEly-oYil:neep} and \citet{wGut-bPel:wwpms}.}
Thus, whether a materialist configuration is stable will depend on
whether those incumbents' post-entry strategies are nearly unchanged after rare mutants enter.
We show that a Nash equilibrium of an objective game can be supported by stable materialist preferences if it is
a strict Nash equilibrium or a \emph{completely mixed} Nash equilibrium or the unique Nash equilibrium of the objective game.
Additionally,
the indirect evolutionary approach with no observability can be viewed as a refinement of the Nash equilibrium concept,
weaker than the concepts of \emph{quasi-strict equilibria} and \emph{strictly perfect equilibria}.\footnote{%
Quasi-strict equilibria and strictly perfect equilibria were introduced in \citet{jHar:grdp} and in \citet{aOka:spep}, respectively.}

Finally, we consider the case of partial observability for robustness checks, focused on pure-strategy outcomes.
In the single-population model of \citet{eDek-jEly-oYil:ep},
efficiency is a necessary condition for pure-strategy outcomes to be stable when observability is almost perfect.
However, we provide a counterexample illustrating that the necessity result of the observable case in our multi-population model is not robust.
Efficiency of stable outcomes would be negatively affected by perturbations on observability.
Unlike the efficiency defined for single-population settings, a Pareto improvement is not a change leaving everyone strictly better off.
Therefore, a Pareto-dominated outcome in our model may not be destabilized if preferences are not perfectly observed.
We show that weak Pareto efficiency, instead of Pareto efficiency, is a necessary condition for pure-strategy outcomes to be stable under almost-perfect observability.
Even so, this result still reveals that materialist preferences may have no fitness advantage when preferences are almost perfectly observed.\footnote{%
This is in the same spirit as \citet{rSet-eSom:per} and \citet{aHei-cSha-ySpi:wmiym}.}
In contrast, the necessity result under no observability is robust:
a pure-strategy outcome is stable under almost-no observability only if it is a Nash equilibrium of the objective game.

Regarding the sufficient conditions for stability,
we first show that a Pareto-efficient strict Nash equilibrium of the two-player objective game remains stable for any degree of observability.
When the number of populations is greater than two,
the sufficiency result for the case of observable preferences is also robust:
all strict union Nash equilibria are stable for all degrees of observability.
However, for pure-strategy outcomes,
the sufficiency result of the unobservable case is not robust.\footnote{%
This result is consistent with that in \citet{eDek-jEly-oYil:ep},
who show that strict Nash equilibria might cease to be evolutionarily stable even when preferences are almost unobservable.
\citet{oPar:nep} extends their model by introducing a probability that the observed preferences are not the opponents' actual preferences.
This guarantees that all strict Nash equilibria are stable if the signal a player receives on his opponent's preferences is noisy enough.}
We give an example of a prisoner's dilemma situation in which the outcome of mutual defection,
the strict Nash equilibrium (which is unique),
is not stable for any positive probability of observing preferences.
The use of entrants' cooperative strategies in this example indicates that efficiency would play a role in preference evolution
as long as preferences are not completely unobservable.
In addition, unless preferences are completely unobservable, materialist preferences may not prevail.

The paper proceeds as follows.
We introduce and motivate our multi-population stability concept in Section~\ref{sec:model}.
We analyze the cases in which preferences are observable and unobservable in Sections~\ref{sec:observable} and~\ref{sec:unobservable}, respectively.
In Section~\ref{sec:imp-observ},
we investigate the robustness of the preceding results to the cases of almost-perfect and almost-no observability.
Our results are compared with those of \citet{eDek-jEly-oYil:ep} in the concluding Section~\ref{sec:conclusion},
and all proofs are collected in Appendix~\ref{sec:proofs}.


\section{The Model}
\label{sec:model}


\subsection*{Objective and Subjective Games}

Suppose that $G$ is an $n$-player game with the player set $N = \{1, \dots, n\}$.
For each $i\in N$, denote by $A_i$ the finite set of actions available to player~$i$, and define $A = \prod_{i\in N} A_i$.
Let $\pi_i\colon A\to \mathbb{R}$ be the material payoff function, also called the \emph{fitness function}, for player~$i$.
When an action profile $a\in A$ is played, we interpret the material payoff $\pi_i(a)$ as the \emph{reproductive fitness} received by player~$i$,
which determines the evolutionary success.
We write the set of mixed strategies of player~$i$ as $\Delta(A_i)$, and denote the set of correlated strategies by $\Delta(A)$.
Each $\pi_i$ can be extended by taking expectations to a continuous function defined on the set $\prod_{i\in N}\Delta(A_i)$, or on the set $\Delta(A)$.

By combining the $n$ fitness functions,
we obtain the vector-valued fitness function $\pi\colon A\to\mathbb{R}^n$ that assigns to each action profile $a$ the $n$-tuple $(\pi_1(a), \dots, \pi_n(a))$ of fitness values.
This fitness function $\pi$ can also be extended to the set $\prod_{i\in N}\Delta(A_i)$, or to the set $\Delta(A)$, through $\pi_1$, \dots,~$\pi_n$.
(Note that in the indirect evolutionary approach, behavior of players is independent of the fitness function $\pi$,
although the realized fitnesses are assigned by it.)
We call this underlying game $G$ the \emph{objective game}.


In contrast to the objective game,
a \emph{subjective game} is used to describe strategic interactions among players.
There are $n$ distinct populations, and the number of individuals in each population is imagined to be infinite.
In every game round, players are drawn independently from the $n$ populations, one from each population uniformly at random.
Each player~$i$ chooses an optimal action from the set $\Delta(A_i)$ based on his own preferences and the information about his opponents' preferences.
The actual fitness received by player~$i$ is $\pi_i(\sigma)$ if an action profile $\sigma\in \prod_{i\in N}\Delta(A_i)$ is played.
Let such a game be repeated infinitely many times independently;
then it is plausible that a player will not take into account the effect of his current behavior on the opponents' future behavior.

We assume that the subjective preferences of player~$i$ can be represented by a von Neumann--Morgenstern utility function
which may be different from the material payoff function $\pi_i$.
Let $\Theta = \mathbb{R}^A$, representing the set of von Neumann--Morgenstern utility functions on $A$.
We refer to a utility function in $\Theta$ as a \emph{preference type}, or simply as a \emph{type};
we identify it with a group of players who have such a preference relation and make the same decisions.
Different individuals having the same preference relation may adopt different strategies when they are indifferent among some alternatives.
In such a case, any two of these preferences can be represented by different utility functions congruent modulo a positive affine transformation.

Assume that there are only a finite number of preference types in each population.
Denote by $\mathcal{M}(\Theta^n)$ the set of joint distributions of $n$ independent random variables defined on the same sample space $\Theta$ with finite supports.
Let $\mu\in \mathcal{M}(\Theta^n)$ represent the distribution of types in the $n$ populations.
Then the \emph{support} of $\mu$ can be written as $\supp\mu = \prod_{i\in N}\supp\mu_i$, where $\mu_i$ is the marginal distribution over all types of player~$i$.
For a matched $n$-tuple $\theta\in \supp\mu$ and for $k\in N$, the conditional probability $\mu(\theta_{-k}|\theta_k)$ is equal to $\prod_{i\neq k}\mu_i(\theta_i)$.
For notational simplicity, we write $\mu_{-k}(\theta_{-k})$ and $\supp\mu_{-k}$ for $\mu(\theta_{-k}|\theta_k)$ and $\prod_{i\neq k}\supp\mu_i$, respectively.
Similarly, we also write $\supp\mu_{-T}$ for $\prod_{i\notin T}\supp\mu_i$, where $T$ is a nonempty proper subset of $N$.

Regarding the information about opponents' types,
we follow \citet{eDek-jEly-oYil:ep}, hereafter~DEY;
we assume that each player~$i$ observes the opponents' preferences with probability $p\in [0, 1]$,
and knows only the joint distribution $\mu_{-i}$ over the opponents' preferences with probability $1 - p$.\footnote{%
Here, partial observability is used to model the noise in the two extreme cases, perfect observability and no observability.
For simplicity, we ignore the possibility that an individual
has complete information about the preferences of some of the opponents and has incomplete information about the preferences of the others.
We emphasize that the difference in the two kinds of noise settings does not affect our results.}
The degree of observability $p$ is an exogenous parameter indicating the level of observation, and is common knowledge among all players.
But the two realizations, called \emph{perfect observability} and \emph{no observability}, are private and independent across players.
In every match, each player chooses his best response to the expected behavior of others under a given degree of observability.
Such an $n$-player Bayesian game (which we will use to derive behavioral strategies)
is denoted by $\Gamma_p(\mu)$ and called a \emph{subjective game}; the pair $(G,\Gamma_p(\mu))$ is referred to as an \emph{environment}.

\subsection*{The Stability Concept}

As in much of the evolutionary literature,
we present a static solution concept to capture the stable states of the multi-population evolutionary processes.
According to the principle of the ``survival of the fittest'', only the preference types earning the highest average fitness will survive.
Hence, in a given environment $(G,\Gamma_p(\mu))$,
a necessary condition for evolutionary stability is that for every $i\in N$,
all \emph{incumbents} in the $i$-th population, which constitute the set $\supp\mu_i$, should receive the same average fitness.
Besides, it is necessary to verify whether the incumbents are immune to competition from rare entrants.
To do so, we assume that at the same time, there exists at most one mutant type arising in each of the $n$ populations.
Let $J = \{i_1,\dots,i_k\}$ be an arbitrary nonempty subset of $N$. A \emph{mutant sub-profile} for $J$, denoted by $\witilde{\theta}_J$,
refers to a $k$-tuple of preference types $(\witilde{\theta}_{i_1},\dots,\witilde{\theta}_{i_k})$ derived from $\prod_{j\in J} \cplm{(\supp\mu_j)}$,
where $\cplm{(\supp\mu_j)}$ is the complement of $\supp\mu_j$.\footnote{%
It indicates that we allow for all preference relations, which satisfy the von Neumann--Morgenstern axioms,
and that mutants are distinguishable from the incumbents in the post-entry populations,
although both may have the same preference relation.}
The vector $(\varepsilon_{i_1},\dots,\varepsilon_{i_k})$ of the population shares of the $k$ mutant types is often denoted simply by $\varepsilon$,
and we define its \emph{norm} to be $\|\varepsilon\| = \max\{ \varepsilon_{i_1}, \dots, \varepsilon_{i_k} \}$.\footnote{%
Since each population is assumed to be infinite,
the population share of a mutant type can hypothetically take on any positive value,
no matter how small it may be.}
After the mutants have entered, the resulting $n$ populations can be described by the post-entry type distribution $\witilde{\mu}^{\varepsilon}$:
\[
\witilde{\mu}^{\varepsilon}_i =
\begin{cases}
    (1-\varepsilon_i)\mu_i+\varepsilon_i\delta_{\witilde{\theta}_i} & \text{if $i\in J$,}\\
    \mu_i & \text{if $i\in N\setminus J$,}
\end{cases}
\]
where $\delta_{\witilde{\theta}_i}$ is the degenerate distribution concentrated at $\witilde{\theta}_i$.

In a single-population evolutionary model,
the key requirement for static stability is that invading mutants do not outperform the incumbents.
For multi-population settings,
it would be natural to extend the stability definition by allowing the entrant to be a mutant sub-profile,
which we regard as a \emph{unit} of mutation.
We say that a mutant sub-profile fails to invade if one of the mutant types will become extinct.
In other words, a multi-population stability criterion would be fulfilled if
for any entry of a mutant sub-profile, there are mutants earning a lower average fitness than the incumbents in at least one population.
The reason for this is that if some of the mutants gain fitness advantages at the expense of some other mutants,
then these fitness advantages will eventually disappear when the latter go extinct.
This concept is consistent with that of the multi-population ESS formulated by \citet{rCre:scegt},
and yet distinct from the concept introduced in \citet{pTay:essttp}.\footnote{%
For introductions to the popular multi-population ESS concepts introduced in \citet{pTay:essttp} and \citet{rCre:scegt},
see \citet[p.~166]{jWei:egt} and \citet[p.~280]{wSan:pged}.}

More precisely,
our stability criteria are defined for a \emph{configuration} consisting of a joint type distribution and a Bayesian--Nash equilibrium.
In addition to the incumbents in the same population earning the same average fitness,
a stable multi-population configuration should satisfy: after a rare mutant sub-profile appears,
\begin{itemize}
\item the behavioral outcomes remain unchanged or nearly unchanged;
\item the mutant sub-profile fails to invade, or the incumbents coexist with the mutants in every population.
\end{itemize}
We assume that the incumbents in a post-entry environment
would tend to adopt the strategies as close as possible to the pre-entry ones
if each would believe that the others in a match are likely to do the same.
Therefore, when their opponents are very likely to be other incumbents,
they attempt to keep their strategies unchanged.
We also allow mutants to coexist with the incumbents in a stable configuration,
because there are no restrictions on the preference relations of entrants and the best-response correspondences for different preferences may coincide.\footnote{%
The idea of coexistence is consistent with the concept of a \emph{neutrally stable strategy},
which was introduced in \citet[p.~107]{jMay:etg}.}
According to our multi-population stability criterion,
the failure of a mutant sub-profile can be determined solely by one mutant type among them.
Even so,
such multi-population extension quite satisfies the appropriate stability conditions as introduced in DEY:
there is no single mutant sub-profile that can obviously destabilize the configuration including
the behavioral outcomes and the distribution of preference types (see Remarks~\ref{rem:170815} and~\ref{rem:170816}).

We shall formally define stability criteria for various information structures in the following sections;
these definitions follow the same principles of multi-population evolutionary stability.
The criteria defined under perfect observability and no observability are the two limiting cases of those defined under partial observability,
as the degree of observability approaches $1$ and $0$, respectively.
It will be confirmed that the degree of observability can decisively affect the outcome of preference evolution.


\section{Perfect Observability}
\label{sec:observable}


In this section we discuss the case where the degree of observability is equal to one, that is,
players can always observe one another's preferences.
Therefore,
the interactions among players in each game round can be seen as an $n$-player game in normal form.

Let $\mu\in \mathcal{M}(\Theta^n)$.
A \emph{strategy} for player~$i$ under perfect observability is a function $b_i\colon\supp\mu\to\Delta(A_i)$.\footnote{%
As discussed in Section~\ref{sec:model},
it is admitted throughout the paper that players having the same preference relation may adopt different strategies.}
The function $b\colon \supp\mu\to \prod_{i\in N}\Delta(A_i)$
defined by $b(\theta)=(b_1(\theta),\dots,b_n(\theta))$ is an \emph{equilibrium} in the subjective game $\Gamma_1(\mu)$
if for each $\theta\in\supp\mu$, the action profile $b(\theta)$ is a Nash equilibrium of the corresponding normal-form game played by $\theta$,
that is, for each $i\in N$,
\[
b_i(\theta)\in\argmax_{\sigma_i\in\Delta(A_i)}\theta_i\big( \sigma_i, b(\theta)_{-i} \big).
\]
Let $B_1(\mu)$ denote the set of all such equilibria in $\Gamma_1(\mu)$.
A pair $(\mu, b)$ consisting of a type distribution $\mu\in \mathcal{M}(\Theta^n)$ and an equilibrium $b\in B_1(\mu)$ is called a \emph{configuration}.
We define the \emph{aggregate outcome} of a configuration $(\mu,b)$ to be the probability distribution $\varphi_{\mu,b}$ over the set of pure-strategy profiles:
\[
\varphi_{\mu,b}(a_1, \dots, a_n) = \sum_{\theta\in \supp\mu} \mu(\theta) \prod_{i\in N} b_i(\theta)(a_i).
\]
Then the aggregate outcome $\varphi_{\mu,b}$ can be regarded as a \emph{correlated strategy} belonging to $\Delta(A)$.
A strategy profile $\sigma\in\prod_{i\in N}\Delta(A_i)$ is called an \emph{aggregate outcome}
if the correlated strategy $\varphi_{\sigma}$ induced by $\sigma$ is the aggregate outcome of some configuration,
where $\varphi_{\sigma}$ is defined by
$\varphi_{\sigma}(a_1, \dots, a_n) = \prod_{i\in N} \sigma_i(a_i)$ for all $(a_1, \dots, a_n)\in A$.\footnote{%
An aggregate outcome generated by a strategy profile is used in the case of \emph{monomorphic} configurations
for which individuals in the same population are endowed with the same type.}

According to the law of large numbers,
the \emph{average fitness} of a preference type $\theta_i\in\supp\mu_i$ with respect to $(\mu,b)$ is measured by
\[
\varPi_{\theta_i}(\mu;b)=\sum_{\theta'_{-i}\in\supp\mu_{-i}}
\mu_{-i}(\theta'_{-i})\pi_i\big( b(\theta_i,\theta'_{-i}) \big),
\]
on which the evolution of preferences depends.\footnote{%
The equation for average fitness indicates that the type distribution is unchanged in the process of learning to play an equilibrium.
To justify this representation,
we assume as in most related literature that the evolution of preferences is infinitely slower than the process of learning, which is supported by \citet[p.~21]{rSel:eleb}.}
A configuration $(\mu,b)$ is said to be \emph{balanced} if all incumbents in the same population have the same average fitness, that is,
for each $i\in N$,
the equality $\varPi_{\theta_i}(\mu;b)=\varPi_{\theta'_i}(\mu;b)$ holds for every $\theta_i$,~$\theta'_i\in \supp\mu_i$.

Since incumbents will maintain their behavioral strategies as much as possible in a familiar environment,
we assume that after an entry,
incumbents have no intention of changing their behavior against incumbent opponents, called the \emph{focal property}.
However,
it seems implausible that we can say which equilibrium will be played when
not all players in a matched tuple are incumbents.
Hence, in such a match, there are no restrictions on the players' equilibrium behavior.

\begin{Def}\label{def:focal-p=1}
Let $(\mu,b)$ be a configuration under perfect observability,
and suppose that $\witilde{\mu}^{\varepsilon}$ is a post-entry type distribution.
In $\Gamma_1(\witilde{\mu}^{\varepsilon})$,
an equilibrium $\witilde{b}\in B_1(\witilde{\mu}^{\varepsilon})$ is said to be \emph{focal} relative to $b$
if $\witilde{b}(\theta)=b(\theta)$ for every $\theta\in\supp\mu$.
Let $B_1(\witilde{\mu}^{\varepsilon};b)$ denote the set of all focal equilibria relative to $b$ in $\Gamma_1(\witilde{\mu}^{\varepsilon})$,
called the \emph{focal set}.
\end{Def}

\begin{Rem}\label{rem:170815}
When preferences are observable,
the focal set $B_1(\witilde{\mu}^{\varepsilon};b)$ is always nonempty,
and its elements are independent of the population share vector $\varepsilon$.
In addition, the desired property that $\varphi_{\witilde{\mu}^{\varepsilon},\witilde{b}}\to \varphi_{\mu,b}$ as $\|\varepsilon\|\to 0$
is naturally held for any equilibrium $\witilde{b}\in B_1(\witilde{\mu}^{\varepsilon};b)$,
which means that rare mutants cannot cause the behavioral outcomes to move far away.
\end{Rem}

Now the stability criterion for a perfectly observable environment can be defined.

\begin{Def}\label{def:stable-p=1}
In $(G,\Gamma_1(\mu))$,
a configuration $(\mu,b)$ is said to be \emph{stable} if
for any nonempty subset $J\subseteq N$ and any mutant sub-profile $\witilde{\theta}_J\in \prod_{j\in J}\cplm{(\supp\mu_j)}$,
there exists some $\bar{\epsilon}\in (0,1)$ such that for every $\varepsilon\in (0,1)^{|J|}$ with $\|\varepsilon\| \in (0,\bar{\epsilon})$
and every $\witilde{b}\in B_1(\witilde{\mu}^{\varepsilon};b)$, either Condition~\ref{C1:wiped-out} or Condition~\ref{C1:coexist} is satisfied:
\begin{enumerate}[label=(\roman*)]
\item $\varPi_{\theta_j}(\witilde{\mu}^{\varepsilon};\witilde{b}) > \varPi_{\witilde{\theta}_j}(\witilde{\mu}^{\varepsilon};\witilde{b})$ for some $j\in J$ and for every $\theta_j\in\supp\mu_j$;\label{C1:wiped-out}
\item $\varPi_{\theta_i}(\witilde{\mu}^{\varepsilon};\witilde{b}) = \varPi_{\wihat{\theta}_i}(\witilde{\mu}^{\varepsilon};\witilde{b})$
      for every $i\in N$ and for every $\theta_i$,~$\wihat{\theta}_i\in \supp\witilde{\mu}^{\varepsilon}_i$.\label{C1:coexist}
\end{enumerate}
A strategy profile $\sigma\in\prod_{i\in N}\Delta(A_i)$ is \emph{stable} if it is the aggregate outcome of some stable configuration.
\end{Def}

Condition~\ref{C1:wiped-out} indicates the case where one of the mutant types will become extinct,
and thus this mutant sub-profile fails to invade.
Condition~\ref{C1:coexist} indicates the case where the incumbents and the mutants will continue to coexist in every population.
Here, the ``balanced'' condition for incumbents in the pre-entry configuration is not explicitly stated,
since a stable configuration satisfying Definition~\ref{def:stable-p=1} must be balanced, which will be shown in Theorem~\ref{prop:120706}.

\smallskip

Let us for a moment study the multi-population stability criterion for preference evolution
based on the two-species ESS concept of \citet{pTay:essttp},
where the fitness comparison between incumbents and mutants is done in the aggregate.
For simplicity,
consider a monomorphic configuration $(\mu, b)$,
and suppose that an $n$-tuple $(\witilde{\theta}_1, \dots, \witilde{\theta}_n)$ enters with its population share vector $\varepsilon$.
Then, in this view,
Condition~\ref{C1:wiped-out} should be replaced by
\begin{equation}\label{eq:20190925}
\sum_{i\in N} \varPi_{\theta_i}(\witilde{\mu}^{\varepsilon};\witilde{b})> \sum_{i\in N} \varPi_{\witilde{\theta}_i}(\witilde{\mu}^{\varepsilon};\witilde{b}).
\end{equation}
Unlike extending the definition of an ESS to a multi-population setting,
such a stability criterion is indeed stronger than ours under perfect observability.
The key difference is the ability to adjust one's actions depending on the opponents;
see the discussion after Theorem~\ref{prop:0526} for details.
Besides, as in \citet{rSel:nessaac}, asymmetric interactions can also be captured by embedding interactions in a larger symmetric game
in which individuals are randomly assigned player roles, with all role assignments being equally likely.
In this way, the single-population stability criterion of DEY can be extended to general games with asymmetric contests.
An individual of type $\theta$ in the symmetrized game is represented by a type pair $(\theta_{\mathrm{I}}, \theta_{\mathrm{II}})$,
and the fitness to $\theta$ when playing against an opponent of type $\witilde{\theta}$ is
\[
\frac{1}{2} \big[ \pi_1\big( \witilde{b}(\theta_{\mathrm{I}}, \witilde{\theta}_{\mathrm{II}}) \big) +
\pi_2\big( \witilde{b}(\witilde{\theta}_{\mathrm{I}}, \theta_{\mathrm{II}}) \big) \big],
\]
where $\witilde{b}$ is a post-entry equilibrium.
Let the population share of the mutant type $\witilde{\theta}$ be $\varepsilon$.
Then after $\witilde{\theta}$ has entered,
the condition that the average fitness of $\theta$ is greater than that of $\witilde{\theta}$ can be written as
\begin{multline*}
\big[ (1 - \varepsilon) \pi_1\big( \witilde{b}(\theta_{\mathrm{I}}, \theta_{\mathrm{II}}) \big) +
\varepsilon \pi_1\big( \witilde{b}(\theta_{\mathrm{I}}, \witilde{\theta}_{\mathrm{II}}) \big) \big] +
\big[ (1 - \varepsilon) \pi_2\big( \witilde{b}(\theta_{\mathrm{I}}, \theta_{\mathrm{II}}) \big) +
\varepsilon \pi_2\big( \witilde{b}(\witilde{\theta}_{\mathrm{I}}, \theta_{\mathrm{II}}) \big) \big]\\
>
\big[ (1 - \varepsilon) \pi_1\big( \witilde{b}(\witilde{\theta}_{\mathrm{I}}, \theta_{\mathrm{II}}) \big) +
\varepsilon \pi_1\big( \witilde{b}(\witilde{\theta}_{\mathrm{I}}, \witilde{\theta}_{\mathrm{II}}) \big) \big] +
\big[ (1 - \varepsilon) \pi_2\big( \witilde{b}(\theta_{\mathrm{I}}, \witilde{\theta}_{\mathrm{II}}) \big) +
\varepsilon \pi_2\big( \witilde{b}(\witilde{\theta}_{\mathrm{I}}, \witilde{\theta}_{\mathrm{II}}) \big) \big],
\end{multline*}
which is almost exactly the same as \eqref{eq:20190925}.
Therefore, the two criteria derived from \citet{pTay:essttp} and \citet{rSel:nessaac} are almost exactly equivalent.\footnote{%
If the population share vector $\varepsilon$ appearing in \eqref{eq:20190925} satisfies $\varepsilon_1 = \dots =\varepsilon_n$,
then these two stability criteria are equivalent in the case $n = 2$.}

\begin{Rem}
We can apply the concept of neutral stability to the multi-population ESS concepts
introduced in \citet{pTay:essttp} and \citet{rCre:scegt}, respectively, as follows.
A strategy profile $\sigma^*$ is a \emph{Taylor NSS} if for every strategy profile $\sigma$,
the weak inequality
$\sum_{i\in N} u_i(\sigma^*_i, x_{-i})\geq \sum_{i\in N} u_i(\sigma_i, x_{-i})$
holds for all sufficiently small $\varepsilon> 0$, where $x = (1 - \varepsilon)\sigma^* + \varepsilon\sigma$.
We call $\sigma^*$ a \emph{Cressman NSS} if for every $\sigma$,
either $u_i(\sigma^*_i, x_{-i})> u_i(\sigma_i, x_{-i})$ for some $i\in N$
or $u_i(\sigma^*_i, x_{-i}) = u_i(\sigma_i, x_{-i})$ for all $i\in N$.
Since it is well known that the two multi-population ESS concepts are consistent,
it is easy to see that the two definitions of Taylor NSS and Cressman NSS coincide.
\end{Rem}

\smallskip

The following remarks describe the basic characteristics of each of our multi-population stability criteria,
which are defined separately for each information assumption.

\begin{Rem}\label{rem:140306}
In our stability criteria, the invasion barrier $\bar{\epsilon}$ seems to depend on the mutant sub-profile.
In fact, the existence of such a barrier is equivalent to the existence of a uniform barrier against all mutant sub-profiles.
Consider an \emph{indifferent} mutant sub-profile $\witilde{\theta}^0_J$ for a nonempty subset $J$ of $N$.\footnote{%
A preference type is said to be \emph{indifferent} if it is a constant utility function;
a mutant sub-profile is said to be \emph{indifferent} if all its preference types are indifferent.}
Since an indifferent type $\witilde{\theta}^0_j$ is indifferent among all actions, all available actions will be dominant for him.
Hence, by the condition that all possible equilibrium strategies for the entrants are admitted in a post-entry environment,
the barrier $\bar{\epsilon}_J$ which works for $\witilde{\theta}^0_J$ is certainly a uniform barrier against all potential mutant sub-profiles for the subset $J$.
Thus, since the number of all subsets of $N$ is finite, we have a uniform invasion barrier that can work for all potential mutant sub-profiles.
\end{Rem}

\begin{Rem}\label{rem:170816}
When the multi-population stability criterion is satisfied, no incumbent would be wiped out after an entry of a mutant sub-profile,
although Condition~\ref{C1:wiped-out} may be determined just by one of the mutant types.
The intuition behind this is as follows.
Let $(\mu,b)$ be a stable configuration in the sense of Definition~\ref{def:stable-p=1},
and suppose that Condition~\ref{C1:wiped-out} holds after a mutant sub-profile $\witilde{\theta}_J$ is introduced.
Then those mutants earning the lowest average fitness will be wiped out.
Such a trend can make $\witilde{\theta}_J$ converge to a smaller sub-profile $\witilde{\theta}_{J'}$.
Meanwhile, there may be slight changes in the population shares of the incumbents during this process.
One could regard the next entire population status as this:
the mutant sub-profile $\witilde{\theta}_{J'}$ tries its luck in the incumbent configuration which may be slightly perturbed just recently.
Recall that in the paper, a stable configuration is defined by considering all possible mutant sub-profiles.
Thus, as after the entry of $\witilde{\theta}_{J'}$, either Condition~\ref{C1:wiped-out} or Condition~\ref{C1:coexist} should be satisfied,
provided that the population shares of those mutants forming $\witilde{\theta}_{J'}$ are sufficiently small
and that the slight changes in the population shares of the incumbents do not affect the rank order of their average fitness values.
Such a repeated process would lead to the desired goal.\footnote{%
Especially in a stable polymorphic configuration,
occasional mutations may cause the population to drift between the incumbents.
For more on the idea of drift, see \citet{kBin-lSam:d}.}
\end{Rem}

The Battle of the Sexes described in \citet{rDaw:sg} is an example concerning the male-female conflict over parental care of offspring.
Males can be either faithful or philandering, and females can be either coy or fast.
In this game, there is a unique Nash equilibrium in completely mixed strategies,
and thus there is no ESS, as discussed in \citet[p.~243]{vDam:spne}.
Furthermore, we can see that the game admits no neutrally stable strategy in the sense of \citet{rSel:nessaac},
and therefore no \emph{uniform limit ESS}.\footnote{%
\citet{yHel:steg} introduced the concept of a uniform limit ESS.
In fact, Dawkins' Battle of the Sexes game also admits no \emph{limit ESS}, which was introduced in \citet{rSel:esetg}.}
However, if the Battle of the Sexes is modeled by means of our multi-population indirect evolutionary approach,
then an evolutionarily stable outcome may appear in this game, as shown in Example~\ref{exam:20171114}.
It suggests that when preferences are perfectly observable,
the Pareto-efficient strategy pair $(\text{Faithful}, \text{Fast})$ is stable
with two monomorphic populations consisting, respectively, of male and female individuals of the following types.
In the male population,
a male has no choice but to be faithful as long as his opponent may be a coy female.
In the female population,
a female is never fast unless her opponent is a faithful male with probability $1$.\footnote{%
\citet{jMcN-lFro-zBar-aHou:ocg} develop a comprehensive game theory model
that combines an explicit model of future behavior with a model of optimal female coyness.
They show that if brood success without male help is very low,
or if the ratio of males to females is high enough,
then there exists a unique stable outcome in which all males are helpful and all females are fast.}

\begin{example}\label{exam:20171114}
Let preferences be observable,
and let the fitness assignment be represented by Dawkins' Battle of the Sexes game as shown below.

\begin{table}[!h]
\centering
\renewcommand{\arraystretch}{1.2}
    \begin{tabular}{r|c|c|}
      \multicolumn{1}{c}{} & \multicolumn{1}{c}{Coy} & \multicolumn{1}{c}{Fast}\\ \cline{2-3}
      Faithful &     $2$, $2$ & $\phantom{1}5$, $\phantom{-}5$ \\  \cline{2-3}
      Philandering & $0$, $0$ &           $15$,           $-5$ \\  \cline{2-3}
    \end{tabular}
\end{table}

\noindent
We imagine that the first population consists of all males, and the second population consists of all females.
Suppose that all males have preferences such that
they are faithful if a female is coy,
and they are indifferent between being faithful and being philandering if a female is fast.
Next, suppose that females' preferences prompt them to be indifferent between being coy and being fast if a male is faithful,
and to be coy if a male is a philanderer.
Then the pair $(\text{Faithful}, \text{Fast})$ is a Nash equilibrium for such males and females.
Let this strategy pair be chosen.
We claim that the outcome $(\text{Faithful}, \text{Fast})$ can be stable in this setting.

To see this, let $(\mu, b)$ be the configuration constructed as described above.
It is obvious that if mutants appear in only one of the two populations, their average fitnesses cannot be greater than what the incumbents have.
Consider two types of mutants $\witilde{\theta}_1$ and $\witilde{\theta}_2$ entering the first and second populations
with population shares $\varepsilon_1$ and $\varepsilon_2$, respectively.
Let $\witilde{b}$ be the played focal equilibrium, and suppose that for $i = 1$,~$2$ and for $\theta_{-i}\in \supp\mu_{-i}$,
the action $\witilde{b}_i(\witilde{\theta}_i, \theta_{-i})$ is $(q_i, 1 - q_i)$ where $0\leq q_i\leq 1$.\footnote{%
For two-player games throughout the paper,
the subscript $-i$ on $\theta_{-i}$, $\witilde{\theta}_{-i}$, or $\varepsilon_{-i}$
denotes the population index $j$ with $j\in \{1, 2\}\setminus \{i\}$,
and the pair $(\witilde{\theta}_i, \theta_{-i})$, for example,
refers to the ordered pair consisting of two preference types arranged in ascending order according to their population indices.}
If $q_1\neq 1$, then the post-entry average fitnesses of $\theta_1$ and $\witilde{\theta}_1$ satisfy
$\varPi_{\theta_1}(\witilde{\mu}^{\varepsilon};\witilde{b})\geq 5(1 - \varepsilon_2) + 2\varepsilon_2$ and
$\varPi_{\witilde{\theta}_1}(\witilde{\mu}^{\varepsilon};\witilde{b})\leq 2q_1(1 - \varepsilon_2) + 15\varepsilon_2$, respectively.
Hence we have
$\varPi_{\theta_1}(\witilde{\mu}^{\varepsilon};\witilde{b})> \varPi_{\witilde{\theta}_1}(\witilde{\mu}^{\varepsilon};\witilde{b})$ whenever $\varepsilon_2< 3/16$.
If $q_1 = 1$ and $q_2\neq 0$,
then the post-entry average fitness of $\theta_2$ is strictly greater than that of $\witilde{\theta}_2$ as long as $\varepsilon_1$ is small enough.
Finally, suppose that $q_1 = 1$ and $q_2 = 0$.
If there are male or female incumbents earning a fitness not equal to $5$ when matched against the mutants,
then there are mutants earning strictly lower average fitness as long as the mutants are rare enough.
Thus we only have to let the fitness to each individual in each match be $5$ except in the pair $(\witilde{\theta}_1, \witilde{\theta}_2)$.
When $\witilde{\theta}_1$ and $\witilde{\theta}_2$ are matched together,
the Pareto efficiency of the strategy pair $(\text{Faithful}, \text{Fast})$ implies that
the conditions affording mutants a fitness advantage in the first population can result in a loss to mutants in the second population.
To finish the proof of the claim, we have to find a uniform invasion barrier $\bar{\epsilon}$ which can work for all possible focal equilibria,
even though those actions yield the mutants' fitnesses arbitrarily close to $5$ when matched against the incumbents.
Now let us consider the two cases.
\vspace{5pt}

\noindent
\textbf{Case 1}: $q_1 = 1$ and $q_2\neq 0$.
Let $\witilde{b}_2(\witilde{\theta}_1, \theta_2) = (s_2, 1 - s_2)$ where $0\leq s_2\leq 1$.
For the pair $(\witilde{\theta}_1, \witilde{\theta}_2)$,
it is enough to consider the mutants' fitnesses on the Pareto frontier of the payoff region,
that is, we suppose that $\witilde{b}(\witilde{\theta}_1, \witilde{\theta}_2) = ((r_1, 1 - r_1), (0,1))$ where $0\leq r_1\leq 1$.
Then $\varPi_{\theta_1}(\witilde{\mu}^{\varepsilon};\witilde{b})> \varPi_{\witilde{\theta}_1}(\witilde{\mu}^{\varepsilon};\witilde{b})$
if and only if
$3s_2> [ 3s_2 + 3q_2 + 10(1 - r_1)]\varepsilon_2$, and
$\varPi_{\theta_2}(\witilde{\mu}^{\varepsilon};\witilde{b})> \varPi_{\witilde{\theta}_2}(\witilde{\mu}^{\varepsilon};\witilde{b})$
if and only if
$3q_2> [ 3s_2 + 3q_2 - 10(1 - r_1)]\varepsilon_1$.
When $s_2 = 0$, it is clear that
$\varPi_{\theta_2}(\witilde{\mu}^{\varepsilon};\witilde{b})> \varPi_{\witilde{\theta}_2}(\witilde{\mu}^{\varepsilon};\witilde{b})$ for all $\varepsilon_1< 1$.
Let $s_2\neq 0$, and suppose that we need $\varepsilon_1$ arbitrarily close to $0$ to get
$\varPi_{\theta_2}(\witilde{\mu}^{\varepsilon};\witilde{b})> \varPi_{\witilde{\theta}_2}(\witilde{\mu}^{\varepsilon};\witilde{b})$
as $q_2$ tends to $0$.
This implies that $3s_2> 10(1 - r_1)$ and $q_2/s_2$ tends to $0$ when $q_2$ tends to $0$.
Note that the inequality
$\varPi_{\theta_1}(\witilde{\mu}^{\varepsilon};\witilde{b})> \varPi_{\witilde{\theta}_1}(\witilde{\mu}^{\varepsilon};\witilde{b})$ holds whenever
\[
\varepsilon_2< \frac{3s_2}{3s_2 + 3q_2 + 10(1 - r_1)} = \frac{1}{1 + \frac{q_2}{s_2} + \frac{10(1 - r_1)}{3s_2}}.
\]
Thus to get
$\varPi_{\theta_1}(\witilde{\mu}^{\varepsilon};\witilde{b})> \varPi_{\witilde{\theta}_1}(\witilde{\mu}^{\varepsilon};\witilde{b})$ in this case,
it does not simultaneously compel the population share $\varepsilon_2$ to be arbitrarily close to $0$ as $q_2$ tends to $0$.
\vspace{5pt}

\noindent
\textbf{Case 2}: $q_1 = 1$ and $q_2 = 0$.
For $i = 1$,~$2$ and for $\theta_i\in \supp\mu_i$, let $\witilde{b}_i(\theta_i, \witilde{\theta}_{-i}) = (s_i, 1 - s_i)$ where $0\leq s_i\leq 1$.
For the pair $(\witilde{\theta}_1, \witilde{\theta}_2)$, it is enough to consider
$\witilde{b}(\witilde{\theta}_1, \witilde{\theta}_2) = ((r_1, 1 - r_1), (0,1))$ where $0\leq r_1\leq 1$.
Then $\varPi_{\theta_1}(\witilde{\mu}^{\varepsilon};\witilde{b})> \varPi_{\witilde{\theta}_1}(\witilde{\mu}^{\varepsilon};\witilde{b})$
if and only if
$3s_2> ( 3s_2 + 10s_1 - 10r_1 )\varepsilon_2$, and
$\varPi_{\theta_2}(\witilde{\mu}^{\varepsilon};\witilde{b})> \varPi_{\witilde{\theta}_2}(\witilde{\mu}^{\varepsilon};\witilde{b})$
if and only if
$10 - 10s_1> ( 3s_2 + 10r_1 - 10s_1 )\varepsilon_1$.
When $s_2 = 0$, it is easy to see that
$\varPi_{\theta_2}(\witilde{\mu}^{\varepsilon};\witilde{b})> \varPi_{\witilde{\theta}_2}(\witilde{\mu}^{\varepsilon};\witilde{b})$ for all $\varepsilon_1< 1$.
When $s_1 = 1$ and $s_2\neq 0$, the inequality
$\varPi_{\theta_1}(\witilde{\mu}^{\varepsilon};\witilde{b})> \varPi_{\witilde{\theta}_1}(\witilde{\mu}^{\varepsilon};\witilde{b})$
holds as long as $\varepsilon_2< \frac{3s_2}{3s_2 + 10(1 - r_1)}$.
Furthermore, if we let $3s_2\geq 10(1 - r_1)$,
then the population share $\varepsilon_2$
is not compelled to be arbitrarily close to $0$, although $s_2$ tends to $0$.
If $3s_2< 10(1 - r_1)$,
then
$\varPi_{\theta_2}(\witilde{\mu}^{\varepsilon};\witilde{b})> \varPi_{\witilde{\theta}_2}(\witilde{\mu}^{\varepsilon};\witilde{b})$
for all $\varepsilon_1< 1$.
Finally, let $s_1\neq 1$ and $s_2\neq 0$, and suppose that we need $\varepsilon_1$ arbitrarily close to $0$ to get
$\varPi_{\theta_2}(\witilde{\mu}^{\varepsilon};\witilde{b})> \varPi_{\witilde{\theta}_2}(\witilde{\mu}^{\varepsilon};\witilde{b})$
as $s_1$ tends to $1$.
Then it must be that $3s_2> 10(s_1 - r_1)$ when $s_1$ tends to $1$.
Note that $\varPi_{\theta_1}(\witilde{\mu}^{\varepsilon};\witilde{b})> \varPi_{\witilde{\theta}_1}(\witilde{\mu}^{\varepsilon};\witilde{b})$
whenever $1> ( 1 + \frac{10(s_1 - r_1)}{3s_2} )\varepsilon_2$.
Thus to get
$\varPi_{\theta_1}(\witilde{\mu}^{\varepsilon};\witilde{b})> \varPi_{\witilde{\theta}_1}(\witilde{\mu}^{\varepsilon};\witilde{b})$ in this case,
it is not necessary to simultaneously compel the population share $\varepsilon_2$ to be arbitrarily close to $0$ as $s_1$ tends to $1$.
\end{example}

The stable outcome $(\text{Faithful}, \text{Fast})$ in Dawkins' Battle of the Sexes game is a Pareto-efficient strategy profile.
Our first result will show that this remains true for general $n$-player games:
if a configuration is stable under perfect observability,
then the outcome of any match played by $n$ incumbents must be Pareto efficient with respect to the fitness function $\pi$.
The reason is simple.
If the outcome is not Pareto efficient, a tuple of mutant types with the ``secret handshake'' flavor can destabilize this inefficient outcome.
These mutants have preferences that enable them to maintain the pre-entry outcome when playing against the incumbents,
and achieve a more efficient outcome when playing against themselves.
Accordingly, the observability of preferences plays a key role in obtaining the ``stable only if efficient'' result.
To avoid confusion, we list the definitions concerning Pareto efficiency.

\begin{Def}
Let $(N, A, \pi)$ be a finite normal-form game.
A strategy profile $\sigma$ \emph{strongly Pareto dominates} a strategy profile $\sigma'$ if $\pi_i(\sigma) > \pi_i(\sigma')$ for all $i\in N$.
A strategy profile $\sigma$ is \emph{weakly Pareto efficient} if there does not exist another strategy profile that strongly Pareto dominates $\sigma$.
A strategy profile $\sigma$ \emph{Pareto dominates} a strategy profile $\sigma'$ if
$\pi_i(\sigma)\geq \pi_i(\sigma')$ for all $i\in N$ and $\pi_j(\sigma)>\pi_j(\sigma')$ for some $j\in N$.
A strategy profile $\sigma$ is \emph{Pareto efficient} if there does not exist another strategy profile that Pareto dominates $\sigma$.
\end{Def}

\begin{theoremEnd}[no link to theorem, restate]{thm}\label{prop:003}
Let $(\mu,b)$ be a stable configuration in $(G,\Gamma_1(\mu))$.
Then for each $\theta\in\supp\mu$, the outcome $b(\theta)$ is Pareto efficient with respect to $\pi$.
\end{theoremEnd}

\begin{proofEnd}
Suppose that there exists $\bar{\theta}\in\supp\mu$ such that $b(\bar{\theta})$ is not Pareto efficient, that is,
there exists $\sigma\in\prod_{i\in N}\Delta(A_i)$ such that
$\pi_i(\sigma)\geq \pi_i\big(b(\bar{\theta})\big)$ for all $i\in N$ and
$\pi_j(\sigma) > \pi_j\big(b(\bar{\theta})\big)$ for some $j\in N$.
Let an indifferent mutant profile $\witilde{\theta}^0 = (\witilde{\theta}^0_1, \dots, \witilde{\theta}^0_n)$ be introduced
with its population share vector $\varepsilon = (\varepsilon_1, \dots, \varepsilon_n)$.
Let $\witilde{b}\in B_1(\witilde{\mu}^{\varepsilon};b)$ be the played equilibrium satisfying
(1) $\witilde{b}(\witilde{\theta}^0) = \sigma$;
(2) for any proper subset $T\varsubsetneq N$ and any $\theta_{-T}\in \supp\mu_{-T}$,
we have $\witilde{b}(\witilde{\theta}^0_T, \theta_{-T}) = b(\bar{\theta}_T, \theta_{-T})$.\footnote{%
In the proofs, the indifferent types enable us easily to make a general argument about the mutants' particular interactions.
In fact, we can substitute non-indifferent types for the indifferent types in most cases.}
Then for every $i\in N$, the difference between the average fitnesses of $\witilde{\theta}^0_i$ and $\bar{\theta}_i$ is
\[
\varPi_{\witilde{\theta}^0_i}(\witilde{\mu}^{\varepsilon};\witilde{b}) - \varPi_{\bar{\theta}_i}(\witilde{\mu}^{\varepsilon};\witilde{b}) =
\witilde{\mu}^{\varepsilon}_{-i}(\witilde{\theta}^0_{-i}) \big[ \pi_i(\sigma) - \pi_i\big(b(\bar{\theta})\big) \big].
\]
Thus, for any vector $\varepsilon\in (0,1)^n$, we have $\varPi_{\witilde{\theta}^0_i}(\witilde{\mu}^{\varepsilon};\witilde{b})\geq \varPi_{\bar{\theta}_i}(\witilde{\mu}^{\varepsilon};\witilde{b})$ for every $i\in N$,
and $\varPi_{\witilde{\theta}^0_j}(\witilde{\mu}^{\varepsilon};\witilde{b}) > \varPi_{\bar{\theta}_j}(\witilde{\mu}^{\varepsilon};\witilde{b})$ for some $j\in N$.
This means that the configuration $(\mu,b)$ is not stable.
\end{proofEnd}

The existing single-population model based on the indirect evolutionary approach, as in DEY,
has demonstrated that if a configuration is stable and preferences are observed,
then the fitness each player receives in any match between incumbents is efficient, with a unique value in a symmetric game.
Unlike strategic interactions in single-population settings,
here a player in a multi-population match will only meet opponents coming from the other populations.
Thus, for a given symmetric objective game,
the single- and multi-population settings may yield quite different stable aggregate outcomes.

\begin{example}\label{exam:20160617}
Let the following anti-coordination game denote the fitness assignment, where $\nu> 0$, $\omega> 10$, and $\nu + \omega> 20$.
Suppose that preferences are observable,
and that each player~$i$ has the same action set $\{C, D\}$.

\begin{table}[!h]
\centering
\renewcommand{\arraystretch}{1.2}
    \begin{tabular}{r|c|c|}
      \multicolumn{1}{c}{} & \multicolumn{1}{c}{$C$} & \multicolumn{1}{c}{$D$}\\ \cline{2-3}
      $C$ &     $10$,  $10$ & $\nu$, $\omega$ \\  \cline{2-3}
      $D$ & $\omega$, $\nu$ &   $0$,      $0$ \\  \cline{2-3}
    \end{tabular}
\end{table}

\noindent
The efficient strategy $\sigma^*$ in this symmetric game is such that $\sigma^*(C) = \frac{\nu + \omega}{2(\nu + \omega - 10)}$
with the efficient fitness $\frac{(\nu + \omega)^2}{4(\nu + \omega - 10)}$.
When considered in the single-population model of DEY,
the unique efficient strategy profile $(\sigma^*, \sigma^*)$ is DEY-stable if and only if the equality $\nu = \omega$ holds.
Therefore, if the DEY-stable outcome exists, the efficient fitness is $\frac{\nu^2}{2\nu - 10}$
($= \frac{\omega^2}{2\omega - 10}$), which is strictly less than $\nu$ and $\omega$.

In the case when $\nu = \omega$, all Pareto-efficient strategy profiles in this objective game are $(C, D)$ and $(D, C)$,
and so Theorem~\ref{prop:003} implies that the strategy profile $(\sigma^*, \sigma^*)$ cannot be stable in the sense of multi-population stability.
The reason for the difference between the single- and multi-population models is that when the interaction takes place between two mutants from distinct populations,
they can play a suitable action profile, $(C, D)$ or $(D, C)$, to gain fitness advantages;
however, this cannot happen in single-population matching settings, where the mutant type a mutant can encounter is himself.
By Theorem~\ref{prop:0526},
it will be guaranteed that the Pareto-efficient strict Nash equilibria $(C, D)$ and $(D, C)$ can be stable if the multi-population setting is applied to this game.
\end{example}

Theorem~\ref{prop:003} says that configurations in our multi-population model should tend towards Pareto efficiency when preferences are observable.
Unlike the efficient fitness for a symmetric game,
Pareto-efficient fitness vectors are generally not unique in an arbitrary game.
Nevertheless, we can show that if a configuration is stable under perfect observability,
then the incumbents from the same population always earn the same fitness in all matches among incumbents,
no matter what their preferences may be.
(The fitness vectors for all matched tuples of incumbents are of the same Pareto-efficient form, no matter what incumbent members of the tuples are.)

\begin{theoremEnd}[no link to theorem, restate]{lem}\label{prop:170921}
Let $(\mu,b)$ be a stable configuration in $(G,\Gamma_1(\mu))$. Then the equality $\pi\big(b(\theta)\big) = \pi\big(b(\theta')\big)$ holds
for every $\theta$,~$\theta'\in \supp\mu$.
\end{theoremEnd}

\begin{proofEnd}
Let $(\mu,b)$ be a stable configuration.
We claim that for any $j\in N$ and any $\theta'_j$,~$\theta''_j\in \supp\mu_j$, the equality
$\pi\big( b(\theta'_j, \theta_{-j}) \big) = \pi\big( b(\theta''_j, \theta_{-j}) \big)$
holds for all $\theta_{-j}\in \supp\mu_{-j}$.
To see this, we first suppose that there exist
$\theta'_j$,~$\theta''_j\in \supp\mu_j$ for some $j\in N$ such that
$\pi_j\big(b(\theta'_j, \bar{\theta}_{-j})\big)> \pi_j\big(b(\theta''_j, \bar{\theta}_{-j})\big)$
for some $\bar{\theta}_{-j}\in \supp\mu_{-j}$.
Let $\witilde{\theta}^0_j\in \cplm{(\supp\mu_j)}$
be an indifferent type entering the $j$-th population.
Let the played equilibrium $\witilde{b}\in B_1(\witilde{\mu}^{\varepsilon};b)$ satisfy
$\witilde{b}(\witilde{\theta}^0_j, \bar{\theta}_{-j}) = b(\theta'_j, \bar{\theta}_{-j})$ and
$\witilde{b}(\witilde{\theta}^0_j, \theta_{-j}) = b(\theta''_j, \theta_{-j})$ for all $\theta_{-j}\neq \bar{\theta}_{-j}$.
Then for an arbitrary population share $\varepsilon$ of $\witilde{\theta}^0_j$,
the difference between the average fitnesses of $\witilde{\theta}^0_j$ and $\theta''_j$ is
\[
\varPi_{\witilde{\theta}^0_j}(\witilde{\mu}^{\varepsilon};\witilde{b}) - \varPi_{\theta''_j}(\witilde{\mu}^{\varepsilon};\witilde{b}) =
\mu_{-j}(\bar{\theta}_{-j})\big[ \pi_j\big(b(\theta'_j,\bar{\theta}_{-j})\big) - \pi_j\big(b(\theta''_j,\bar{\theta}_{-j})\big) \big]> 0,
\]
which means that the configuration $(\mu,b)$ is not stable.
This shows that for any $j\in N$ and any $\theta'_j$,~$\theta''_j\in \supp\mu_j$,
we have
\begin{equation}\label{eq:0401}
\pi_j\big(b(\theta'_j, \theta_{-j})\big) = \pi_j\big(b(\theta''_j, \theta_{-j})\big)
\end{equation}
for all $\theta_{-j}\in \supp\mu_{-j}$.

Next suppose that there exist
$\theta'_j$,~$\theta''_j\in \supp\mu_j$ and $\bar{\theta}_{-j}\in \supp\mu_{-j}$ for some $j\in N$ such that
$\pi_k\big(b(\theta'_j, \bar{\theta}_{-j})\big)> \pi_k\big(b(\theta''_j, \bar{\theta}_{-j})\big)$
for some $k\in N$ with $k\neq j$.
Consider suitable mutant types $\witilde{\theta}^0_j$ and $\witilde{\theta}_k$ entering the populations $j$ and $k$, respectively.
Let the played focal equilibrium $\witilde{b}$ satisfy:
(1) $\witilde{b}(\witilde{\theta}^0_j, \theta_{-j}) = b(\theta''_j, \theta_{-j})$ for all $\theta_{-j}\in \supp\mu_{-j}$,
and $\witilde{b}(\witilde{\theta}_k, \theta_{-k}) = b(\bar{\theta}_k, \theta_{-k})$ for all $\theta_{-k}\in \supp\mu_{-k}$;
(2) $\witilde{b}(\witilde{\theta}^0_j, \witilde{\theta}_k, \bar{\theta}_{-j -k}) = b(\theta'_j, \bar{\theta}_{-j})$
and $\witilde{b}(\witilde{\theta}^0_j, \witilde{\theta}_k, \theta_{-j -k}) = b(\theta''_j, \bar{\theta}_k, \theta_{-j -k})$
for any $\theta_{-j -k}\neq \bar{\theta}_{-j -k}$.
Then by~\eqref{eq:0401}, the equality
$\varPi_{\witilde{\theta}^0_j}(\witilde{\mu}^{\varepsilon};\witilde{b}) =
\varPi_{\theta''_j}(\witilde{\mu}^{\varepsilon};\witilde{b})$ is true for any vector $\varepsilon$ of the population shares of the two mutant types.
Moreover, our assumptions about $\witilde{b}$ also lead to
$\varPi_{\witilde{\theta}_k}(\witilde{\mu}^{\varepsilon};\witilde{b}) > \varPi_{\bar{\theta}_k}(\witilde{\mu}^{\varepsilon};\witilde{b})$,
and hence the configuration $(\mu,b)$ is not stable.
The claim follows.
Now for any $(\theta_1, \dots, \theta_n)$,~$(\theta'_1, \dots, \theta'_n)\in \supp\mu$,
by applying the claim at most $n$ times, we obtain
\begin{align*}
\pi\big( b(\theta_1, \theta_2, \dots, \theta_n) \big)
&= \pi\big( b(\theta'_1, \theta_2, \dots, \theta_n) \big)\\
&= \cdots = \pi\big( b(\theta'_1, \theta'_2, \dots, \theta'_n) \big),
\end{align*}
as desired.
\end{proofEnd}

When preferences are observable, Theorem~\ref{prop:003} and Lemma~\ref{prop:170921} imply that although populations are polymorphic,
a stable configuration yields a unique fitness profile which can be regarded as 
a point lying on the Pareto frontier of the \emph{noncooperative payoff region} of the objective game.\footnote{%
The noncooperative payoff region of an $n$-player game $(N, A, \pi)$ refers to the $n$-dimensional range $\pi\big( \prod_{i\in N}\Delta(A_i) \big)$.}
In addition, it follows directly from Lemma~\ref{prop:170921} that any stable configuration is balanced.
More precisely,
the average fitness of an incumbent is equal to the fitness value that any incumbent from the same population can earn in each of his matches against incumbent opponents.
It also follows that the fitness vector of a stable aggregate outcome consists of the fitness values obtained in any match of the incumbents.

\begin{theoremEnd}[no link to theorem, restate]{thm}\label{prop:120706}
Let $(\mu,b)$ be a stable configuration in $(G,\Gamma_1(\mu))$, and let $\varphi_{\mu,b}$ be the aggregate outcome of $(\mu,b)$.
Then $(\mu, b)$ is balanced, and for each $i\in N$,
\[
\varPi_{\bar{\theta}_i}(\mu;b) = \pi_i\big( b(\theta) \big) = \pi_i(\varphi_{\mu,b})
\]
for any $\bar{\theta}_i\in \supp\mu_i$ and any $\theta\in \supp\mu$.
\end{theoremEnd}

\begin{proofEnd}
Since $(\mu,b)$ is stable, by Lemma~\ref{prop:170921}, we let $v^* = \pi\big( b(\theta) \big)$ for $\theta\in \supp\mu$.
Then for each $i\in N$ and any $\bar{\theta}_i\in \supp\mu_i$, the equality $\varPi_{\bar{\theta}_i}(\mu;b) = v^*_i$ is obvious,
and thus $(\mu, b)$ is balanced.
Finally, for any $i\in N$, we have
\[
\pi_i(\varphi_{\mu,b}) = \sum_{a\in A}\varphi_{\mu,b}(a) \pi_i(a) =
\sum_{\theta\in\supp\mu}\mu(\theta) \sum_{a\in A} \bigg( \prod_{s\in N}b_s(\theta)(a_s) \bigg)\pi_i(a) = v^*_i
\]
since $\pi_i\big( b(\theta) \big) = \sum_{a\in A} \big( \prod_{s\in N}b_s(\theta)(a_s) \big) \pi_i(a)$ for all $\theta\in \supp\mu$.
\end{proofEnd}


In the single-population model of DEY with perfect observability,
it is shown that efficient strict Nash equilibria of a \emph{symmetric} two-player game are stable; see also \citet{aPos:tsespsg}.
In what follows,
we first give a sufficient condition for stability in the two-population case: in an arbitrary two-player game,
any Pareto-efficient strict Nash equilibrium would be an evolutionarily stable outcome.
At first glance, it seems easy to understand.
Let a configuration be given,
and suppose that each strategy in the strict Nash equilibrium is supported by preferences for which it is strictly dominant.
Consider any pair of mutant types with sufficiently small population shares.
Then if the mutants adopt any other action against the incumbents, they will be wiped out.
Furthermore, when two mutants from distinct populations are matched,
Pareto efficiency implies that the fitness of one mutant type cannot be improved without worsening the fitness of the other.
All this seems quite straightforward.
But as can be seen in Example~\ref{exam:20171114} (or in Example~\ref{exam:002}),
it is far from trivial to find a uniform invasion barrier valid for all focal equilibria in a multi-population matching setting.\footnote{%
Note that perfect observability is one limiting case of partial observability, which will be studied in Section~\ref{sec:imp-observ}.
In the other limiting case where preferences are unobservable,
the joint distribution over all types is one of the determinants of the adoption of strategies.
To accommodate varying assumptions on observability,
the stability criterion can only be defined by taking a uniform invasion barrier for all equilibria which are close enough to the original,
rather than just an invasion barrier for a specific equilibrium.}

\begin{theoremEnd}[no link to theorem, restate]{thm}\label{prop:0526}
Let $G$ be a two-player game,
and suppose that $(a^*_1,a^*_2)$ is Pareto efficient with respect to $\pi$.
If $(a^*_1,a^*_2)$ is a strict Nash equilibrium of $G$,
then $(a^*_1,a^*_2)$ is stable in $(G,\Gamma_1(\mu))$ for some $\mu\in \mathcal{M}(\Theta^2)$.
\end{theoremEnd}

\begin{proofEnd}
Let $(a^*_1,a^*_2)$ be a strict Nash equilibrium of $G$, and suppose that it is not a stable strategy profile.
We shall show that $(a^*_1,a^*_2)$ is not Pareto efficient with respect to $\pi$.
To see this, consider a monomorphic configuration $(\mu, b)$ where each $i$-th population consists of $\theta^*_i$ for which $a^*_i$ is the strictly dominant strategy.
Then $(a^*_1, a^*_2)$ is the aggregate outcome of $(\mu,b)$, and hence this configuration is not stable under our assumptions on $(a^*_1,a^*_2)$.
This means that there exists a mutant sub-profile $\witilde{\theta}_J$ for some $J\subseteq \{1, 2\}$ such that for every $\bar{\epsilon}\in (0,1)$, these mutants,
with some population share vector $\varepsilon\in (0,1)^{|J|}$ satisfying $\|\varepsilon\| \in (0,\bar{\epsilon})$,
can play an equilibrium $\witilde{b}\in B_1(\witilde{\mu}^{\varepsilon};b)$ to outperform the incumbents, that is, $\varPi_{\witilde{\theta}_j}(\witilde{\mu}^{\varepsilon};\witilde{b})\geq \varPi_{\theta^*_j}(\witilde{\mu}^{\varepsilon};\witilde{b})$ for all $j\in J$,
with strict inequality for some $j$.

In the case when $|J|=1$, it is clear that mutants have no fitness advantage since $(a^*_1,a^*_2)$ is a strict Nash equilibrium. Let $J = \{1,2\}$,
and suppose that $(\witilde{\theta}_1, \witilde{\theta}_2)$ is a mutant pair having an evolutionary advantage.
For $\varepsilon\in (0,1)^2$ and $\witilde{b}\in B_1(\witilde{\mu}^{\varepsilon};b)$,
the post-entry average fitnesses of $\theta^*_i$ and $\witilde{\theta}_i$ are, respectively,
\[
\varPi_{\theta^*_i}(\witilde{\mu}^{\varepsilon};\witilde{b}) = (1-\varepsilon_{-i}) \pi_i(a^*_1,a^*_2) +
\varepsilon_{-i} \pi_i\big( a^*_i, \witilde{b}_{-i}(\theta^*_i, \witilde{\theta}_{-i}) \big)
\]
and
\[
\varPi_{\witilde{\theta}_i}(\witilde{\mu}^{\varepsilon};\witilde{b}) = (1-\varepsilon_{-i}) \pi_i\big( \witilde{b}_i(\witilde{\theta}_i, \theta^*_{-i}), a^*_{-i} \big) + \varepsilon_{-i} \pi_i\big( \witilde{b}(\witilde{\theta}_1,\witilde{\theta}_2) \big).
\]
Using the instability assumption on $(a^*_1,a^*_2)$,
we gradually reduce $\bar{\epsilon}$ to $0$, and then the sequence of the norms of the corresponding population share vectors converges to $0$.
We can choose a sequence $\{\witilde{b}^t\}$ from the corresponding focal equilibria such that one of the following three cases occurs.
To complete the proof, we will show that $(a^*_1,a^*_2)$ is Pareto dominated in any one of these cases.
\vspace{5pt}

\noindent
\textbf{Case 1}:
$\witilde{b}^t_i(\witilde{\theta}_i, \theta^*_{-i}) = a^*_i$ for each $i$ and each $t$.
Since $(\witilde{\theta}_1, \witilde{\theta}_2)$ has an evolutionary advantage,
it follows that each $\witilde{b}^t(\witilde{\theta}_1, \witilde{\theta}_2)$ Pareto dominates $(a^*_1, a^*_2)$.
\vspace{5pt}

\noindent
\textbf{Case 2}:
$\witilde{b}^t_i(\witilde{\theta}_i, \theta^*_{-i})\neq a^*_i$ and $\witilde{b}^t_{-i}(\theta^*_i, \witilde{\theta}_{-i}) = a^*_{-i}$ for fixed $i$ and for every $t$.
Without loss of generality, suppose that $i = 1$.
Let $\witilde{b}^t_1(\witilde{\theta}_1,\theta^*_2) = (1-\zeta^t_1)a^*_1 + \zeta^t_1\sigma^t_1$,
where $\sigma^t_1\in\Delta( A_1\setminus \{a^*_1\} )$ and $\zeta^t_1\in (0,1]$ for all $t$.
Since $(\witilde{\theta}_1, \witilde{\theta}_2)$ has an evolutionary advantage and $(a^*_1,a^*_2)$ is a strict Nash equilibrium, we have
\[
\frac{\varepsilon^t_2}{1-\varepsilon^t_2}\geq
\frac{\zeta^t_1 [ \pi_1(a^*_1,a^*_2) - \pi_1(\sigma^t_1,a^*_2) ]}{\pi_1\big( \witilde{b}^t(\witilde{\theta}_1, \witilde{\theta}_2) \big) - \pi_1(a^*_1,a^*_2)}> 0
\]
in which $\pi_1\big( \witilde{b}^t(\witilde{\theta}_1, \witilde{\theta}_2) \big)> \pi_1(a^*_1,a^*_2)$ and
\[
\zeta^t_1[ \pi_2(a^*_1,a^*_2) - \pi_2(\sigma^t_1,a^*_2) ]\geq \pi_2(a^*_1,a^*_2) - \pi_2\big( \witilde{b}^t(\witilde{\theta}_1, \witilde{\theta}_2) \big)
\]
for every $t$.
If we let
$\pi_2(a^*_1,a^*_2)> \pi_2\big( \witilde{b}^t(\witilde{\theta}_1, \witilde{\theta}_2) \big)$ for all $t$,
then
$\pi_2(a^*_1,a^*_2)> \pi_2(\sigma^t_1,a^*_2)$ for all $t$;
otherwise, $(a^*_1, a^*_2)$ is Pareto dominated by $\witilde{b}^t(\witilde{\theta}_1, \witilde{\theta}_2)$ for some $t$, as desired.
Now, in the case when
$\pi_2(a^*_1,a^*_2)> \pi_2\big( \witilde{b}^t(\witilde{\theta}_1, \witilde{\theta}_2) \big)$ for all $t$,
we define
\[
\kappa =  \min \left\{\, \frac{\pi_1(a^*_1,a^*_2) - \pi_1(\sigma^t_1,a^*_2)}{\pi_2(a^*_1,a^*_2) - \pi_2(\sigma^t_1,a^*_2)} \biggm|
\sigma^t_1\in\Delta( A_1\setminus \{a^*_1\} ), \ t\in \mathbb{Z}^{+} \,\right\},
\]
and we can deduce
\[
\frac{\varepsilon^t_2}{1-\varepsilon^t_2}\geq \frac{\kappa \big[ \pi_2(a^*_1,a^*_2) - \pi_2\big( \witilde{b}^t(\witilde{\theta}_1, \witilde{\theta}_2) \big) \big]}{\pi_1\big( \witilde{b}^t(\witilde{\theta}_1, \witilde{\theta}_2) \big) - \pi_1(a^*_1,a^*_2)}> 0
\]
for all $t$. Since $\bar{\epsilon}$ converges to $0$, we obtain
\[
\lim_{t\to \infty}
\frac{\pi_2(a^*_1,a^*_2) - \pi_2\big( \witilde{b}^t(\witilde{\theta}_1, \witilde{\theta}_2) \big)}{\pi_1\big( \witilde{b}^t(\witilde{\theta}_1, \witilde{\theta}_2) \big) - \pi_1(a^*_1,a^*_2)} = 0.
\]
If $\pi\big( \witilde{b}^t(\witilde{\theta}_1, \witilde{\theta}_2) \big)$ does not converge to $\pi(a^*_1,a^*_2)$,
then by applying the fact that the payoff region $\pi\big( \prod_{i=1}^{2}\Delta(A_i) \big)$ is compact,
there exists a strategy profile $(\witilde{\sigma}_1, \witilde{\sigma}_2)$ such that $\pi(\witilde{\sigma}_1, \witilde{\sigma}_2)$ is a limit point of the set
$\{\, \pi\big( \witilde{b}^t(\witilde{\theta}_1, \witilde{\theta}_2) \big) \mid t\in \mathbb{Z}^{+} \,\}$,
and it Pareto dominates $(a^*_1,a^*_2)$ in terms of
$\pi_1(\witilde{\sigma}_1, \witilde{\sigma}_2)> \pi_1(a^*_1,a^*_2)$ and
$\pi_2(\witilde{\sigma}_1, \witilde{\sigma}_2) = \pi_2(a^*_1,a^*_2)$, as desired.
Otherwise, by summing up the above, there exists a sequence
$\{ \pi\big( \witilde{b}^t(\witilde{\theta}_1, \witilde{\theta}_2) \big) \}$ converging to $\pi(a_1^*, a_2^*)$ with
$\pi_1\big( \witilde{b}^t(\witilde{\theta}_1, \witilde{\theta}_2) \big)> \pi_1(a_1^*, a_2^*)$ and
$\pi_2(a_1^*, a_2^*)> \pi_2\big( \witilde{b}^t(\witilde{\theta}_1, \witilde{\theta}_2) \big)$ for all $t$;
moreover, the curve connecting the sequence has a horizontal tangent line at $\pi(a_1^*, a_2^*)$.
If $\pi(a_1^*, a_2^*)$ lies on the Pareto frontier of $\pi\big( \prod_{i=1}^{2}\Delta(A_i) \big)$,
then intuitively it seems that 
there is a part of the boundary of $\pi\big( \prod_{i=1}^{2}\Delta(A_i) \big)$ 
that looks like a part of the boundary of a strictly convex set, 
which contradicts the shape of a noncooperative payoff region.
Therefore $(a_1^*, a_2^*)$ should not be a Pareto-efficient strategy profile.
This can be formally proved using the properties of extreme points of a noncooperative payoff region; see \citet{2017arXiv170501454T}.
\vspace{5pt}

\noindent
\textbf{Case 3}:
$\witilde{b}^t_i(\witilde{\theta}_i, \theta^*_{-i})\neq a^*_i$ for all $i$ and all $t$. For each $i\in \{1,2\}$, let $\witilde{b}^t_i(\witilde{\theta}_i,\theta^*_{-i}) = (1-\xi^t_i)a^*_i + \xi^t_i\sigma^t_i$, where $\sigma^t_i\in \Delta( A_i\setminus \{a^*_i\} )$ and $\xi^t_i\in (0,1]$ for all $t$.
Note that $(a^*_1,a^*_2)$ is a strict Nash equilibrium,
that the mutant pair $(\witilde{\theta}_1, \witilde{\theta}_2)$ has an evolutionary advantage,
and that the corresponding norm $\|\varepsilon^t\|$ converges to $0$.
Thus, by comparing the post-entry average fitnesses of $\theta^*_i$ and $\witilde{\theta}_i$, we obtain
\[
\frac{\varepsilon^t_{-i}}{1 - \varepsilon^t_{-i}}\geq
\frac{\xi^t_i[\pi_i(a^*_1, a^*_2) - \pi_i(\sigma^t_i, a^*_{-i})]}{\pi_i\big( \witilde{b}^t(\witilde{\theta}_1, \witilde{\theta}_2) \big) - \pi_i(a^*_1, a^*_2) + \xi^t_{-i}[\pi_i(a^*_1, a^*_2) - \pi_i(a^*_i, \sigma^t_{-i})]}> 0
\]
for all $i$ and all $t$, and it follows that
\[
\frac{1}{\xi^t_i}\big[ \pi_i\big( \witilde{b}^t(\witilde{\theta}_1, \witilde{\theta}_2) \big) - \pi_i(a^*_1, a^*_2) \big]\to \infty
\quad \text{or} \quad
\frac{\xi^t_{-i}}{\xi^t_i}[\pi_i(a^*_1, a^*_2) - \pi_i(a^*_i,\sigma^t_{-i})]\to \infty
\]
for each $i\in \{1,2\}$. We discuss all the possibilities. If
\[
\frac{1}{\xi^t_1}\big[ \pi_1\big( \witilde{b}^t(\witilde{\theta}_1, \witilde{\theta}_2) \big) - \pi_1(a^*_1, a^*_2) \big]\to \infty
\quad \text{and} \quad
\frac{1}{\xi^t_2}\big[ \pi_2\big( \witilde{b}^t(\witilde{\theta}_1, \witilde{\theta}_2) \big) - \pi_2(a^*_1, a^*_2) \big]\to \infty
\]
occur simultaneously, then $(a^*_1, a^*_2)$ is Pareto dominated. Next, because $\xi^t_2/\xi^t_1\to \infty$ and $\xi^t_1/\xi^t_2\to \infty$ cannot occur simultaneously, the remaining possibility is that
\[
\frac{1}{\xi^t_i}\big[ \pi_i\big( \witilde{b}^t(\witilde{\theta}_1, \witilde{\theta}_2) \big) - \pi_i(a^*_1, a^*_2) \big]\to \infty
\quad \text{and} \quad
\frac{\xi^t_i}{\xi^t_{-i}}[\pi_{-i}(a^*_1, a^*_2) - \pi_{-i}(\sigma^t_i, a^*_{-i})]\to \infty,
\]
where either $i = 1$ or $i = 2$. In each case, we have $\xi^t_1\to 0$ and $\xi^t_2\to 0$.

Without loss of generality, let $i = 1$.
It is enough to consider the sequence $\{ \witilde{b}^t \}$ satisfying
$\pi_1\big( \witilde{b}^t(\witilde{\theta}_1, \witilde{\theta}_2) \big)> \pi_1(a^*_1, a^*_2)$.
If the inequality
$\pi_2\big( \witilde{b}^t(\witilde{\theta}_1, \witilde{\theta}_2) \big)\geq \pi_2(a^*_1, a^*_2)$
also holds for some $t$,
then
$(a^*_1, a^*_2)$ is Pareto dominated by some $\witilde{b}^t(\witilde{\theta}_1, \witilde{\theta}_2)$, as desired.
Hence we only have to suppose that
$\pi_2(a^*_1, a^*_2)> \pi_2\big( \witilde{b}^t(\witilde{\theta}_1, \witilde{\theta}_2) \big)$ for all $t$.
From comparing the post-entry average fitness of $\theta^*_2$ with that of $\witilde{\theta}_2$, we know that for each $t$,
\[
\pi_2\big( \witilde{b}^t(\witilde{\theta}_1, \witilde{\theta}_2) \big) - \pi_2(a^*_1, a^*_2) + \xi^t_1[\pi_2(a^*_1, a^*_2) - \pi_2(\sigma^t_1, a^*_2)]> 0.
\]
Then
\[
\frac{\xi^t_1[\pi_2(a^*_1, a^*_2) - \pi_2(\sigma^t_1, a^*_2)]}{\pi_1\big( \witilde{b}^t(\witilde{\theta}_1, \witilde{\theta}_2) \big) - \pi_1(a^*_1, a^*_2)}>
\frac{\pi_2(a^*_1, a^*_2) - \pi_2\big( \witilde{b}^t(\witilde{\theta}_1, \witilde{\theta}_2) \big)}{\pi_1\big( \witilde{b}^t(\witilde{\theta}_1, \witilde{\theta}_2) \big) - \pi_1(a^*_1, a^*_2)}> 0
\]
for all $t$.
Since we let $i = 1$ and the term $\pi_2(a^*_1, a^*_2) - \pi_2(\sigma^t_1, a^*_2)$ is bounded, we conclude that
\[
\lim_{t\to \infty}
\frac{\pi_2(a^*_1,a^*_2) - \pi_2\big( \witilde{b}^t(\witilde{\theta}_1, \witilde{\theta}_2) \big)}{\pi_1\big( \witilde{b}^t(\witilde{\theta}_1, \witilde{\theta}_2) \big) - \pi_1(a^*_1,a^*_2)} = 0.
\]
Then, as discussed in Case~2 of this proof, the strategy profile $(a^*_1, a^*_2)$ cannot be Pareto efficient.
\end{proofEnd}

\smallskip

We now verify the claim made after Definition~\ref{def:stable-p=1}, which concerns the different concepts of multi-population stability.
Let $a^*$ be a Pareto-efficient strict Nash equilibrium of a two-player game,
and suppose that the \emph{utilitarian social welfare function} is not maximized at $a^*$, that is, there is a strategy pair $\bar{a}$ such that
\[
\pi_1(\bar{a}) + \pi_2(\bar{a})> \pi_1(a^*) + \pi_2(a^*).
\]
Then $a^*$ is still a stable aggregate outcome by the above theorem.
On the other hand,
if the chosen stability criterion is based on \citet{pTay:essttp} (or \citet{rSel:nessaac}), as described after Definition~\ref{def:stable-p=1},
a configuration supporting the aggregate outcome $a^*$ would be displaced by a mutant pair with the ``secret handshake'' flavor.
Such mutants can destabilize $a^*$
by maintaining $a^*$ when matched against the incumbents and by selecting $\bar{a}$ when matched against themselves.
This says that $a^*$ cannot be stable under the stability criterion where evolutionary success depends on the sum of the average fitnesses.
Therefore, unlike extending the ESS concept to a general $n$-player game with $n$ species,
the two criteria derived from \citet{pTay:essttp} and \citet{rCre:scegt}, respectively, are not consistent
in the multi-population model of preference evolution with perfect observability.\footnote{%
If the alternative stability criterion based on \citet{rSel:nessaac} is chosen,
then an outcome is stable under perfect observability if it is a strict union Nash equilibrium (which will be defined in Definition~\ref{def:sNEagg}).
Moreover, in the case of no observability, all of our results would hold under this alternative criterion.}

\smallskip

Regarding the sufficient conditions for multi-population stability,
it is important to note that when the number of populations increases,
there may exist various possibilities of multilateral deviations such that mutants could have evolutionary advantages over the incumbents.
(As we shall show below,
a Pareto-efficient strict Nash equilibrium in a three-player game may not be stable.)
These multilateral deviations motivate us to consider the concept of a \emph{strong Nash equilibrium}; see \citet{rAum:apgcng}.
An equilibrium is strong if no multilateral deviation can benefit all of the deviators.
The formal definition is as follows.

\begin{Def}\label{def:sNE}
In an $n$-player game $(N, A, \pi)$,
a strategy profile $x$ is a \emph{strong Nash equilibrium}
if for any nonempty $J\subseteq N$ and for any $\sigma_J\in \prod_{j\in J}\Delta(A_j)$,
\[
\pi_k(x)\geq \pi_k(\sigma_J, x_{-J})
\]
for some $k\in J$.
If the weak inequality can be replaced by a strict inequality for every $\sigma_J\neq x_J$,
then $x$ is called a \emph{strictly strong Nash equilibrium}.\footnote{%
The strategy profile $(\sigma_J, x_{-J})$ results from $x$ by replacing $x_j$ with $\sigma_j$ for each $j\in J$.
The notation $\sigma_J\neq x_J$ means that $\sigma_j\neq x_j$ for some $j\in J$.}
\end{Def}

However, such a refinement is not sufficient to ensure multi-population stability,
even though all deviators for any pure-strategy multilateral deviation will be harmed.
We construct examples
which exhibit novel strategic interactions not found in two-player games
to illustrate some of these possibilities.

\begin{example}\label{exam:20171130}

Suppose that preferences are observable.
Let the objective game $G$ be the following three-player game,
and let $\{a_{i1}, a_{i2}, a_{i3}\}$ be the set of possible actions available to player~$i$ for $i = 1$,~$2$,~$3$.

\begin{table}[!h]
\centering
\renewcommand{\arraystretch}{1.2}
    \begin{tabular}{r|c|c|c|}
      \multicolumn{1}{c}{} & \multicolumn{1}{c}{$a_{21}$} & \multicolumn{1}{c}{$a_{22}$}
      & \multicolumn{1}{c}{$a_{23}$}\\ \cline{2-4}
      $a_{11}$ & $7$, $7$, $7$ & $3$, $0$, $3$ & $3$, $0$, $3$ \\  \cline{2-4}
      $a_{12}$ & $0$, $3$, $3$ & $7$, $7$, $0$ & $1$, $1$, $1$ \\  \cline{2-4}
      $a_{13}$ & $0$, $3$, $3$ & $1$, $1$, $1$ & $1$, $1$, $1$ \\  \cline{2-4}
      \multicolumn{1}{c}{} & \multicolumn{3}{c}{$a_{31}$}
    \end{tabular} \qquad
    \begin{tabular}{r|c|c|c|}
      \multicolumn{1}{c}{} & \multicolumn{1}{c}{$a_{21}$} & \multicolumn{1}{c}{$a_{22}$}
      & \multicolumn{1}{c}{$a_{23}$}\\ \cline{2-4}
      $a_{11}$ & $3$, $3$, $0$ & $1$, $1$, $1$ & $1$, $1$, $1$ \\  \cline{2-4}
      $a_{12}$ & $1$, $1$, $1$ & $7$, $7$, $0$ & $1$, $1$, $1$ \\  \cline{2-4}
      $a_{13}$ & $7$, $0$, $7$ & $7$, $0$, $7$ & $7$, $0$, $7$ \\  \cline{2-4}
      \multicolumn{1}{c}{} & \multicolumn{3}{c}{$a_{32}$}
    \end{tabular} \qquad
    \begin{tabular}{r|c|c|c|}
      \multicolumn{1}{c}{} & \multicolumn{1}{c}{$a_{21}$} & \multicolumn{1}{c}{$a_{22}$}
      & \multicolumn{1}{c}{$a_{23}$}\\ \cline{2-4}
      $a_{11}$ & $3$, $3$, $0$ & $1$, $1$, $1$ & $0$, $7$, $7$ \\  \cline{2-4}
      $a_{12}$ & $1$, $1$, $1$ & $7$, $7$, $0$ & $0$, $7$, $7$ \\  \cline{2-4}
      $a_{13}$ & $1$, $1$, $1$ & $1$, $1$, $1$ & $0$, $7$, $7$ \\  \cline{2-4}
      \multicolumn{1}{c}{} & \multicolumn{3}{c}{$a_{33}$}
    \end{tabular}
\end{table}

\noindent
For a given configuration $(\mu, b)$, let its aggregate outcome be $(a_{11},a_{21},a_{31})$.
Then $b(\theta) = (a_{11},a_{21},a_{31})$ for all $\theta\in \supp\mu$.
In this game $G$,
although the strategy profile $(a_{11},a_{21},a_{31})$ is a Pareto-efficient strict Nash equilibrium,
there are bilateral deviations in which two players can maintain the high fitness values,
and their opponent will incur a loss no matter what the opponent does.
Consider three indifferent types $\witilde{\theta}^0_1$, $\witilde{\theta}^0_2$, and $\witilde{\theta}^0_3$ entering the three populations, respectively.
Suppose that the players choose the focal equilibrium $\witilde{b}$ satisfying
\begin{gather*}
\witilde{b}(\witilde{\theta}^0_1,\theta_2,\theta_3) =
\witilde{b}(\theta_1,\witilde{\theta}^0_2,\theta_3) =
\witilde{b}(\theta_1,\theta_2,\witilde{\theta}^0_3) =
\witilde{b}(\witilde{\theta}^0_1,\witilde{\theta}^0_2,\witilde{\theta}^0_3) = (a_{11},a_{21},a_{31}),\\
\witilde{b}_1(\witilde{\theta}^0_1,\witilde{\theta}^0_2,\theta_3)=a_{12},\qquad
\witilde{b}_2(\witilde{\theta}^0_1,\witilde{\theta}^0_2,\theta_3)=a_{22},\\
\witilde{b}_1(\witilde{\theta}^0_1,\theta_2,\witilde{\theta}^0_3)=a_{13},\qquad
\witilde{b}_3(\witilde{\theta}^0_1,\theta_2,\witilde{\theta}^0_3)=a_{32},\\
\witilde{b}_2(\theta_1,\witilde{\theta}^0_2,\witilde{\theta}^0_3)=a_{23},\qquad
\witilde{b}_3(\theta_1,\witilde{\theta}^0_2,\witilde{\theta}^0_3)=a_{33},
\end{gather*}
for any $(\theta_1,\theta_2,\theta_3)\in \supp\mu$.
Let $\varepsilon_i = \varepsilon_0$ for all $i$.
Then by comparing the post-entry average fitness of the mutants with that of the incumbents in each population, respectively, we obtain
\[
\varPi_{\witilde{\theta}^0_i}(\witilde{\mu}^{\varepsilon}; \witilde{b}) - \varPi_{\theta_i}(\witilde{\mu}^{\varepsilon}; \witilde{b}) = 7\varepsilon_0^2> 0
\]
for $i = 1$,~$2$,~$3$.
Therefore, the strategy profile $(a_{11}, a_{21}, a_{31})$ cannot be stable,
and hence we easily conclude from Theorem~\ref{prop:003} that there is no stable aggregate outcome in this game.
\end{example}

The strategy profile $(a_{11},a_{21},a_{31})$ in Example~\ref{exam:20171130} is obviously a strong Nash equilibrium of $G$,
and it may be unsurprising that the strong Nash equilibrium concept is too weak to prevent multilateral deviations.
In fact,
evolutionary advantages of mutants also can come from the averages over their interactions,
although any multilateral deviation will decrease the fitness for at least one deviator.

\begin{example}\label{exam:20150603}
Suppose that preferences are observable.
Let the following three-player game be the objective game,
and let $\{a_{i1}, a_{i2}\}$ be the action set of player~$i$ for $i = 1$,~$2$,~$3$.

\begin{table}[!h]
\centering
\renewcommand{\arraystretch}{1.2}
    \begin{tabular}{r|c|c|}
      \multicolumn{1}{c}{} & \multicolumn{1}{c}{$a_{21}$} & \multicolumn{1}{c}{$a_{22}$}\\ \cline{2-3}
      $a_{11}$ & $7$, $7$, $7$ & $9$, $6$, $0$  \\  \cline{2-3}
      $a_{12}$ & $6$, $0$, $9$ & $6$, $0$, $9$ \\  \cline{2-3}
      \multicolumn{1}{c}{} & \multicolumn{2}{c}{$a_{31}$}
    \end{tabular} \qquad
    \begin{tabular}{r|c|c|}
      \multicolumn{1}{c}{} & \multicolumn{1}{c}{$a_{21}$} & \multicolumn{1}{c}{$a_{22}$}\\ \cline{2-3}
      $a_{11}$ & $0$, $9$, $6$ & $9$, $6$, $0$ \\  \cline{2-3}
      $a_{12}$ & $0$, $9$, $6$ & $1$, $1$, $1$ \\  \cline{2-3}
      \multicolumn{1}{c}{} & \multicolumn{2}{c}{$a_{32}$}
    \end{tabular}
\end{table}

\noindent
In this game, there are additional conditions imposed on the strictly strong Nash equilibrium $(a_{11},a_{21},a_{31})$:
the fitness values of all deviators will be reduced for any pure-strategy multilateral deviation.
Even so, the strategy profile $(a_{11},a_{21},a_{31})$ is not stable.
To verify this,
let a configuration $(\mu,b)$ with the aggregate outcome $(a_{11},a_{21},a_{31})$ be given.
Then $b(\theta_1,\theta_2,\theta_3) = (a_{11},a_{21},a_{31})$ for all $(\theta_1,\theta_2,\theta_3)\in \supp\mu$.
For each $i$, consider an indifferent mutant type $\witilde{\theta}^0_i$ entering the $i$-th population with its population share $\varepsilon_i$,
and suppose that the played focal equilibrium $\witilde{b}$ satisfies the conditions: for any $(\theta_1,\theta_2,\theta_3)\in \supp\mu$,
\begin{gather*}
\witilde{b}(\witilde{\theta}^0_1,\theta_2,\theta_3) =
\witilde{b}(\theta_1,\witilde{\theta}^0_2,\theta_3) =
\witilde{b}(\theta_1,\theta_2,\witilde{\theta}^0_3) =
\witilde{b}(\witilde{\theta}^0_1,\witilde{\theta}^0_2,\witilde{\theta}^0_3) = (a_{11},a_{21},a_{31}),\\
\witilde{b}_1(\witilde{\theta}^0_1,\witilde{\theta}^0_2,\theta_3)=a_{11},\qquad
\witilde{b}_2(\witilde{\theta}^0_1,\witilde{\theta}^0_2,\theta_3)=a_{22},\\
\witilde{b}_1(\witilde{\theta}^0_1,\theta_2,\witilde{\theta}^0_3)=a_{12},\qquad
\witilde{b}_3(\witilde{\theta}^0_1,\theta_2,\witilde{\theta}^0_3)=a_{31},\\
\witilde{b}_2(\theta_1,\witilde{\theta}^0_2,\witilde{\theta}^0_3)=a_{21},\qquad
\witilde{b}_3(\theta_1,\witilde{\theta}^0_2,\witilde{\theta}^0_3)=a_{32}.
\end{gather*}
Let $\varepsilon_i = \varepsilon_0$ for all $i$.
Then for any $\varepsilon_0> 0$ and for $i = 1$,~$2$,~$3$,
the difference between the average fitnesses of $\witilde{\theta}^0_i$ and $\theta_i$ is
\[
\varPi_{\witilde{\theta}^0_i}(\witilde{\mu}^{\varepsilon}; \witilde{b}) - \varPi_{\theta_i}(\witilde{\mu}^{\varepsilon}; \witilde{b}) = \varepsilon_0 + 6\varepsilon_0^2> 0,
\]
and hence $(a_{11},a_{21},a_{31})$ is not a stable strategy profile.
\end{example}

So far, we have seen that a group of mutants may act as if they play by forming coalitions.
Examples~\ref{exam:20171130} and~\ref{exam:20150603} reveal
how mutants can multilaterally deviate under complete information such that
no matter how the incumbents behave,
every mutant can achieve a strictly higher average fitness than the incumbents.
In order to provide a succinct and convenient sufficient condition for stability in the $n$-population case,
we shall define a special type of Nash equilibrium, describing a strategy profile in which
no subgroup of individuals can maintain the sum of the payoffs whenever a deviation occurs in the subgroup.
This new equilibrium concept could imply that
not all losses caused by deviations can be compensated from other matches.

\begin{Def}\label{def:sNEagg}
In an $n$-player game $(N, A, \pi)$, a strategy profile $x$ is a \emph{strict union Nash equilibrium} if
for any nonempty $J\subseteq N$ and for any $\sigma_J\in \prod_{j\in J}\Delta(A_j)$ with $\sigma_J\neq x_J$,
\[
\sum_{j\in J}\pi_j(x)> \sum_{j\in J}\pi_j(\sigma_J, x_{-J}).
\]
\end{Def}

Clearly, every strict union Nash equilibrium is a strictly strong Nash equilibrium,
and therefore a strong Nash equilibrium.
The payoff profile of a strict union Nash equilibrium is obviously represented as a point lying on the Pareto frontier of the payoff region,
and furthermore, it maximizes the sum of the players' payoffs in the game.
We are now in a position to state the sufficient condition for an outcome to be stable in any $n$-player game.

\begin{theoremEnd}[no link to theorem, restate]{thm}\label{prop:20190731}
Let $(a^*_1, \dots, a^*_n)$ be a strict union Nash equilibrium in an $n$-player game $G$.
Then $(a^*_1, \dots, a^*_n)$ is stable in $(G,\Gamma_1(\mu))$ for some $\mu\in \mathcal{M}(\Theta^n)$.
\end{theoremEnd}

\begin{proofEnd}
Let $a^* = (a^*_1, \dots, a^*_n)$ be a strict union Nash equilibrium of $G$,
and let $(\mu, b)$ be a monomorphic configuration where each $i$-th population consists of $\theta_i^*$ for which $a^*_i$ is the strictly dominant strategy,
so that $a^*$ is the aggregate outcome of $(\mu, b)$.
For any nonempty subset $J$ of $N$, consider a mutant sub-profile $\witilde{\theta}_J$ entering with its population share vector $\varepsilon$.
To verify stability, we shall show that there exists a uniform invasion barrier $\bar{\epsilon}\in (0,1)$ such that
for every $\varepsilon\in (0,1)^{|J|}$ with $\|\varepsilon\| \in (0,\bar{\epsilon})$, the inequality
\[
\sum_{i\in J} \varepsilon_i
\Big(
\varPi_{\theta_i^*}(\witilde{\mu}^{\varepsilon};\witilde{b}) - \varPi_{\witilde{\theta}_i}(\witilde{\mu}^{\varepsilon};\witilde{b})
\Big) \geq 0
\]
holds for all $\witilde{b}\in B_1(\witilde{\mu}^{\varepsilon};b)$.
By direct calculation,
\begin{multline}\label{eq:20190804}
\sum_{i\in J} \varepsilon_i
\Big(
\varPi_{\theta_i^*}(\witilde{\mu}^{\varepsilon};\witilde{b}) - \varPi_{\witilde{\theta}_i}(\witilde{\mu}^{\varepsilon};\witilde{b})
\Big) =
\sum_{i\in J} \varepsilon_i \prod_{j\in J_{-i}}(1 - \varepsilon_j) \big[ \pi_i(a^*) - \pi_i\big( \witilde{b}(\witilde{\theta}_i, \theta^*_{-i}) \big) \big]\\
+ \sum_{i\in J} \sum_{\substack{|T|=1\\ T\subseteq J_{-i}}} \varepsilon_i \varepsilon_T \prod_{j\in J_{-i}\setminus T}(1 - \varepsilon_j)
\big[ \pi_i\big( \witilde{b}(\theta^*_i, \witilde{\theta}_T, \theta^*_{-i -T}) \big) -
\pi_i\big( \witilde{b}(\witilde{\theta}_i, \witilde{\theta}_T, \theta^*_{-i -T}) \big) \big] \\
+ \dots + \sum_{i\in J} \varepsilon_i \varepsilon_{J_{-i}}
\big[ \pi_i\big( \witilde{b}(\theta^*_i, \witilde{\theta}_{J_{-i}}, \theta^*_{-J}) \big) -
\pi_i\big( \witilde{b}(\witilde{\theta}_i, \witilde{\theta}_{J_{-i}}, \theta^*_{-J}) \big) \big],
\end{multline}
where $\varepsilon_T$ denotes $\prod_{j\in T}\varepsilon_j$, and it continues to be used.

On the right-hand side of~\eqref{eq:20190804}, for any $i\in J$ and any $T\subseteq J_{-i}$,
write the difference between the two fitnesses as
\begin{multline*}
\pi_i\big( \witilde{b}(\theta^*_i, \witilde{\theta}_T, \theta^*_{-i -T}) \big) -
\pi_i\big( \witilde{b}(\witilde{\theta}_i, \witilde{\theta}_T, \theta^*_{-i -T}) \big)\\
=
\big[ \pi_i\big( \witilde{b}(\theta^*_i, \witilde{\theta}_T, \theta^*_{-i -T}) \big) - \pi_i(a^*) \big] +
\big[ \pi_i(a^*) - \pi_i\big( \witilde{b}(\witilde{\theta}_i, \witilde{\theta}_T, \theta^*_{-i -T}) \big) \big].
\end{multline*}
Then classify the sum in~\eqref{eq:20190804} according to the number of mutants occurring in those differences between
$\pi_i(a^*)$ and $\pi_i\big( \witilde{b}(\witilde{\theta}_H, \theta^*_{-H}) \big)$, where $i\in J$ and $H\subseteq J$.
In this way, any classified part with $k$ mutants in matched tuples is either of the form
\begin{multline}\label{eq:20190805}
\sum_{i\in J} \sum_{\substack{|T|=k-1\\ T\subseteq J_{-i}}} \varepsilon_i \varepsilon_T \prod_{j\in J_{-i}\setminus T}(1 - \varepsilon_j)
\big[ \pi_i(a^*) - \pi_i\big( \witilde{b}(\witilde{\theta}_i, \witilde{\theta}_T, \theta^*_{-i -T}) \big) \big]\\
+
\sum_{i\in J} \sum_{\substack{|T|=k\\ T\subseteq J_{-i}}} \varepsilon_i \varepsilon_T \prod_{j\in J_{-i}\setminus T}(1 - \varepsilon_j)
\big[ \pi_i\big( \witilde{b}(\theta^*_i, \witilde{\theta}_T, \theta^*_{-i -T}) \big) - \pi_i(a^*) \big]
\end{multline}
where $1\leq k\leq |J| - 1$, or of the form
$\sum_{i\in J} \varepsilon_J \big[ \pi_i(a^*) - \pi_i\big( \witilde{b}(\witilde{\theta}_J, \theta^*_{-J}) \big) \big]$
whenever $k = |J|$.
First, it is clear that
\[
\sum_{i\in J} \varepsilon_J \big[ \pi_i(a^*) - \pi_i\big( \witilde{b}(\witilde{\theta}_J, \theta^*_{-J}) \big) \big] \geq 0
\]
for any $\witilde{b}\in B_1(\witilde{\mu}^{\varepsilon};b)$, since $a^*$ is a strict union Nash equilibrium of $G$.

Next, for a given nonempty proper subset $S$ of $J$,
the sum of the terms involving differences between
$\pi_i(a^*)$ and $\pi_i\big( \witilde{b}(\witilde{\theta}_S, \theta^*_{-S}) \big)$, $i\in J$, in~\eqref{eq:20190805}
can also be written as
\begin{multline}\label{eq:20190806}
\sum_{i\in S} \varepsilon_S \prod_{j\in J\setminus S}(1 - \varepsilon_j)
\big[ \pi_i(a^*) - \pi_i\big( \witilde{b}(\witilde{\theta}_S, \theta^*_{-S}) \big) \big]\\
+
\sum_{i\in J\setminus S} \varepsilon_i \varepsilon_S \prod_{j\in J_{-i}\setminus S}(1 - \varepsilon_j)
\big[ \pi_i\big( \witilde{b}(\witilde{\theta}_S, \theta^*_{-S}) \big) - \pi_i(a^*) \big].
\end{multline}
We claim that there exists a uniform invasion barrier $\bar{\epsilon}_S$ such that the sum in~\eqref{eq:20190806} is greater than or equal to $0$
for any $\varepsilon$ with $\|\varepsilon\| \in (0,\bar{\epsilon}_S)$ and for any $\witilde{b}\in B_1(\witilde{\mu}^{\varepsilon};b)$.
To show it, suppose that for every $j\in S$,
$\witilde{b}_j(\witilde{\theta}_S,\theta^*_{-S}) = (1-\gamma_j)a^*_j + \gamma_j\sigma_j$,
where $\sigma_j\in\Delta( A_j\setminus \{a^*_j\} )$ and $\gamma_j\in [0,1]$.
Then the term in the expansion of the fitness
$\pi_i\big( \witilde{b}(\witilde{\theta}_S, \theta^*_{-S}) \big)$ of each player~$i$ can be represented as follows:
\[
\prod_{j\in S} c_j(\gamma_j, \alpha_j) \pi_i\Big( \big(x_j(\gamma_j, \alpha_j)\big)_{j\in S}, a^*_{-S} \Big),
\]
where the index $\alpha_j$ takes the value $0$ or $1$;
for any given $\alpha_j$, the coefficient function $c_j$ is defined by letting
$c_j(\gamma_j, \alpha_j)$ be $(1 - \gamma_j)^{\alpha_j} \gamma_j^{1 - \alpha_j}$ if $0< \gamma_j< 1$ and be $1$ otherwise;
the strategy function $x_j$ is defined by
\[
x_j(\gamma_j, \alpha_j) =
\begin{cases}
    a^*_j & \text{if $\gamma_j = 0$,}\\
    \sigma_j & \text{if $\gamma_j = 1$,}\\
    \sigma_j & \text{if $\gamma_j\in (0, 1)$ and $\alpha_j = 0$,}\\
    a^*_j & \text{if $\gamma_j\in (0, 1)$ and $\alpha_j = 1$.}
\end{cases}
\]
Thus, for any one term in the expansion of the difference $\pi_i(a^*) - \pi_i\big( \witilde{b}(\witilde{\theta}_S, \theta^*_{-S}) \big)$,
we can write it as
\[
\prod_{j\in S} c_j(\gamma_j, \alpha_j) \Big[ \pi_i(a^*) - \pi_i\Big( \big(x_j(\gamma_j, \alpha_j)\big)_{j\in S}, a^*_{-S} \Big) \Big]
\]
for some $(\alpha_j)_{j\in S}\in \{0, 1\}^{|S|}$.
Similarly for the expansion of the difference $\pi_i\big( \witilde{b}(\witilde{\theta}_S, \theta^*_{-S}) \big) - \pi_i(a^*)$.
Collecting the terms that have the same coefficient $\prod_{j\in S} c_j(\gamma_j, \alpha_j)$ in~\eqref{eq:20190806},
we obtain the sum
\begin{multline}\label{eq:20190807}
\varepsilon_S \prod_{j\in J\setminus S}(1 - \varepsilon_j) \prod_{j\in S} c_j(\gamma_j, \alpha_j) \Bigg\{
\sum_{i\in S}
\Big[ \pi_i(a^*) - \pi_i\Big( \big(x_j(\gamma_j, \alpha_j)\big)_{j\in S}, a^*_{-S} \Big) \Big]\\
+
\sum_{i\in J\setminus S} \Big( \frac{\varepsilon_i}{1 - \varepsilon_i} \Big)
\Big[ \pi_i\Big( \big(x_j(\gamma_j, \alpha_j)\big)_{j\in S}, a^*_{-S} \Big) - \pi_i(a^*) \Big] \Bigg\}.
\end{multline}
Of course,
the sum in~\eqref{eq:20190807} is equal to $0$ if $x_j(\gamma_j, \alpha_j) = a^*_j$ for all $j\in S$.
Otherwise, since $a^*$ is a strict union Nash equilibrium of $G$,
there exists $m > 0$ such that for every $\sigma_j$ satisfying $\sigma_j(a^*_j) = 0$,
\[
\sum_{i\in S} \Big[ \pi_i(a^*) - \pi_i\Big( \big(x_j(\gamma_j, \alpha_j)\big)_{j\in S}, a^*_{-S} \Big) \Big]\geq m.
\]
Furthermore,
the difference $\pi_i\Big( \big(x_j(\gamma_j, \alpha_j)\big)_{j\in S}, a^*_{-S} \Big) - \pi_i(a^*)$ is bounded for every $i\in J\setminus S$,
and we know that $\{0, 1\}^{|S|}$ is a finite set.
Therefore, we can find a uniform invasion barrier $\bar{\epsilon}_S$ such that
for any $\varepsilon$ with $\|\varepsilon\| \in (0,\bar{\epsilon}_S)$ and any $\witilde{b}\in B_1(\witilde{\mu}^{\varepsilon};b)$,
the sum in~\eqref{eq:20190806} is greater than or equal to $0$.
This completes the proof since the number of subsets of $J$ is finite.\footnote{%
If we try to prove Theorem~\ref{prop:0526} using the method of proving Theorem~\ref{prop:20190731},
then we will see that it is unclear whether a uniform invasion barrier can be found for all focal equilibria.}
\end{proofEnd}

The conditions for the existence of a strict union Nash equilibrium are used to eliminate the possibility that
multilateral deviations could bring about a higher average fitness for every mutant, with at least one strict inequality.
It is unsurprising that such a sufficient condition satisfies the so-called \emph{best response property},
introduced in \citet{dRay-rVoh:eba} for games with coalition structures.
This gives us an insight into the relationship between the core of a normal-form game and the stable outcome of preference evolution.


\section{No Observability}
\label{sec:unobservable}


We now turn to the case where the degree of observability is equal to zero.
In this section, each player is ignorant of the preferences of his opponents,
but knows his own preference type and the joint distribution over the opponents' types.
The interactions among players can then be analyzed as an $n$-player Bayesian game.

For a given $\mu\in \mathcal{M}(\Theta^n)$,
a \emph{strategy} for player~$i$ under no observability is a function $s_i\colon\supp\mu_i\to\Delta(A_i)$.
The function $s\colon\supp\mu\to\prod_{i\in N}\Delta(A_i)$ defined by $s(\theta) = (s_1(\theta_1),\dots,s_n(\theta_n))$
is a \emph{Bayesian--Nash equilibrium} of the subjective game $\Gamma_0(\mu)$
if for each $i\in N$ and each $\theta_i\in\supp\mu_i$,
\[
s_i(\theta_i) \in \argmax_{\sigma_i\in\Delta(A_i)} \sum_{\theta'_{-i}\in\supp\mu_{-i}} \mu_{-i}(\theta'_{-i}) \theta_i\big( \sigma_i, s_{-i}(\theta'_{-i}) \big),
\]
where we write $s_{-i}(\theta'_{-i})$ for $s(\theta_i, \theta'_{-i})_{-i}$. Let $B_0(\mu)$ denote the set of all Bayesian--Nash equilibria of $\Gamma_0(\mu)$.
We call the pair $(\mu,s)$ of $\mu\in \mathcal{M}(\Theta^n)$ and $s\in B_0(\mu)$ a \emph{configuration}.
Here the \emph{aggregate outcome} of $(\mu,s)$, defined as a correlated strategy, can be written
\[
\varphi_{\mu,s}(a_1, \dots, a_n) =
\sum_{\theta\in \supp\mu} \mu(\theta) \prod_{i\in N} s_i(\theta_i)(a_i) =
\prod_{i\in N} \sum_{\theta_i\in \supp\mu_i} \mu_i(\theta_i) s_i(\theta_i)(a_i)
\]
for every $(a_1, \dots, a_n)\in A$.
Indeed, this correlated strategy $\varphi_{\mu,s}$ can be induced by the mixed-strategy profile
\[
x(\mu,s) = \Big( \sum_{\theta_1\in\supp\mu_1} \mu_1(\theta_1) s_1(\theta_1), \dots, \sum_{\theta_n\in\supp\mu_n} \mu_n(\theta_n) s_n(\theta_n) \Big).
\]
Therefore, under incomplete information, we also refer to the mixed-strategy profile $x(\mu,s)$ as the aggregate outcome of $(\mu,s)$.

For $i\in N$, the \emph{average fitness} of a type $\theta_i\in\supp\mu_i$ with respect to $(\mu,s)$ can be simply represented in terms of the aggregate outcome $x(\mu,s)$:
\[
\varPi_{\theta_i}(\mu;s) = \sum_{\theta'_{-i}\in\supp\mu_{-i}} \mu_{-i}(\theta'_{-i}) \pi_i\big( s(\theta_i,\theta'_{-i}) \big) = \pi_i\big( s_i(\theta_i), x(\mu,s)_{-i} \big).
\]
As in the case of perfect observability, we say that a configuration $(\mu,s)$ is \emph{balanced} if
the equality $\varPi_{\theta_i}(\mu;s) = \varPi_{\theta'_i}(\mu;s)$ holds for every $i\in N$ and every $\theta_i$,~$\theta'_i\in\supp\mu_i$.

For a balanced configuration under no observability, the average fitness of any type in a given population is equal to
the fitness value assigned to the aggregate outcome by the corresponding real-valued fitness function.
This is analogous to the result in Theorem~\ref{prop:120706}, without resorting to the stability condition.

\begin{theoremEnd}[no link to theorem, restate]{lem}\label{prop:011}
Let $(\mu,s)$ be a balanced configuration in $(G,\Gamma_0(\mu))$ with the aggregate outcome $x$. Then
\[
\varPi_{\theta_i}(\mu;s) = \pi_i(x)
\]
for every $i\in N$ and every $\theta_i\in\supp\mu_i$.
\end{theoremEnd}

\begin{proofEnd}
Since $(\mu,s)$ is a balanced configuration, we have that for each $i\in N$ and for any fixed $\theta_i\in\supp\mu_i$,
\[
\pi_i\big(s_i(\theta_i),x_{-i}\big) = \varPi_{\theta_i}(\mu;s) = \varPi_{\theta'_i}(\mu;s) = \pi_i\big(s_i(\theta'_i),x_{-i}\big)
\]
for every $\theta'_i\in\supp\mu_i$. This implies that
\[
\pi_i(x)=\sum_{\theta'_i\in\supp\mu_i}\mu_i(\theta'_i)\pi_i\big(s_i(\theta'_i),x_{-i}\big)
=\pi_i\big(s_i(\theta_i),x_{-i}\big)=\varPi_{\theta_i}(\mu;s),
\]
and the proof is complete.
\end{proofEnd}

When preferences are unobservable, a focal equilibrium,
defined for incumbents satisfying the focal property, does not always exist,
since after an entry incumbents may not be able to ignore the minor perturbation in the distribution of preference types.
For this case, we consider \emph{nearby equilibria} which can be chosen arbitrarily close to the pre-entry one
if the population shares of mutants are sufficiently small.

\begin{Def}
Let $(\mu,s)$ be a configuration under no observability, and let $\witilde{\mu}^{\varepsilon}$ be a post-entry type distribution.
In $\Gamma_0(\witilde{\mu}^{\varepsilon})$, for a given $\eta\geq 0$,
an equilibrium $\witilde{s}\in B_0(\witilde{\mu}^{\varepsilon})$ is called a \emph{nearby equilibrium} relative to $s$ within $\eta$
if $d(\witilde{s}(\theta), s(\theta))\leq \eta$ for all $\theta\in \supp\mu$,
where $d(\witilde{s}(\theta), s(\theta)) = \max_{i\in N} d^*(\witilde{s}_i(\theta_i), s_i(\theta_i))$ with $d^*$ denoting the Euclidean metric.
Let $B_0^{\eta}(\witilde{\mu}^{\varepsilon};s)$ be the set of all nearby equilibria relative to $s$ within $\eta$,
called a \emph{nearby set}.
\end{Def}

By definition, a nearby equilibrium $\witilde{s}$ at zero distance from $s$ is just a \emph{focal equilibrium} relative to $s$.
Thus, the nearby set with $\eta = 0$ is the \emph{focal set}, and will be denoted by $B_0(\witilde{\mu}^{\varepsilon};s)$.
Additionally, the relation $B_0^{\eta_1}(\witilde{\mu}^{\varepsilon};s)\subseteq B_0^{\eta_2}(\witilde{\mu}^{\varepsilon};s)$ holds whenever $0\leq \eta_1< \eta_2$.

The stability criterion under incomplete information is defined in the same way as in Definition~\ref{def:stable-p=1},
except that the focal set may be replaced by a nearby set.
But what nearby sets are relevant if the focal set is empty?
The requirements for our stability are that as long as the entering mutants are rare enough,
(1) there exists an appropriate set of nearby equilibria, which are sufficiently close to the pre-entry one;
(2) no incumbent will be wiped out when the played equilibrium belongs to this nearby set.
It is worth noting that the way of describing an appropriate nearby set as shown below is precise but very simplified.
We will compare it with another nearby set which is considered in DEY; see Remark~\ref{rem:140622} and Example~\ref{exam:20160522}.

\begin{Def}\label{def:stable-p=0}
In $(G,\Gamma_0(\mu))$, a configuration $(\mu,s)$ is \emph{stable} if it is balanced,
and if for any nonempty $J\subseteq N$, any $\witilde{\theta}_J\in\prod_{j\in J}\cplm{(\supp\mu_j)}$, and any $\eta > 0$,
there exist $\bar{\epsilon}\in (0,1)$ and $\bar{\eta}\in [0, \eta)$
such that for every $\varepsilon\in (0,1)^{|J|}$ with $\|\varepsilon\|\in (0,\bar{\epsilon})$,
the nearby set $B_0^{\bar{\eta}}(\witilde{\mu}^{\varepsilon};s)$ is nonempty,
and either~\ref{C2:wiped-out} or~\ref{C2:coexist} is satisfied for every $\witilde{s}\in B_0^{\bar{\eta}}(\witilde{\mu}^{\varepsilon};s)$. 
\begin{enumerate}[label=(\roman*)]
\item $\varPi_{\theta_j}(\witilde{\mu}^{\varepsilon};\witilde{s}) > \varPi_{\witilde{\theta}_j}(\witilde{\mu}^{\varepsilon};\witilde{s})$ for some $j\in J$ and for every $\theta_j\in\supp\mu_j$.\label{C2:wiped-out}
\item $\varPi_{\theta_i}(\witilde{\mu}^{\varepsilon};\witilde{s}) = \varPi_{\wihat{\theta}_i}(\witilde{\mu}^{\varepsilon};\witilde{s})$ for every $i\in N$ and
      for every $\theta_i$,~$\wihat{\theta}_i\in \supp\witilde{\mu}^{\varepsilon}_i$.\label{C2:coexist}
\end{enumerate}
A strategy profile $\sigma\in \prod_{i\in N}\Delta(A_i)$ is said to be \emph{stable} if there exists a stable configuration $(\mu,s)$ such that $\sigma = x(\mu,s)$.
\end{Def}

After introducing a mutant sub-profile $\witilde{\theta}_J$,
if the focal set $B_0(\witilde{\mu}^{\varepsilon};s)$ is nonempty for all sufficiently small $\varepsilon> 0$,
then the post-entry equilibria that we must consider are the focal equilibria.
If there are no focal equilibria,
then this definition enables us to decide which nearby set is most appropriate
whenever there are too many equilibria satisfying the requirement of a given $\eta$.
In addition,
the stability of $(\mu,s)$ ensures that the post-entry aggregate outcomes get arbitrarily close to the pre-entry one,
as described in Remark~\ref{rem:170815};
formally, for any $\xi> 0$,
there exist $\bar{\epsilon}\in (0,1)$ and $\bar{\eta}> 0$ such that for every $\varepsilon$ with $\|\varepsilon\|\in (0,\bar{\epsilon})$,
the relation $d(x(\witilde{\mu}^{\varepsilon},\witilde{s}), x(\mu,s))< \xi$ holds for all $\witilde{s}\in B_0^{\bar{\eta}}(\witilde{\mu}^{\varepsilon};s)$.

\begin{Rem}\label{rem:140622}
When there are no focal equilibria,
the stability definition of DEY requires that for any $\eta > 0$ and for any sufficiently small $\varepsilon> 0$,
the entrants never outperform the incumbents for every nearby equilibrium in $B_0^{\eta}(\witilde{\mu}^{\varepsilon};s)$.
However, the nearby set $B_0^{\eta}(\witilde{\mu}^{\varepsilon};s)$ for a given $\eta$ may include equilibria
which are far from the pre-entry equilibrium $s$,
no matter how small $\varepsilon$ is.
These outlier equilibria, also satisfying the requirement of the given $\eta$,
may lead to diverging conclusions,
contrary to the original intent of DEY; see Example~\ref{exam:20160522}.\footnote{%
\citet{yHel-eMoh:cdp} also have very similar viewpoints, but they do not give a concrete example.}
It should be stressed that, unlike in DEY, we can avoid unnecessary equilibria by creating an adequate barrier $\bar{\eta}$ for any given $\eta$.
\end{Rem}

\begin{example}\label{exam:20160522}
Let $(G,\Gamma_0(\mu))$ be an environment with incomplete information,
in which $G$ is a two-player game with $A_i = \{a_{i1}, a_{i2}, a_{i3}\}$ available to player~$i$,
and each $i$-th population consists of a single preference type $\theta_i$. Suppose that
\[
\theta_1(a) = \theta_2(a) =
\begin{cases}
    2 & \text{if $a = (a_{11},a_{21})$ or $(a_{12},a_{22})$,}\\
    1 & \text{if $a = (a_{11},a_{22})$ or $(a_{12},a_{21})$,}\\
    0 & \text{otherwise.}
\end{cases}
\]
Let $s(\theta_1, \theta_2) = ((0.5,0.5,0), (0.5,0.5,0))$; then $s$ is a Bayesian--Nash equilibrium of $\Gamma_0(\mu)$.
We claim that under the stability criterion in Definition~\ref{def:stable-p=0},
the configuration $(\mu,s)$ is stable provided that the objective game $G$ can be defined as follows.

\begin{table}[!h]
\centering
\renewcommand{\arraystretch}{1.2}
    \begin{tabular}{r|c|c|c|}
      \multicolumn{1}{c}{} & \multicolumn{1}{c}{$a_{21}$} & \multicolumn{1}{c}{$a_{22}$}
                           & \multicolumn{1}{c}{$a_{23}$}\\ \cline{2-4}
      $a_{11}$ & $4$, $4$ & $2$, $2$ & $0$, $5$ \\  \cline{2-4}
      $a_{12}$ & $2$, $2$ & $4$, $4$ & $0$, $0$ \\  \cline{2-4}
      $a_{13}$ & $5$, $0$ & $0$, $0$ & $2$, $2$ \\  \cline{2-4}
    \end{tabular}
\end{table}

\noindent
To see why, for $i = 1$,~$2$,
we introduce an arbitrary mutant type $\witilde{\theta}_i$ entering the $i$-th population with a population share $\varepsilon_i\in (0, 1)$,
and let $\witilde{s}\in B_0(\witilde{\mu}^{\varepsilon})$ be the played equilibrium with
$\witilde{s}_i(\theta_i) = (p_{i1}, p_{i2}, 1-p_{i1}-p_{i2})$ and $\witilde{s}_i(\witilde{\theta}_i) = (q_{i1}, q_{i2}, 1-q_{i1}-q_{i2})$.
Suppose that $\witilde{s}$ is a nearby equilibrium sufficiently close to $s$.
Then $p_{11}$,~$p_{12}$,~$p_{21}$,~$p_{22}\neq 0$,
and it is true that for each $i$,
\[
\theta_i(a_{i1}, \witilde{x}_{-i}) = \theta_i(a_{i2}, \witilde{x}_{-i})\geq \theta_i(a_{i3}, \witilde{x}_{-i}),
\]
where $\witilde{x}_{-i}$ refers to the action $(1 - \varepsilon_j)\witilde{s}_j(\theta_j) + \varepsilon_j\witilde{s}_j(\witilde{\theta}_j)$ for $j\neq i$.
This further implies that for $i = 1$,~$2$, we have
$p_{i1} = \frac{1 - \varepsilon_i(1 + q_{i1} - q_{i2})}{2(1 - \varepsilon_i)}$,
$p_{i2} = \frac{1 - \varepsilon_i(1 - q_{i1} + q_{i2})}{2(1 - \varepsilon_i)}$, and of course $1-p_{i1}-p_{i2} = 0$, which are well defined if $\varepsilon_i< 1/2$.
Applying Kakutani's fixed point theorem to the mutants' best response correspondences,
such a nearby equilibrium can be found,
and it ensures that there exist post-entry equilibria arbitrarily close to $s$ as $\|\varepsilon\|$ tends to $0$.
For each $i$, let $j\neq i$. Then it can be deduced that
\[
\pi_i(a_{i1}, \witilde{x}_{-i}) = \pi_i(a_{i2}, \witilde{x}_{-i}) = 3[1 - \varepsilon_j(1 - q_{j1} - q_{j2})]>
\pi_i(a_{i3}, \witilde{x}_{-i}) = \frac{1}{2}[5 - \varepsilon_j(1 - q_{j1} - q_{j2})]
\]
whenever $\varepsilon_j< 1/5$.
Thus, entrants receive no higher average fitnesses than the incumbents for all nearby equilibria in $B_0^{\bar{\eta}}(\witilde{\mu}^{\varepsilon};s)$
as long as $\bar{\eta}$ and $\varepsilon$ are sufficiently small, which proves the claim.\footnote{%
\citet[p.~234]{aRob-lSam} conclude that
``\,The indirect evolutionary approach with unobservable preferences then gives us an alternative description of the evolutionary process,
one that is perhaps less reminiscent of biological determinism, but leads to no new results.''
However, for this symmetric game $G$,
the strategy $(0.5, 0.5, 0)$ is not a neutrally stable strategy because it can be displaced by $(1, 0, 0)$.
Unlike pre-programmed strategies,
the minor strategic adjustments depending on players' beliefs under incomplete information
would still prevent them from being eliminated.}

It must be noted that another stability criterion derived from DEY can lead to a different conclusion.
Consider the pair $(\witilde{\theta}_1, \witilde{\theta}_2)$ of mutant types, each of which satisfies:
$\witilde{\theta}_i(a)$ equals $1$ if $a = (a_{11},a_{21})$, $(a_{11},a_{23})$, $(a_{13},a_{21})$, $(a_{13},a_{23})$,
equals $2$ if $a = (a_{12},a_{22})$, and equals $0$ otherwise.
In this case, it is not difficult to check that the focal set $B_0(\witilde{\mu}^{\varepsilon};s)$ is empty whenever $\|\varepsilon\|< 1/3$.
Now let $\witilde{s}'$ be the strategy pair defined by
$\witilde{s}'_1(\theta_1) = \witilde{s}'_2(\theta_2) = (1, 0, 0)$ and $\witilde{s}'_1(\witilde{\theta}_1) = \witilde{s}'_2(\witilde{\theta}_2) = (0.5, 0, 0.5)$.
Then $\witilde{s}'$ is also a Bayesian--Nash equilibrium of $\Gamma_0(\witilde{\mu}^{\varepsilon})$, no matter what $\varepsilon$ may be.
Thus if $\eta$, as described in Remark~\ref{rem:140622}, is too large,
then the nearby set $B_0^{\eta}(\witilde{\mu}^{\varepsilon};s)$ would include $\witilde{s}'$ for every $\varepsilon$,
and hence the ``nearby'' equilibrium $\witilde{s}'$ should be used to check whether or not the configuration $(\mu,s)$ is stable in the sense of DEY.
However, for an arbitrary population share vector $\varepsilon$, the inequality $\varPi_{\witilde{\theta}_i}(\witilde{\mu}^{\varepsilon};\witilde{s}')> \varPi_{\theta_i}(\witilde{\mu}^{\varepsilon};\witilde{s}')$ holds for $i = 1$,~$2$.
This means that the configuration $(\mu,s)$ is not stable according to the stability criterion proposed by DEY.\footnote{%
Clearly, this discussion is entirely applicable to the single-population matching setting of DEY.}
\end{example}

The previous example demonstrates clearly that as we let the population shares of the mutants continue to decline,
even if there are nearby equilibria arbitrarily close to the original,
another equilibrium far from the original still exists at the same time.
To be consistent with the concept of a focal equilibrium,
it is natural to assume that incumbents will adopt strategies as close as possible to the original;
equilibria relatively far from the original should not be considered.
To achieve this, we construct the barrier $\bar{\eta}$ that can be adjusted accordingly to exclude the unnecessary post-entry equilibria;
it is the key factor leading to a different conclusion.

When preferences are not observed,
we first show that a stable aggregate outcome must be a Nash equilibrium of the objective game.
The intuition for this is simple.
Suppose that the aggregate outcome of a balanced configuration is not a Nash equilibrium.
Then entrants can gain a fitness advantage by adopting another strategy,
provided that the entrants are rare enough and
the incumbents' post-entry strategies are nearly unchanged.

\begin{theoremEnd}[no link to theorem, restate]{thm}\label{prop:005}
If a configuration $(\mu,s)$ is stable in $(G,\Gamma_0(\mu))$, then the aggregate outcome of $(\mu,s)$ is a Nash equilibrium of $G$.
\end{theoremEnd}

\begin{proofEnd}
Let $x$ be the aggregate outcome of a stable configuration $(\mu,s)$, and suppose that $x$ is not a Nash equilibrium.
Then there exist $k\in N$ and $a_k\in A_k$ such that $\pi_k(a_k,x_{-k}) > \pi_k(x)$.
We have seen in the proof of Lemma~\ref{prop:011} that a balanced configuration implies that
$\pi_i(x) = \pi_i\big(s_i(\theta_i), x_{-i}\big)$ for all $i\in N$ and all $\theta_i\in \supp\mu_i$, and thus we obtain
\begin{equation*}
\pi_k(a_k, x_{-k}) > \pi_k\big(s_k(\theta_k), x_{-k}\big)
\end{equation*}
for all $\theta_k\in \supp\mu_k$.
Consider a mutant type $\witilde{\theta}_k\in \cplm{(\supp\mu_k)}$ appearing in the $k$-th population with its population share $\varepsilon$,
for which $a_k$ is the strictly dominant strategy. Then for any post-entry equilibrium $\witilde{s}$, we get $\witilde{s}_k(\witilde{\theta}_k) = a_k$, and the average fitnesses of $\witilde{\theta}_k$ and any $\theta_k\in\supp\mu_k$ are, respectively,
\[
\varPi_{\witilde{\theta}_k}(\witilde{\mu}^{\varepsilon};\witilde{s}) = \pi_k(a_k,\witilde{x}_{-k})
\text{\quad and \quad}
\varPi_{\theta_k}(\witilde{\mu}^{\varepsilon};\witilde{s}) = \pi_k\big( \witilde{s}_k(\theta_k),\witilde{x}_{-k} \big),
\]
where $\witilde{x} = x(\witilde{\mu}^{\varepsilon},\witilde{s})$.
Since $(\mu,s)$ is a stable configuration and $\pi_k$ is a continuous function on $\prod_{i\in N}\Delta(A_i)$,
it follows that as long as $\bar{\eta}$ and $\varepsilon$ are sufficiently small, the inequality
\[
\pi_k(a_k,\witilde{x}_{-k})> \pi_k \big( \witilde{s}_k(\theta_k),\witilde{x}_{-k} \big)
\]
holds for all $\witilde{s}\in B_0^{\bar{\eta}}(\witilde{\mu}^{\varepsilon};s)$ and all $\theta_k\in \supp\mu_k$
(see the discussion after Definition~\ref{def:stable-p=0}).
Thus we have arrived at a contradiction.
\end{proofEnd}

If the types of individuals cannot be distinguished,
non-materialist preferences will no longer have an effect as making a commitment.
In our model, it is easy to see that materialist preferences have an obvious fitness advantage when preferences are unobservable.\footnote{%
Under incomplete information,
it is possible that non-materialist preferences earn higher fitness than materialist preferences in certain strategic environments.
\citet{iAlg-jWei:hmpeiiam} consider assortative matching and show that
pure selfishness, acting so as to maximize one's own material payoffs,
should perhaps be replaced by a blend of selfishness and morality.
Our result, under incomplete information and uniform random matching,
is a special case of their study with the index of assortativity being zero.}
But since the existence of appropriate nearby equilibria is not guaranteed,
it is not necessarily true that a materialist configuration must be stable.
For example, suppose that the fitness assignment is described by the following non-generic game,
and that all populations consist of materialist preferences only.

\begin{table}[!h]
\centering
\renewcommand{\arraystretch}{1.2}
\begin{tabular}{r|c|c|c|}
\multicolumn{1}{c}{} & \multicolumn{1}{c}{$a_{21}$} & \multicolumn{1}{c}{$a_{22}$} & \multicolumn{1}{c}{$a_{23}$}\\ \cline{2-4}
      $a_{11}$ & $1$, $1$ & $1$, $0$ & $0$, $0$ \\ \cline{2-4}
      $a_{12}$ & $1$, $1$ & $0$, $0$ & $1$, $0$ \\ \cline{2-4}
\end{tabular}
\end{table}

\noindent
Then every strategy pair in which player~$2$ plays $a_{21}$ is an equilibrium.
Consider a mutant type appearing in the second population.
No matter what the pre-entry equilibrium outcome is,
player~$1$ should play $a_{11}$ if he expects that the mutant type will play $a_{22}$ more often;
player~$1$ should play $a_{12}$ if he expects that the mutant type will play $a_{23}$ more often.
This implies that the post-entry equilibrium far from the pre-entry one always exists if suitable mutants are introduced.
Thus, materialist preferences in such an environment cannot form a stable configuration.

Similar reasoning leads to the result that the above non-generic game possesses no strictly perfect equilibria; see \citet[p.~16]{vDam:spne}.
Motivated by this, it can be shown that a strictly perfect equilibrium actually ensures that
materialist players with no observability will earn the highest average fitness by choosing strategies arbitrarily close to the pre-entry ones.

For convenience,
we give here a brief description of the strictly perfect equilibrium.
This concept introduced in \citet{aOka:spep}
is based on the idea that an equilibrium should be robust against all possible small mistakes.
Let $G = (N, A, \pi)$ be a finite normal-form game,
and for $t = 1$,~$2, \dots$,
let $\eta^t = (\eta^t_1, \dots, \eta^t_n)$ be an error function for all players,
where each $\eta^t_i$ is a function from $A_i$ to $(0, 1)$ with $\eta^t_i\rightarrow 0$ as $t\rightarrow \infty$.
A strategy profile $\sigma$ is a \emph{strictly perfect equilibrium} if
for every sequence $\{ G(\eta^t) \}$ of perturbed games,
there exists a sequence $\{ \sigma^t \}$ with each $\sigma^t$ being a Nash equilibrium of $G(\eta^t)$ such that
$\sigma^t$ converges to $\sigma$ as $t$ tends to infinity.

\begin{theoremEnd}[no link to theorem, restate]{thm}\label{prop:20230316}
If $\sigma^*$ is a strictly perfect equilibrium of $G$,
then it is stable in $(G,\Gamma_0(\mu))$ for some $\mu\in \mathcal{M}(\Theta^n)$.
\end{theoremEnd}

\begin{proofEnd}
Let $\sigma^*$ be supported by a monomorphic configuration $(\mu, s)$ with materialist preferences.
Then $s(\theta) = \sigma^*$ for $\theta\in \supp\mu$.
Suppose that $(\mu, s)$ is not a stable configuration under no observability.
Then,
since each population consists of materialist preferences having a fitness advantage,
there exist a mutant sub-profile $\witilde{\theta}_J$ for some $J\subseteq N$ and a positive number $\bar{\eta}$ for which
we can choose a sequence $\{ \varepsilon^t \}$ with $\| \varepsilon^t \|$ converging to $0$ such that the corresponding nearby set
$B_0^{\bar{\eta}}(\witilde{\mu}^{\varepsilon^t};s) = \varnothing$ for all $t$.
This means that for any $t$ and any $\witilde{s}^t\in B_0(\witilde{\mu}^{\varepsilon^t})$,
we have $d(\witilde{s}^t(\theta), s(\theta))> \bar{\eta}$ for $\theta\in \supp\mu$.
Then for every $t$, we can conclude that there exist $\witilde{x}^t_J\in \prod_{j\in J}\Delta(A_j)$ and
a mutant sub-profile $\witilde{\theta}^t_J$
defined by always adopting $\witilde{x}^t_J$ with its population share vector $\varepsilon^t$
for which
the corresponding nearby set within $\bar{\eta}$ is still empty.
This would imply that for any $\witilde{s}^t\in B_0(\witilde{\mu}^{\witilde{x}^t_J})$,
where $\witilde{\mu}^{\witilde{x}^t_J}$ denotes the type distribution after the entry of $\witilde{\theta}^t_J$,
there exist $i\in N$ and $a_i\in \supp\sigma^*_i$ such that
\begin{equation}\label{eq:20230321}
\pi_i(a_i, \wihat{y}^t_{-i}) < \pi_i\big(\witilde{s}^t_i(\theta_i), \wihat{y}^t_{-i} \big),
\end{equation}
where $\wihat{y}^t$ is defined by
$\wihat{y}^t_i = (1 - \varepsilon^t_i)\witilde{s}^t_i(\theta_i) + \varepsilon^t_i \witilde{x}^t_i$ if $i\in J$,
and $\wihat{y}^t_i = \witilde{s}^t_i(\theta_i)$ if $i\in N\setminus J$.

Next, for each $t$ and each $i\in N$, define
$\eta^t_i\colon A_i\to \mathbb{R}$ by
\[
\eta^t_i(a_i) =
\begin{cases}
    \varepsilon^t_i \witilde{x}^t_i(a_i) & \text{if $i\in J$ and $\witilde{x}^t_i(a_i)> 0$,}\\
    0 & \text{otherwise.}
\end{cases}
\]
Let $\eta^t = (\eta^t_1, \dots, \eta^t_n)$ be an error function for all players.
Suppose that $\sigma^*$ is a strictly perfect equilibrium of $G$,
and consider the sequence $\{ G(\eta^t) \}$ of the perturbed games.
Then it is also true that
there exists a sequence $\{ \sigma^t \}$, where each $\sigma^t$ is a Nash equilibrium of $G(\eta^t)$, such that
$\sigma^t$ converges to $\sigma^*$ as $t$ tends to infinity.
Therefore for sufficiently large $t$ and for each $i\in N$, we have
\begin{equation}\label{eq:20230322}
\pi_i(a_i, \sigma^t_{-i})\geq \pi_i(b_i, \sigma^t_{-i})
\end{equation}
for all $a_i\in \supp\sigma^*_i$ and $b_i\in A_i$.
Note that
for any Nash equilibrium $\sigma^t$ of the perturbed game $G(\eta^t)$,
there exists $\witilde{s}^t\in B_0(\witilde{\mu}^{\witilde{x}^t_J})$ such that
$\sigma^t_i = (1 - \varepsilon^t_i)\witilde{s}^t_i(\theta_i) + \varepsilon^t_i \witilde{x}^t_i$ if $i\in J$,
and $\sigma^t_i = \witilde{s}^t_i(\theta_i)$ if $i\in N\setminus J$.
This implies that \eqref{eq:20230322} is contradictory to \eqref{eq:20230321}, as desired.
\end{proofEnd}

\citet{aOka:spep} has proved that a strict Nash equilibrium is strictly perfect,
that a completely mixed Nash equilibrium is strictly perfect,
and that the unique Nash equilibrium is strictly perfect.
From Theorem~\ref{prop:20230316}, it immediately follows that
each of the three equilibrium outcomes can be stable supported by a materialist configuration with no observability.

\begin{theoremEnd}[]{cor}\label{prop:20180530}
Under incomplete information, a strategy profile $\sigma$ is stable if it is one of the following equilibria of $G$:
\begin{enumerate}[label=\upshape (\arabic*)]
\item a strict Nash equilibrium;
\item a completely mixed Nash equilibrium;
\item the unique Nash equilibrium.
\end{enumerate}
\end{theoremEnd}

In addition to the concept of a strict Nash equilibrium,
\citet{jHar:grdp} also defined the concept of a \emph{quasi-strict equilibrium} as
a strategy profile $\sigma$ such that for each player~$i$,
the set of pure best replies against $\sigma_{-i}$ coincides with the support of $\sigma_i$.
We show below that any quasi-strict equilibrium outcome is stable under no observability.\footnote{%
We are grateful to an anonymous referee for this suggestion.}
Note that in a generic game, all Nash equilibria are quasi-strict.
Hence under incomplete information,
any Nash equilibrium of a generic game would be a stable outcome.

\begin{theoremEnd}[no link to theorem, restate]{thm}\label{prop:20230328}
If $\sigma^*$ is a quasi-strict equilibrium of $G$,
then it is stable in $(G,\Gamma_0(\mu))$ for some $\mu\in \mathcal{M}(\Theta^n)$.
\end{theoremEnd}

\begin{proofEnd}
Let $\sigma^*$ be a quasi-strict equilibrium of $G = (N, A, \pi)$.
Then $\supp\sigma^*_i$ is the set of all pure best replies of player~$i$ against $\sigma^*_{-i}$.
Define the game $\bar{G} = (N, A, \bar{\pi})$ by
\[
\bar{\pi}_i(a) =
\begin{cases}
    \pi_i(a) & \text{if $a\in \supp\sigma^*$,}\\
    l_i & \text{otherwise}
\end{cases}
\]
for all $i\in N$ and all $a\in A$,
where $\supp\sigma^* = \prod_{i\in N} \supp \sigma_i^*$ and $l_i$ is a constant strictly smaller than $\pi_i(a)$ for all $a\in A$.
Then it is obvious that $\sigma^*$ is a strictly perfect equilibrium of $\bar{G}$.

Consider a monomorphic configuration $(\mu, s)$ consisting of $\theta_1$, \dots,~$\theta_n$
with the preferences $\bar{\pi}_1$, \dots,~$\bar{\pi}_n$, respectively, for which $s(\theta) = \sigma^*$.
Since $\sigma^*$ is strictly perfect in $\bar{G}$,
we see that for any entry of $\witilde{\theta}_J$ with $\varepsilon\in (0,1)^{|J|}$ and for any $\eta> 0$,
as the claim in the proof of Theorem~\ref{prop:20230316},
there exists $\bar{\epsilon}\in (0, 1)$ such that
$B_0^{\eta}(\witilde{\mu}^{\varepsilon};s)$ is nonempty for all $\varepsilon$ with $\| \varepsilon \|\in (0, \bar{\epsilon})$.
For $\witilde{s}\in B_0^{\bar{\eta}}(\witilde{\mu}^{\varepsilon};s)$, where $\bar{\eta}\in [0, \eta)$,
define $\witilde{x}\in \prod_{i\in N}\Delta(A_i)$ by
$\witilde{x}_i = (1 - \varepsilon_i)\witilde{s}_i(\theta_i) + \varepsilon_i \witilde{s}_i(\witilde{\theta}_i)$ if $i\in J$,
and $\witilde{x}_i = \witilde{s}_i(\theta_i)$ if $i\in N\setminus J$.
Then $\witilde{x}$ can be made arbitrarily close to $\sigma^*$ by making $\bar{\eta}$ and $\varepsilon$ sufficiently small.
Since $\sigma^*$ is a quasi-strict equilibrium of $G$,
we can conclude that if there exists $b_j\in A_j\setminus \supp\sigma^*_j$ for $j\in N$,
then when $\bar{\eta}$ and $\varepsilon$ are small enough,
\[
\pi_j(b_j, \witilde{x}_{-j}) < \pi_j(a_j, \witilde{x}_{-j})
\]
for all $a_j\in \supp\sigma^*_j$.
Furthermore,
since $\supp \witilde{s}_i(\theta_i)\subseteq \supp\sigma^*_i$ here for any $i\in N$,
it follows that $\pi_j(b_j, \witilde{x}_{-j})< \pi_i(\witilde{s}_j(\theta_j), \witilde{x}_{-j})$
as long as $\bar{\eta}$ and $\varepsilon$ are small enough.
Therefore, if $\supp \witilde{s}_i(\witilde{\theta}_i)\nsubseteq \supp\sigma^*_i$ for some $i\in J$,
the rare mutants would fail to invade for all suitable nearby equilibria.

Next, suppose that $\supp \witilde{s}_i(\witilde{\theta}_i)\subseteq \supp\sigma^*_i$ for all $i\in J$.
Then the game we have to consider is $\hat{G} = (N, \supp\sigma^*, \hat{\pi})$,
where $\hat{\pi}$ is the restriction of $\pi$ to $\supp\sigma^*$.
Now, the incumbents can be viewed as individuals endowed with materialist preferences with respect to $\hat{\pi}$,
and therefore such invading mutants cannot outperform them.
\end{proofEnd}

For the non-generic objective game described after Theorem~\ref{prop:005},
we have shown that those materialist configurations cannot be stable under no observability.
However, a stable aggregate outcome does exist in this objective game.
For example, let each $i$-th population consist of preferences for which $a_{i1}$ is the strictly dominant strategy.
Then a focal equilibrium exists for any entry, and mutants in the second population adopting strategies distinct from $a_{21}$ will be wiped out.
It can easily be verified that the strategy profile $(a_{11}, a_{21})$ is a stable aggregate outcome supported by these non-materialist preferences.
It is interesting to note that
a non-materialist configuration may induce a stable outcome which cannot be induced by materialist preferences,
although there is no fitness advantage in having non-materialist preferences.

In view of Theorem~\ref{prop:005},
the indirect evolutionary approach with incomplete information yields a strict refinement of the Nash equilibrium concept,
since not every Nash equilibrium is a stable strategy profile under no observability.\footnote{%
Consider a modification of the game in Example~\ref{exam:20160617} by letting $\nu = \omega = 0$.
For $(D, D)$, we introduce mutants for which $C$ is the strictly dominant strategy.
Then it is easily verified that the Nash equilibrium $(D, D)$ cannot be a stable outcome under no observability.}
Furthermore, by Theorems~\ref{prop:20230316} and~\ref{prop:20230328},
this equilibrium refinement is weaker than the concepts of strictly perfect equilibria and quasi-strict equilibria,
since the stable outcome $(a_{11}, a_{21})$ in the above discussion is neither strictly perfect, nor quasi-strict.


\section{Partial Observability}
\label{sec:imp-observ}


In this section, we assume that each player observes the opponents' types with a fixed probability $p\in (0,1)$,
and knows only the joint distribution over the opponents' types with the remaining probability $1-p$.
The two realizations are independently distributed among the players, and these statuses are private information.
Additionally, players always recognize their own types, and the degree of observability $p$ is common knowledge.

For $\mu\in \mathcal{M}(\Theta^n)$ and $i\in N$,
let $b^p_i\colon \supp\mu\to\Delta(A_i)$ and $s^p_i\colon \supp\mu_i\to\Delta(A_i)$ be two strategy functions of player~$i$,
corresponding to the two observation statuses, respectively.
We can then define two functions $b^p$,~$s^p\colon \supp\mu\to\prod_{i\in N}\Delta(A_i)$ by
$b^p(\theta) = \big(b^p_1(\theta),\dots,b^p_n(\theta)\big)$ and $s^p(\theta) = \big(s^p_1(\theta_1),\dots,s^p_n(\theta_n)\big)$.
For a matched $n$-tuple $\theta\in \supp\mu$,
the notations $b^p_i(\theta)$ and $s^p_i(\theta_i)$ denote, respectively,
the action of $\theta_i$ against the opponents $\theta_{-i}$ under perfect observability and
the action of $\theta_i$ with the belief $\mu_{-i}$ under no observability.
Moreover, let
\[
T = \{\, j\in N \mid \theta_j \text{ is a player in $\theta$ and $\theta_j$ is ignorant of the opponents' types} \,\}.
\]
Then the action profile played by this $n$-tuple $\theta$ is $\big( s^p(\theta)_T, b^p(\theta)_{-T} \big)$.

The pair of strategy functions $(b^p, s^p)$ is an \emph{equilibrium} in the subjective game $\Gamma_p(\mu)$
if for each $\theta\in \supp\mu$ and each $i\in N$,
the actions $b^p_i(\theta)$ and $s^p_i(\theta_i)$ satisfy the following properties.
First, when $\theta_i$ observes the $(n-1)$-tuple $\theta_{-i}$ of the opponents' types,
\[
b^p_i(\theta_i, \theta_{-i})\in \argmax_{\sigma_i\in\Delta(A_i)} \sum_{T\subseteq N\setminus\{i\}} p^{n-1-|T|}(1-p)^{|T|}
\theta_i\big( \sigma_i, (s^p_{-i}(\theta_{-i})_T, b^p(\theta_i, \theta_{-i})_{-i-T}) \big).
\]
Here we note that the subset $T$ might be the empty set, and the cardinal number of the empty set is zero.
Second, when $\theta_i$ knows only the distribution $\mu_{-i}$ over the opponents' types,
\begin{multline*}
s^p_i(\theta_i) \in \argmax_{\sigma_i\in\Delta(A_i)} \sum_{\theta'_{-i}\in\supp\mu_{-i}} \bigg( \mu_{-i}(\theta'_{-i})\\
\times \sum_{T\subseteq N\setminus\{i\}} p^{n-1-|T|}(1-p)^{|T|} \theta_i\big( \sigma_i, (s^p_{-i}(\theta'_{-i})_T, b^p(\theta_i, \theta'_{-i})_{-i-T}) \big) \bigg).
\end{multline*}
The set of all such equilibrium pairs in $\Gamma_p(\mu)$ is denoted by $B_p(\mu)$.
For $\mu\in \mathcal{M}(\Theta^n)$ and $(b^p, s^p)\in B_p(\mu)$, we call the triple $(\mu, b^p, s^p)$ a \emph{configuration},
and the \emph{aggregate outcome} of $(\mu, b^p ,s^p)$ refers to the probability distribution $\varphi_{\mu, b^p ,s^p}$ defined on $A$ by
\[
\varphi_{\mu,b^p,s^p}(a_1, \dots, a_n) = \sum_{\theta\in \supp\mu} \mu(\theta) \sum_{T\subseteq N} p^{n-|T|}(1-p)^{|T|} \prod_{i\in N} (s^p(\theta)_T, b^p(\theta)_{-T})_i(a_i).
\]
A strategy profile $\sigma\in\prod_{i\in N}\Delta(A_i)$ is called an \emph{aggregate outcome}
if the induced correlated strategy $\varphi_{\sigma}$ is the aggregate outcome of some configuration.

For $i\in N$, the \emph{average fitness} of a type $\theta_i\in\supp\mu_i$ with respect to $(\mu, b^p, s^p)$ can be calculated by
\[
\varPi_{\theta_i}(\mu;b^p,s^p) = \sum_{\theta'_{-i}\in\supp\mu_{-i}}\mu_{-i}(\theta'_{-i})
\sum_{T\subseteq N} p^{n-|T|}(1-p)^{|T|} \pi_i\big( s^p(\theta_i, \theta'_{-i})_T, b^p(\theta_i, \theta'_{-i})_{-T} \big).
\]
As before, a configuration $(\mu, b^p, s^p)$ is said to be \emph{balanced} if for each $i\in N$,
the equality $\varPi_{\theta_i}(\mu;b^p,s^p) = \varPi_{\theta'_i}(\mu;b^p,s^p)$ holds for every $\theta_i$,~$\theta'_i\in \supp\mu_i$.

Next, we extend the definitions of a nearby equilibrium and a stable configuration to the setting with partial observability.
For notational simplicity, we shall write $(b,s)$ for $(b^p,s^p)$.

\begin{Def}
Suppose that $(\mu,b,s)$ is a configuration under partial observability.
Let $\witilde{\mu}^{\varepsilon}$ be a post-entry type distribution, and let $\eta\geq 0$ be given.
In $\Gamma_p(\witilde{\mu}^{\varepsilon})$,
a pair $(\witilde{b}, \witilde{s})\in B_p(\witilde{\mu}^{\varepsilon})$
is called a \emph{nearby equilibrium pair} relative to $(b, s)$ within $\eta$
if $d(\witilde{b}(\theta), b(\theta))\leq \eta$ and $d(\witilde{s}(\theta), s(\theta))\leq \eta$ for all $\theta\in \supp\mu$.
Let $B_p^{\eta}(\witilde{\mu}^{\varepsilon};b,s)$ be the set of all nearby equilibrium pairs relative to $(b, s)$ within $\eta$,
called a \emph{nearby set}.
\end{Def}

Accordingly,
the nearby set with $\eta = 0$ is the \emph{focal set}, which we also denote by $B_p(\witilde{\mu}^{\varepsilon};b,s)$.
Any element $(\witilde{b}, \witilde{s})$ of the focal set, a \emph{focal equilibrium pair} relative to $(b,s)$, should satisfy
$\witilde{b}(\theta) = b(\theta)$ and $\witilde{s}(\theta) = (\theta)$ for all $\theta\in \supp\mu$.

\begin{Def}\label{def:stable-p=0-1}
In $(G,\Gamma_p(\mu))$ with $p\in (0,1)$, a configuration $(\mu,b,s)$ is \emph{stable} if it is balanced,
and if for any nonempty $J\subseteq N$, any $\witilde{\theta}_J\in\prod_{j\in J}\cplm{(\supp\mu_j)}$, and any $\eta> 0$,
there exist $\bar{\epsilon}\in (0,1)$ and $\bar{\eta}\in [0, \eta)$ such that
for every $\varepsilon\in (0,1)^{|J|}$ with $\|\varepsilon\|\in (0,\bar{\epsilon})$, the nearby set $B_p^{\bar{\eta}}(\witilde{\mu}^{\varepsilon};b,s)$ is nonempty,
and either~\ref{Cp:wiped-out} or~\ref{Cp:coexist} is satisfied for every $(\witilde{b}, \witilde{s})\in B_p^{\bar{\eta}}(\witilde{\mu}^{\varepsilon};b,s)$.
\begin{enumerate}[label=(\roman*)]
\item $\varPi_{\theta_j}(\witilde{\mu}^{\varepsilon};\witilde{b}, \witilde{s}) > \varPi_{\witilde{\theta}_j}(\witilde{\mu}^{\varepsilon};\witilde{b}, \witilde{s})$ for some $j\in J$ and for every $\theta_j\in\supp\mu_j$.\label{Cp:wiped-out}
\item $\varPi_{\theta_i}(\witilde{\mu}^{\varepsilon};\witilde{b}, \witilde{s}) = \varPi_{\wihat{\theta}_i}(\witilde{\mu}^{\varepsilon};\witilde{b}, \witilde{s})$ for every $i\in N$ and for every $\theta_i$,~$\wihat{\theta}_i\in \supp\witilde{\mu}^{\varepsilon}_i$.\label{Cp:coexist}
\end{enumerate}
A strategy profile $\sigma\in \prod_{i\in N}\Delta(A_i)$ is said to be \emph{stable} if it is the aggregate outcome of some stable configuration.
\end{Def}

It is not hard to verify that
the equilibria, the aggregate outcomes, the average fitnesses, and the stability criteria defined in the two extreme cases, $p = 1$ and $p = 0$,
can be regarded as the limiting cases of those defined here as $p\to 1$ and $p\to 0$, respectively.
It allows us to check whether the preceding results in Section~\ref{sec:observable} and Section~\ref{sec:unobservable} remain valid
if the two degrees $p=1$ and $p=0$ are perturbed, respectively.
As in DEY,
when considering the case of partial observability,
we restrict attention to aggregate outcomes which can be interpreted as degenerate distributions, called \emph{pure-strategy outcomes}.

In the single-population model of DEY, they show that
efficiency is a necessary condition for pure-strategy outcomes to be stable when observability is almost perfect.
We begin by providing a counterexample to show that the necessity result of the observable case is not robust:
Pareto efficiency is not a necessary condition for stability whenever $p$ is less than $1$, no matter how close $p$ is to $1$.
This says that Theorem~\ref{prop:003} will not hold when observability is not perfect;
Efficiency can be impaired by arbitrarily small perturbations on observability.
The reason for this difference is that, unlike the efficiency defined in single-population settings,
moving from a Pareto-dominated outcome may not make everyone strictly better off.
Therefore, in our model with partial observability, a Pareto improvement would no longer necessarily represent a chance for a mutant sub-profile
to make some mutant type receive a higher average fitness without any expense to itself.

\begin{example}\label{exam:002}
Let the objective game $G$ be the following asymmetric coordination game, and suppose that the degree of observability $p\in (0, 1)$.
In $G$, the strict Nash equilibrium $(a_{12},a_{22})$ Pareto dominates and risk dominates the strict Nash equilibrium $(a_{11},a_{21})$.

\begin{table}[!h]
\centering
\renewcommand{\arraystretch}{1.2}
    \begin{tabular}{r|c|c|}
      \multicolumn{1}{c}{}  &  \multicolumn{1}{c}{$a_{21}$}  &  \multicolumn{1}{c}{$a_{22}$}\\ \cline{2-3}
      $a_{11}$  &  $5$, $5$  &  $0$, $\phantom{1}0$ \\ \cline{2-3}
      $a_{12}$  &  $0$, $0$  &  $5$, $10$ \\ \cline{2-3}
    \end{tabular}
\end{table}

\noindent
Hence, by Theorem~\ref{prop:003}, the strategy profile $(a_{11},a_{21})$ is not stable for $p=1$.
Even so, it can be stable for any $p< 1$.
To see this, we begin with the assumption that $(a_{11},a_{21})$ is not stable, and then we show that this leads to a contradiction.
Consider a configuration $(\mu,b,s)$ where each $i$-th population consists of types for which $a_{i1}$ is the strictly dominant strategy.
Then $b(\theta) = (a_{11},a_{21})$ and $s(\theta) = (a_{11},a_{21})$ for all $\theta\in\supp\mu$,
and thus the aggregate outcome of $(\mu,b,s)$ is $(a_{11}, a_{21})$.
Under the instability assumption on $(a_{11}, a_{21})$,
there exists a mutant sub-profile $\witilde{\theta}_J$ for some $J\subseteq \{1, 2\}$ such that for every $\bar{\epsilon}\in (0,1)$, these mutants,
with some $\varepsilon\in (0,1)^{|J|}$ satisfying $\|\varepsilon\| \in (0,\bar{\epsilon})$,
can outperform the incumbents by playing an equilibrium pair $(\witilde{b}, \witilde{s})\in B_p(\witilde{\mu}^{\varepsilon};b,s)$, that is,
$\varPi_{\witilde{\theta}_j}(\witilde{\mu}^{\varepsilon};\witilde{b},\witilde{s})\geq \varPi_{\theta_j}(\witilde{\mu}^{\varepsilon};\witilde{b},\witilde{s})$
for all $j\in J$ and all $\theta_j\in \supp\mu_j$,
with at least one strict inequality.

In the case when $|J|=1$, it is clear that mutants have no fitness advantage.
Now let $J = \{1,2\}$,
and suppose that $(\witilde{\theta}_1, \witilde{\theta}_2)$ is a mutant pair having an evolutionary advantage.
For $\varepsilon\in (0,1)^2$ and $(\witilde{b}, \witilde{s})\in B_p(\witilde{\mu}^{\varepsilon};b,s)$,
define a function $\witilde{z}_i\colon \supp\witilde{\mu}^{\varepsilon}\to \Delta(A_i)$ by
$\witilde{z}_i(\wihat{\theta}_1, \wihat{\theta}_2) = p\witilde{b}_{i}(\wihat{\theta}_1, \wihat{\theta}_2) + (1-p)\witilde{s}_i(\wihat{\theta}_i)$ for $i = 1$,~$2$.
Then the post-entry average fitness of an incumbent $\theta_i$ is written
\[
\varPi_{\theta_i}(\witilde{\mu}^{\varepsilon};\witilde{b},\witilde{s}) =
(1-\varepsilon_{-i}) \pi_i(a_{11}, a_{21}) +
\varepsilon_{-i} \pi_i\big( a_{i1}, \witilde{z}_{-i}(\theta_i, \witilde{\theta}_{-i}) \big).
\]
Similarly, the post-entry average fitness of the mutant $\witilde{\theta}_i$ can be written as
\[
\varPi_{\witilde{\theta}_i}(\witilde{\mu}^{\varepsilon};\witilde{b},\witilde{s}) =
(1-\varepsilon_{-i}) \pi_i\big( \witilde{z}_i(\witilde{\theta}_i, \theta_{-i}), a_{-i1} \big) +
\varepsilon_{-i} \pi_i\big( \witilde{z}(\witilde{\theta}_1,\witilde{\theta}_2) \big),
\]
where $a_{-i1}$ denotes the pure action $a_{j1}$ with $j\neq i$.
Next,
we gradually reduce $\bar{\epsilon}$ to $0$, and then the sequence of the norms of the corresponding population share vectors converges to $0$.
We can choose a sequence $\{ (\witilde{b}^t, \witilde{s}^t) \}$ from the corresponding focal equilibrium pairs such that one of the following four cases occurs.
To complete the proof, we will show that each case leads to a contradiction.\footnote{%
A similar argument is used in more detail in the proof of Theorem~\ref{prop:0526}.}
\vspace{5pt}

\noindent
\textbf{Case 1}:
$\witilde{z}^t_i(\witilde{\theta}_i, \theta_{-i}) = a_{i1}$ for all $i$ and all $t$.
Then $\witilde{b}^t_i(\witilde{\theta}_i, \theta_{-i}) = \witilde{s}^t_i(\witilde{\theta}_i) = a_{i1}$,
and hence $\witilde{z}^t_i(\witilde{\theta}_1,\witilde{\theta}_2) \neq a_{i2}$ for each $i$ and each $t$.
For $i = 1$,~$2$, we certainly have
$\varPi_{\theta_i}(\witilde{\mu}^{\varepsilon^t};\witilde{b}^t,\witilde{s}^t) =
\varPi_{\witilde{\theta}_i}(\witilde{\mu}^{\varepsilon^t};\witilde{b}^t,\witilde{s}^t)$
if $\witilde{z}^t(\witilde{\theta}_1,\witilde{\theta}_2) = (a_{11},a_{21})$;
otherwise, $\varPi_{\theta_1}(\witilde{\mu}^{\varepsilon^t};\witilde{b}^t,\witilde{s}^t) >
\varPi_{\witilde{\theta}_1}(\witilde{\mu}^{\varepsilon^t};\witilde{b}^t,\witilde{s}^t)$.
This leads to a contradiction.
\vspace{5pt}

\noindent
\textbf{Case 2}:
$\witilde{z}^t_1(\witilde{\theta}_1, \theta_2) = a_{11}$ and $\witilde{z}^t_2(\theta_1, \witilde{\theta}_2) \neq a_{21}$ for every $t$.
Since $(\witilde{\theta}_1, \witilde{\theta}_2)$ would outperform the incumbents, we can deduce
\[
\frac{\varepsilon^t_1}{1-\varepsilon^t_1}>
\frac{5 - \pi_1\big( \witilde{z}^t(\witilde{\theta}_1, \witilde{\theta}_2) \big)}{\pi_2\big( \witilde{z}^t(\witilde{\theta}_1, \witilde{\theta}_2) \big) - 5}\geq 0,
\]
where $5\geq \pi_1\big( \witilde{z}^t(\witilde{\theta}_1, \witilde{\theta}_2) \big)$ and
$\pi_2\big( \witilde{z}^t(\witilde{\theta}_1, \witilde{\theta}_2) \big)> 5$ for every $t$.
Since $\varepsilon^t_1$ converges to $0$, it follows that
\[
\lim_{t\to \infty}\frac{5 - \pi_1\big( \witilde{z}^t(\witilde{\theta}_1, \witilde{\theta}_2) \big)}
{\pi_2\big( \witilde{z}^t(\witilde{\theta}_1, \witilde{\theta}_2) \big) - 5} = 0.
\]
For any $t$,
we have $\witilde{s}^t_1(\witilde{\theta}_1) = a_{11}$ since $\witilde{z}^t_1(\witilde{\theta}_1, \theta_2) = a_{11}$.
Then we can conclude that $\pi\big( \witilde{z}^t(\witilde{\theta}_1, \witilde{\theta}_2) \big)$ converges to $(5, 5)$
with $5\geq \pi_1\big( \witilde{z}^t(\witilde{\theta}_1, \witilde{\theta}_2) \big)$ and
$\pi_2\big( \witilde{z}^t(\witilde{\theta}_1, \witilde{\theta}_2) \big)> 5$ for every $t$,
which contradicts the shape of the noncooperative payoff region of $G$.
\vspace{5pt}

\noindent
\textbf{Case 3}:
$\witilde{z}^t_1(\witilde{\theta}_1, \theta_2) \neq a_{11}$ and $\witilde{z}^t_2(\theta_1, \witilde{\theta}_2) = a_{21}$ for every $t$.
Let $\witilde{z}^t_1(\witilde{\theta}_1, \theta_2) = (1-\zeta^t_1)a_{11} + \zeta^t_1 a_{12}$, where $\zeta^t_1\in (0,1]$ for all $t$.
Since the incumbents are outperformed, we have
$\varPi_{\witilde{\theta}_1}(\witilde{\mu}^{\varepsilon^t};\witilde{b}^t,\witilde{s}^t)\geq
\varPi_{\theta_1}(\witilde{\mu}^{\varepsilon^t};\witilde{b}^t,\witilde{s}^t)$.
This implies that
\[
\varepsilon^t_2 \big[ \pi_1\big( \witilde{z}^t(\witilde{\theta}_1, \witilde{\theta}_2) \big) - 5 \big] \geq 5 \zeta^t_1 (1 - \varepsilon^t_2) > 0,
\]
a contradiction to $5\geq \pi_1\big( \witilde{z}^t(\witilde{\theta}_1, \witilde{\theta}_2) \big)$.
\vspace{5pt}

\noindent
\textbf{Case 4}:
$\witilde{z}^t_i(\witilde{\theta}_i, \theta_{-i})\neq a_{i1}$ for all $i$ and all $t$.
For $i = 1$,~$2$,
let $\witilde{z}^t_i(\witilde{\theta}_i,\theta_{-i}) = (1-\xi^t_i)a_{i1} + \xi^t_i a_{i2}$,
where $\xi^t_i\in (0,1]$ for all $t$.
Since $(\witilde{\theta}_1, \witilde{\theta}_2)$ would outperform the incumbents, we obtain
\[
\frac{\varepsilon^t_{-i}}{1-\varepsilon^t_{-i}}\geq
\frac{5 \xi^t_i}{\pi_i\big( \witilde{z}^t(\witilde{\theta}_1, \witilde{\theta}_2) \big) - 5 + 5 \xi^t_{-i}}> 0
\]
for all $i$ and all $t$.
Since $\bar{\epsilon}$ converges to $0$, each $\xi^t_i$ converges to $0$, and hence each $\witilde{s}^t_i(\witilde{\theta}_i)$ converges to $a_{i1}$.
It also follows that
\[
\frac{\pi_i\big( \witilde{z}^t(\witilde{\theta}_1, \witilde{\theta}_2) \big) - 5 + 5 \xi^t_{-i}}{\xi^t_i}\to \infty
\]
for each $i$.
Since $5\geq \pi_1\big( \witilde{z}^t(\witilde{\theta}_1, \witilde{\theta}_2) \big)$, the sequence $\{ \xi^t_2/\xi^t_1 \}$ converges to $\infty$,
and hence we must have $\pi_2\big( \witilde{z}^t(\witilde{\theta}_1, \witilde{\theta}_2) \big)> 5$ for sufficiently large $t$.
In addition, the inequality $\varPi_{\witilde{\theta}_1}(\witilde{\mu}^{\varepsilon^t};\witilde{b}^t,\witilde{s}^t)\geq
\varPi_{\theta_1}(\witilde{\mu}^{\varepsilon^t};\witilde{b}^t,\witilde{s}^t)$ implies that
$5\xi^t_2 > 5 - \pi_1\big( \witilde{z}^t(\witilde{\theta}_1, \witilde{\theta}_2) \big)\geq 0$.
Therefore, we have
\[
\lim_{t\to \infty}\frac{5 - \pi_1\big( \witilde{z}^t(\witilde{\theta}_1, \witilde{\theta}_2) \big)}
{\pi_2\big( \witilde{z}^t(\witilde{\theta}_1, \witilde{\theta}_2) \big) - 5} = 0.
\]
Since each $\witilde{s}^t_i(\witilde{\theta}_i)$ converges to $a_{i1}$,
we can conclude that $\pi\big( \witilde{z}^t(\witilde{\theta}_1, \witilde{\theta}_2) \big)$ converges to $(5, 5)$
with $5\geq \pi_1\big( \witilde{z}^t(\witilde{\theta}_1, \witilde{\theta}_2) \big)$ and
$\pi_2\big( \witilde{z}^t(\witilde{\theta}_1, \witilde{\theta}_2) \big)> 5$ for sufficiently large $t$.
However, this contradicts the shape of the noncooperative payoff region of $G$.
\vspace{2pt}

The above discussion shows that
the strategy profile $(a_{11},a_{21})$ is stable for any $p\in (0, 1)$,
even though it is dominated by the equilibrium $(a_{12},a_{22})$.
\end{example}

Instead of Pareto efficiency,
we can show that weak Pareto efficiency is a necessary condition for pure-strategy outcomes to be stable when preferences are almost observable.
If a pure-strategy outcome is strongly Pareto dominated by another strategy profile,
then since the effects of no observability would be ignored in such a case as long as the degree of observability is large enough,
it follows that this outcome can be destabilized by the ``secret handshake'' entrants.

\begin{theoremEnd}[no link to theorem, restate]{thm}\label{prop:008}
Let $a^*$ be a pure-strategy profile in $G$.
If for any $\bar{p}\in (0,1)$ there exists $p\in (\bar{p},1)$ such that $a^*$ is stable for the degree $p$,
then it is weakly Pareto efficient with respect to $\pi$.
\end{theoremEnd}

\begin{proofEnd}
Let $p\in (0, 1)$ be a degree of observability, and suppose that $a^*$ is not weakly Pareto efficient with respect to $\pi$.
Let $(\mu,b,s)$ be a configuration with the aggregate outcome $a^*$. Then $b(\theta)=a^*$ and $s(\theta)=a^*$ for all $\theta\in \supp\mu$.
To prove the theorem, it suffices to show that there exists $\bar{p}\in (0,1)$ such that the configuration $(\mu,b,s)$ is not stable for any $p\in (\bar{p},1)$.
Because $a^*$ is not weakly Pareto efficient, there exists $\sigma\in\prod_{i\in N}\Delta(A_i)$ such that $\pi_i(\sigma)>\pi_i(a^*)$ for all $i\in N$.
Consider an indifferent mutant profile $\witilde{\theta}^0 = (\witilde{\theta}^0_1, \dots, \witilde{\theta}^0_n)$
entering with its population share vector $\varepsilon = (\varepsilon_1, \dots, \varepsilon_n)$.
Suppose that for $(\witilde{b}, \witilde{s})\in B_p(\witilde{\mu}^{\varepsilon})$, the mutants' strategies satisfy:
(1) $\witilde{b}(\witilde{\theta}^0)=\sigma$ and $\witilde{s}(\witilde{\theta}^0)=a^*$;
(2) for any nonempty proper subset $T$ of $N$ and any $\theta_{-T}\in \supp\mu_{-T}$, the equality $\witilde{b}_j(\witilde{\theta}^0_T, \theta_{-T}) = a^*_j$ holds for all $j\in T$.
Then the focal set $B_p(\witilde{\mu}^{\varepsilon};b,s)$ is nonempty for an arbitrary population share vector $\varepsilon\in (0,1)^n$.

Let $(\witilde{b}, \witilde{s})\in B_p(\witilde{\mu}^{\varepsilon};b,s)$ be played with $\witilde{b}(\witilde{\theta}^0_T, \theta_{-T}) = a^*$ for any $T\varsubsetneq N$ and any $\theta_{-T}\in \supp\mu_{-T}$. Then for each $i\in N$, the difference
$\varPi_{\witilde{\theta}^0_i}(\witilde{\mu}^{\varepsilon};\witilde{b}, \witilde{s}) - \varPi_{\theta_i}(\witilde{\mu}^{\varepsilon};\witilde{b}, \witilde{s})$
in average fitnesses between the mutant $\witilde{\theta}^0_i$ and an incumbent $\theta_i$ is
\[
\witilde{\mu}^{\varepsilon}_{-i}(\witilde{\theta}^0_{-i}) \bigg(
p^n [\pi_i(\sigma) - \pi_i(a^*)] +
\sum_{\varnothing\neq T\subseteq N} p^{n-|T|}(1-p)^{|T|} [\pi_i(a^*_T,\sigma_{-T}) - \pi_i(a^*)] \bigg).
\]
Since $\pi_i(\sigma) > \pi_i(a^*)$ for all $i\in N$ and the game $G$ is finite, there exists $\bar{p}\in (0,1)$ such that for any $p\in (\bar{p},1)$,
the inequality $\varPi_{\witilde{\theta}^0_i}(\witilde{\mu}^{\varepsilon};\witilde{b}, \witilde{s}) > \varPi_{\theta_i}(\witilde{\mu}^{\varepsilon};\witilde{b}, \witilde{s})$
holds for all $i\in N$ and all $\theta_i\in\supp\mu_i$, regardless of the population shares of the mutants.
\end{proofEnd}

While the necessity result of the observable case is not robust,
the analysis of this necessary condition still reveals that materialist preferences may have no fitness advantage when observability is almost perfect.

Next, we show that if a pure-strategy profile is not a Nash equilibrium of the objective game, then it is not stable when preferences are almost unobservable.
This means that for the case of pure-strategy outcomes, the result of Theorem~\ref{prop:005} is robust against small perturbations in the degrees of observability.

\begin{theoremEnd}[no link to theorem, restate]{thm}\label{prop:009}
Let $a^*$ be a pure-strategy profile in $G$.
If for any $\bar{p}\in (0,1)$ there exists $p\in (0,\bar{p})$ such that $a^*$ is stable for the degree $p$,
then it is a Nash equilibrium of $G$.
\end{theoremEnd}

\begin{proofEnd}
Suppose that $p\in (0,1)$, and that $a^*$ is not a Nash equilibrium of $G$.
Let $(\mu,b,s)$ be a configuration with the aggregate outcome $a^*$.
Then $b(\theta) = a^*$ and $s(\theta) = a^*$ for all $\theta\in \supp\mu$.
To prove the theorem, it suffices to show that there exists $\bar{p}\in (0,1)$ such that the configuration $(\mu,b,s)$ is not stable for any $p\in (0, \bar{p})$.
Since $a^*$ is not a Nash equilibrium, there exists a strategy $a_k\in A_k$ for some $k\in N$ such that $\pi_k(a_k,a_{-k}^*)> \pi_k(a^*)$.
Consider a mutant type $\witilde{\theta}_k$ entering the $k$-th population with its population share $\varepsilon$,
for which $a_k$ is the strictly dominant strategy.
Let $p\in (0,1)$ be given.
If there exists $\eta> 0$ such that for any $\bar{\epsilon}\in (0, 1)$,
the nearby set $B^{\eta}_p(\witilde{\mu}^{\varepsilon};b,s)$ is empty for some $\varepsilon\in (0, \bar{\epsilon})$,
then $(\mu,b,s)$ is not stable for the degree $p$, as desired.
Now, for any $\eta> 0$, we let $B^{\eta}_p(\witilde{\mu}^{\varepsilon};b,s)$ be nonempty for all sufficiently small $\varepsilon> 0$.

For $(\witilde{b}, \witilde{s})\in B^{\eta}_p(\witilde{\mu}^{\varepsilon};b,s)$,
the post-entry average fitness of a type $\wihat{\theta}_k$ in the $k$-th population is
\begin{multline*}
\varPi_{\wihat{\theta}_k}(\witilde{\mu}^{\varepsilon};\witilde{b}, \witilde{s}) =
(1-p)^n \sum_{\theta'_{-k}\in\supp\mu_{-k}} \mu_{-k}(\theta'_{-k}) \pi_k\big( \witilde{s}(\wihat{\theta}_k, \theta'_{-k}) \big)\\
+ p \sum_{\theta'_{-k}\in\supp\mu_{-k}} \mu_{-k}(\theta'_{-k}) \sum_{T\varsubsetneq N} p^{n-1-|T|}(1-p)^{|T|}
\pi_k\big( \witilde{s}(\wihat{\theta}_k,\theta'_{-k})_T, \witilde{b}(\wihat{\theta}_k,\theta'_{-k})_{-T} \big),
\end{multline*}
where we note that there are no entrants except the mutants in the $k$-th population.
Because $\pi_k(a_k,a_{-k}^*)> \pi_k(a^*)$ and $\pi_k$ is continuous on $\prod_{i\in N}\Delta(A_i)$,
there exist $m> 0$ and $\bar{\eta}> 0$ such that for any $(\witilde{b}, \witilde{s})\in B^{\bar{\eta}}_p(\witilde{\mu}^{\varepsilon};b,s)$,
\[
\pi_k\big( \witilde{s}(\witilde{\theta}_k, \theta'_{-k}) \big) - \pi_k\big( \witilde{s}(\theta_k, \theta'_{-k}) \big) =
\pi_k\big( a_k, \witilde{s}_{-k}(\theta'_{-k}) \big) - \pi_k\big( \witilde{s}(\theta_k, \theta'_{-k}) \big)\geq m
\]
for all $\theta_k\in\supp\mu_k$ and all $\theta'_{-k}\in\supp\mu_{-k}$.
Thus there exist $\bar{p}\in (0,1)$ and $\bar{\eta}> 0$ such that
for any $p\in (0, \bar{p})$ and any $(\witilde{b}, \witilde{s})\in B^{\bar{\eta}}_p(\witilde{\mu}^{\varepsilon};b,s)$,
the inequality
$\varPi_{\witilde{\theta}_k}(\witilde{\mu}^{\varepsilon}; \witilde{b}, \witilde{s})> \varPi_{\theta_k}(\witilde{\mu}^{\varepsilon}; \witilde{b}, \witilde{s})$
holds for all $\theta_k\in\supp\mu_k$.
\end{proofEnd}


We now turn our attention to the robustness of the preceding results concerning sufficient conditions for stability.
First, the following two theorems show that the conclusions of Theorems~\ref{prop:0526} and~\ref{prop:20190731}
still hold in the case when preferences are imperfectly observable,
and so, by Corollary~\ref{prop:20180530}, both of these hold for all degrees of observability.

\begin{theoremEnd}[no link to theorem, restate]{thm}\label{prop:007}
Let $G$ be a two-player game, and suppose that $(a^*_1,a^*_2)$ is Pareto efficient with respect to $\pi$.
If $(a^*_1,a^*_2)$ is a strict Nash equilibrium of $G$, then $(a^*_1,a^*_2)$ is stable for all $p\in (0,1)$.
\end{theoremEnd}

\begin{proofEnd}
Let $(a^*_1,a^*_2)$ be a strict Nash equilibrium of $G$,
and suppose that it is not stable for some $p\in (0, 1)$.
We will show that $(a^*_1,a^*_2)$ is not Pareto efficient with respect to $\pi$.
For each $i\in \{1,2\}$,
consider the $i$-th population consisting of the only type $\theta^*_i$ for which $a^*_i$ is the strictly dominant strategy,
and denote this configuration by $(\mu, b, s)$.
Then the aggregate outcome of $(\mu,b,s)$ is $(a^*_1,a^*_2)$, and hence this configuration $(\mu, b, s)$ is not stable for some $p\in (0, 1)$.
This means that
there exists some chance for a mutant sub-profile $\witilde{\theta}_J$ to gain an evolutionary advantage over the incumbents
in an imperfectly observable environment.
Besides,
it is obvious that after any entry, the focal set must be nonempty.

In the case when $|J| = 1$, it is obvious that no mutant can have a fitness advantage since $(a^*_1,a^*_2)$ is a strict Nash equilibrium.
In the case when $J = \{1, 2\}$, it must be true that for every $\bar{\epsilon}\in (0,1)$,
there exist a population share vector $\varepsilon\in (0, 1)^2$ with $\|\varepsilon\| \in (0,\bar{\epsilon})$
and an equilibrium pair $(\witilde{b}, \witilde{s})\in B_p(\witilde{\mu}^{\varepsilon};b,s)$
such that
$\varPi_{\witilde{\theta}_i}(\witilde{\mu}^{\varepsilon};\witilde{b}, \witilde{s})\geq
\varPi_{\theta^*_i}(\witilde{\mu}^{\varepsilon};\witilde{b}, \witilde{s})$ for all $i\in \{1,2\}$,
with strict inequality for some $i$.
To complete the proof,
we have to show that $(a^*_1, a^*_2)$ is Pareto dominated in any situation where $(\witilde{\theta}_1, \witilde{\theta}_2)$ has an evolutionary advantage.

For $(\witilde{b}, \witilde{s})\in B_p(\witilde{\mu}^{\varepsilon};b,s)$ and $i\in \{1,2\}$,
define $\witilde{z}_i\colon \supp\witilde{\mu}^{\varepsilon}\to \Delta(A_i)$
by $\witilde{z}_i(\wihat{\theta}_1, \wihat{\theta}_2) = p\witilde{b}_{i}(\wihat{\theta}_1, \wihat{\theta}_2) + (1-p)\witilde{s}_i(\wihat{\theta}_i)$.
Then the post-entry average fitness of $\theta^*_i\in \supp\mu_i$ is
\[
\varPi_{\theta^*_i}(\witilde{\mu}^{\varepsilon};\witilde{b}, \witilde{s}) =
(1-\varepsilon_{-i})\pi_i(a_1^*,a_2^*) + \varepsilon_{-i} \pi_i\big( a_i^*, \witilde{z}_{-i}(\theta^*_i, \witilde{\theta}_{-i}) \big).
\]
Similarly, the post-entry average fitness of $\witilde{\theta}_i\in \cplm{(\supp\mu_j)}$ is
\[
\varPi_{\witilde{\theta}_i}(\witilde{\mu}^{\varepsilon};\witilde{b}, \witilde{s}) =
(1-\varepsilon_{-i}) \pi_i\big( \witilde{z}_i(\witilde{\theta}_i, \theta^*_{-i}), a_{-i}^* \big) +
\varepsilon_{-i} \pi_i\big( \witilde{z}_1(\witilde{\theta}_1, \witilde{\theta}_2), \witilde{z}_2(\witilde{\theta}_1, \witilde{\theta}_2) \big).
\]
With this kind of representation,
the two actions $\witilde{z}_i(\witilde{\theta}_i, \theta^*_{-i})$ and $\witilde{z}_i(\witilde{\theta}_1, \witilde{\theta}_2)$
are correlated through $\witilde{s}_i(\witilde{\theta}_i)$.
Even so, the remaining part of the proof is analogous to the proof of Theorem~\ref{prop:0526}.
\end{proofEnd}

\begin{theoremEnd}[no link to theorem, restate]{thm}\label{prop:20190827}
Let $(a^*_1, \dots, a^*_n)$ be a strict union Nash equilibrium in an $n$-player game $G$.
Then $(a^*_1, \dots, a^*_n)$ is stable for all $p\in (0,1)$.
\end{theoremEnd}

\begin{proofEnd}
As Theorem~\ref{prop:007} being proved by a modification of the proof of Theorem~\ref{prop:0526},
the proof of this theorem is similar to that of Theorem~\ref{prop:20190731} with $\witilde{b}$ replaced by $\witilde{z}$,
where each function $\witilde{z}_i\colon \supp\witilde{\mu}^{\varepsilon}\to \Delta(A_i)$ is defined by
$\witilde{z}_i(\wihat{\theta}_i, \wihat{\theta}_{-i}) = p\witilde{b}_{i}(\wihat{\theta}_i, \wihat{\theta}_{-i}) + (1-p)\witilde{s}_i(\wihat{\theta}_i)$.
\end{proofEnd}

When opponents' preferences are unobservable, we know from Corollary~\ref{prop:20180530} that
a pure-strategy outcome can be stable if it is a strict Nash equilibrium or the unique Nash equilibrium of the objective game.
However, this result may not hold in the case of partial observability.
If the degree of observability is positive, even arbitrarily small,
then mutants may earn strictly higher average fitnesses by playing a more efficient action profile,
which occurs only when they see and meet one another without affecting the incumbents' expectations and strategies; see the following Example.
It indicates that efficiency would play a role in preference evolution as long as the probability of observing preferences is positive.

\begin{example}\label{exam:20190918}
Suppose that the objective game $G$ is the following Prisoner's Dilemma game.

\begin{table}[!h]
\centering
\renewcommand{\arraystretch}{1.2}
    \begin{tabular}{r|c|c|}
      \multicolumn{1}{c}{} & \multicolumn{1}{c}{$C_2$} & \multicolumn{1}{c}{$D_2$}\\ \cline{2-3}
      $C_1$ & $2$, $2$ & $0$, $3$ \\  \cline{2-3}
      $D_1$ & $3$, $0$ & $1$, $1$ \\  \cline{2-3}
    \end{tabular}
\end{table}

\noindent
With respect to the payoffs of $G$, the strategy of defection is strictly dominant for each player,
and so $(D_1, D_2)$ is a strict Nash equilibrium, also the unique Nash equilibrium, of the game.
Then by Corollary~\ref{prop:20180530}, $(D_1, D_2)$ is stable in the case of no observability.
However, we show that $(D_1, D_2)$ can be destabilized if preferences are not completely unobservable.

Let $p\in (0,1)$ and let $(\mu,b,s)$ be a configuration with the aggregate outcome $(D_1, D_2)$.
Then $b(\theta)=(D_1, D_2)$ and $s(\theta)=(D_1, D_2)$ for all $\theta\in\supp\mu$.
Let an indifferent mutant pair $(\witilde{\theta}^0_1, \witilde{\theta}^0_2)$ be introduced
with its population share vector $\varepsilon = (\varepsilon_1, \varepsilon_2)$.
Suppose that the strategies of the mutants have the following properties:
(1) $\witilde{b}(\witilde{\theta}^0_1,\witilde{\theta}^0_2) = (C_1,C_2)$ and $\witilde{s}(\witilde{\theta}^0_1,\witilde{\theta}^0_2) = (D_1,D_2)$;
(2) $\witilde{b}_1(\witilde{\theta}^0_1,\theta_2) = D_1$ for all $\theta_2\in \supp\mu_2$ and $\witilde{b}_2(\theta_1,\witilde{\theta}^0_2) = D_2$ for all $\theta_1\in \supp\mu_1$.
This implies that the focal set $B_p(\witilde{\mu}^{\varepsilon};b,s)$ is nonempty no matter what $\varepsilon$ is.
Let the played equilibrium pair $(\witilde{b}, \witilde{s})\in B_p(\witilde{\mu}^{\varepsilon};b,s)$ further satisfy
$\witilde{b}_1(\theta_1,\witilde{\theta}^0_2) = D_1$ and $\witilde{b}_2(\witilde{\theta}^0_1,\theta_2) = D_2$ for all incumbents $\theta_1$ and $\theta_2$.
Then for each $i\in\{1,2\}$, the post-entry average fitness of $\theta_i\in\supp\mu_i$ is equal to $1$.
In addition, the mutant $\witilde{\theta}^0_i$ would earn the average fitness:
\[
\varPi_{\witilde{\theta}^0_i}(\witilde{\mu}^{\varepsilon};\witilde{b}, \witilde{s}) = (1-\varepsilon_{-i})\cdot 1 + \varepsilon_{-i}\big( p^2\cdot 2 + p(1-p)(0 + 3) + (1-p)^2\cdot 1 \big).
\]
Therefore, for each $p\in (0,1)$ and each $i\in \{1,2\}$, we have
$\varPi_{\witilde{\theta}^0_i}(\witilde{\mu}^{\varepsilon};\witilde{b}, \witilde{s}) > 1$ regardless of the population shares of the mutants,
and hence $(D_1, D_2)$ is not stable for any positive degree of observability.
Furthermore, we can conclude from Theorem~\ref{prop:009} that in this game,
there is no stable pure-strategy outcome when preferences are almost unobservable.
\end{example}

According to the no-observability setting in Section~\ref{sec:unobservable},
materialist preferences always have a fitness advantage under incomplete information, although there may be no appropriate nearby equilibria.
We also know from Theorems~\ref{prop:003} and~\ref{prop:008} that
individuals endowed with materialist preferences may be unable to survive if the preference information is sufficient.
In particular, Example~\ref{exam:20190918} further points out that
materialist preferences would have no fitness advantage as long as the degree of observability is positive.
Thus, we conclude that materialist preferences may not prevail unless preferences are completely unobservable.



\section{Conclusion}
\label{sec:conclusion}


While an asymmetric game played by players from distinct populations is a realistic representation of many human interactions,
it is virtually ignored in the literature on the evolution of preferences.
In this paper, we extend the single-population model of DEY to the multi-population model,
where players have fixed roles and may have different action sets.
We define multi-population evolutionary stability by applying the concept of a two-species ESS introduced in \citet{rCre:scegt}.
Our stability criterion is weaker than the one derived from \citet{pTay:essttp} (or \citet{rSel:nessaac}),
and it can accommodate many different degrees of observability.

We study the basic features of preference evolution under multi-population matching,
and analyze whether the analogous results of DEY remain valid.
When preferences are unobservable, the same results can be achieved:
any stable outcome is a Nash equilibrium; a (quasi-)strict Nash equilibrium outcome must be stable.
When preferences are observable, DEY showed that stability implies efficiency; efficient strict Nash equilibria are stable.
For two-player objective games, we can obtain the same results by substituting Pareto efficiency for efficiency.
As the number $n$ of players in a match increases, the necessity result holds for any $n$-player game.
But it is worth noting that the sufficient condition for stability is no longer valid.
To solve this, we define a stronger equilibrium concept, namely, the strict union Nash equilibrium.
As revealed in Examples~\ref{exam:20171130} and~\ref{exam:20150603},
such strengthening is a natural method of ensuring the existence of stable outcomes.
Finally, DEY showed that for pure-strategy outcomes,
their results are robust to the introduction of partial observability,
except that strict Nash equilibria may fail to be stable under almost-no observability.
We also have the same conclusions as DEY have, with the exception that
a weakly Pareto-efficient strategy profile may be a stable outcome when observability is almost perfect.

Our multi-population model generally differs from the established single-population models in two ways.
First, under the assumption that mutation may arise in every population,
the complexity of mutants' interactions grows explosively in multi-population settings with positive degrees of observability.
If we consider multi-player games in single-population matching environments as defined in DEY,
it should be true that a symmetrically strictly strong Nash equilibrium in a symmetric $n$-player game is stable.\footnote{%
To see this,
consider a population consisting of types for which the strictly strong Nash equilibrium strategy is strictly dominant,
and consider any entrant type.
Then any simultaneous deviation from the symmetrically strictly strong Nash equilibrium
will result in losses to all these deviating mutants (who have the same type).}
Yet, Example~\ref{exam:20150603} illustrates that such a condition is not sufficient to ensure stability in our multi-population model.
(Under single-population matching,
such complex interactions can be seen in a symmetric $n$-player game whenever multiple mutations are considered for the stability concept.)

To ensure stability, we introduce the concept of a strict union Nash equilibrium,
with which each stable population consists of types such that these strict union Nash equilibrium strategies are strictly dominant for them, respectively.
This approach, used for constructing stable configurations as in DEY, helps simply to describe the sufficient condition, and make it easy to be comprehended.
However such a sufficient condition for the existence of an evolutionarily stable outcome seems too strong.
Indeed, a game without pure Nash equilibria still may have a stable outcome, as presented in Example~\ref{exam:20171114},
and it suggests that the sufficient condition would be improved if other preference types, not only inducing strictly dominant strategies, are fully utilized.
This deserves further analysis.

Second, efficiency characteristics in the model with single-population matching are not applicable to the model with multi-population matching.
The efficiency defined in standard single-population models of preference evolution, as in DEY, refers particularly to the highest fitness available from symmetric strategies,
and hence the efficiency value is unique.
In contrast, asymmetric Pareto efficiency and Pareto dominance are applied in multi-population models.
Although there are numerous forms of Pareto efficiency in general games,
we show that for a stable polymorphic configuration,
the played action profiles in all matches have the same Pareto-efficient form.

For the case of almost-perfect observability,
we find that stability only implies weak Pareto efficiency, rather than Pareto efficiency.
It points out that efficiency may be impaired by small perturbations on observability.
There seems to be a tendency for efficiency levels to increase with increasing degrees of observability.
Unlike DEY-efficiency,
an outcome that is not Pareto efficient does not mean that there is an alternative which strongly Pareto dominates it.
Therefore, when preferences are not perfectly observed,
a Pareto improvement would no longer enable mutants to destabilize the Pareto-dominated outcome.
Nevertheless, our results are also in line with the argument that
materialist preferences may not survive even when preferences are almost unobservable.


\appendix
\section{Proofs of Theorems}
\label{sec:proofs}


\printProofs



\bibliographystyle{plainnat}
\bibliography{EvlPrf}

@ARTICLE{iAlg-jWei:hmpeiiam,
  author = {Ingela Alger and J{\"{o}}rgen W. Weibull},
  title = {Homo Moralis--Preference Evolution Under Incomplete Information and Assortative Matching},
  journal = {Econometrica},
  year = {2013},
  volume = {81},
  pages = {2269-2302}
}

@ARTICLE{hBes-wGut:iaes,
  author = {Helmut Bester and Werner G{\"{u}}th},
  title = {Is Altruism Evolutionarily Stable?},
  journal = {Journal of Economic Behavior \& Organization},
  year = {1998},
  volume = {34},
  pages = {193-209}
}

@ARTICLE{kBin-lSam:d,
  author = {Ken Binmore and Larry Samuelson},
  title = {Drift},
  journal = {European Economic Review},
  year = {1994},
  volume = {38},
  pages = {859-867}
}

@ARTICLE{eDek-jEly-oYil:ep,
  author = {Eddie Dekel and Jeffrey C. Ely and Okan Yilankaya},
  title = {Evolution of Preferences},
  journal = {Review of Economic Studies},
  year = {2007},
  volume = {74},
  pages = {685-704}
}

@ARTICLE{jEly-oYil:neep,
  author = {Jeffrey C. Ely and Okan Yilankaya},
  title = {{Nash} Equilibrium and the Evolution of Preferences},
  journal = {Journal of Economic Theory},
  year = {2001},
  volume = {97},
  pages = {255-272}
}

@ARTICLE{wGut:eaecbri,
  author = {Werner G{\"{u}}th},
  title = {An Evolutionary Approach to Explaining Cooperative Behavior by Reciprocal Incentives},
  journal = {International Journal of Game Theory},
  year = {1995},
  volume = {24},
  pages = {323-344}
}

@ARTICLE{wGut-bPel:wwpms,
  author = {Werner G{\"{u}}th and Bezalel Peleg},
  title = {When Will Payoff Maximization Survive? An Indirect Evolutionary Analysis},
  journal = {Journal of Evolutionary Economics},
  year = {2001},
  volume = {11},
  pages = {479-499}
}

@ARTICLE{jHar:grdp,
  author = {John C. Harsanyi},
  title = {Games with Randomly Disturbed Payoffs: A New Rationale for Mixed-Strategy Equilibrium Points},
  journal = {International Journal of Game Theory},
  year = {1973},
  volume = {2},
  pages = {1-23}
}

@ARTICLE{aHei-cSha-ySpi:wmiym,
  author = {Aviad Heifetz and Chris Shannon and Yossi Spiegel},
  title = {What to Maximize if You Must},
  journal = {Journal of Economic Theory},
  year = {2007},
  volume = {133},
  pages = {31-57}
}

@ARTICLE{yHel:steg,
  author = {Yuval Heller},
  title = {Stability and Trembles in Extensive-Form Games},
  journal = {Games and Economic Behavior},
  year = {2014},
  volume = {84},
  pages = {132-136}
}

@ARTICLE{yHel-eMoh:cdp,
  author = {Yuval Heller and Erik Mohlin},
  title = {Coevolution of Deception and Preferences: {Darwin} and {Nash} Meet {Machiavelli}},
  journal = {Games and Economic Behavior},
  year = {2019},
  volume = {113},
  pages = {223-247}
}

@ARTICLE{sHuc-jOec:ieaefa,
  author = {Steffen Huck and J{\"{o}}rg Oechssler},
  title = {The Indirect Evolutionary Approach to Explaining Fair Allocations},
  journal = {Games and Economic Behavior},
  year = {1999},
  volume = {28},
  pages = {13-24}
}

@ARTICLE{jMay-gPri:lac,
  author = {Maynard Smith, John and Price, George R.},
  title = {The Logic of Animal Conflict},
  journal = {Nature},
  year = {1973},
  volume = {246},
  pages = {15-18}
}

@ARTICLE{jMcN-lFro-zBar-aHou:ocg,
  author = {John M. McNamara and Lutz Fromhage and Zoltan Barta and Alasdair I. Houston},
  title = {The Optimal Coyness Game},
  journal = {Proceedings: Biological Sciences},
  year = {2009},
  volume = {276},
  pages = {953-960}
}

@ARTICLE{eOk-fVeg:eipiis,
  author = {Efe A. Ok and Fernando Vega-Redondo},
  title = {On the Evolution of Individualistic Preferences: An Incomplete Information Scenario},
  journal = {Journal of Economic Theory},
  year = {2001},
  volume = {97},
  pages = {231-254}
}

@ARTICLE{aOka:spep,
  author = {Akira Okada},
  title = {On Stability of Perfect Equilibrium Points},
  journal = {International Journal of Game Theory},
  year = {1981},
  volume = {10},
  pages = {67-73}
}

@ARTICLE{eOst:caesn,
  author = {Elinor Ostrom},
  title = {Collective Action and the Evolution of Social Norms},
  journal = {Journal of Economic Perspectives},
  year = {2000},
  volume = {14},
  pages = {137-158}
}

@Article{oPar:nep,
  author = {Oliver Pardo},
  title = {A Note on ``Evolution of Preferences''},
  journal = {Journal of Mathematical Economics},
  year= {2017},
  volume = {71},
  pages = {129-134}
}

@ARTICLE{aPos:tsespsg,
  author = {Alex Possajennikov},
  title = {Two-Speed Evolution of Strategies and Preferences in Symmetric Games},
  journal = {Theory and Decision},
  year = {2005},
  volume = {57},
  pages = {227-263}
}

@ARTICLE{dRay-rVoh:eba,
  author = {Debraj Ray and Rajiv Vohra},
  title = {Equilibrium Binding Agreements},
  journal = {Journal of Economic Theory},
  year = {1997},
  volume = {73},
  pages = {30-78}
}

@ARTICLE{aRob:eegdnsh,
  author = {Arthur J. Robson},
  title = {Efficiency in Evolutionary Games: {Darwin}, {Nash} and the Secret Handshake},
  journal = {Journal of Theoretical Biology},
  year = {1990},
  volume = {144},
  pages = {379-396}
}

@ARTICLE{lSam:iep,
  author = {Larry Samuelson},
  title = {Introduction to the Evolution of Preferences},
  journal = {Journal of Economic Theory},
  year = {2001},
  volume = {97},
  pages = {225-230}
}

@ARTICLE{rSel:nessaac,
  author = {Reinhard Selten},
  title = {A Note on Evolutionarily Stable Strategies in Asymmetric Animal Conflicts},
  journal = {Journal of Theoretical Biology},
  year = {1980},
  volume = {84},
  pages = {93-101}
}

@ARTICLE{rSel:esetg,
  author = {Reinhard Selten},
  title = {Evolutionary Stability in Extensive Two-Person Games},
  journal = {Mathematical Social Sciences},
  year = {1983},
  volume = {5},
  pages = {269-363}
}

@ARTICLE{rSel:eleb,
  author = {Reinhard Selten},
  title = {Evolution, Learning, and Economic Behavior},
  journal = {Games and Economic Behavior},
  year = {1991},
  volume = {3},
  pages = {3-24}
}

@ARTICLE{rSet-eSom:per,
  author = {Rajiv Sethi and E. Somanathan},
  title = {Preference Evolution and Reciprocity},
  journal = {Journal of Economic Theory},
  year = {2001},
  volume = {97},
  pages = {273-297}
}

@ARTICLE{jSwi:esee,
  author = {Jeroen M. Swinkels},
  title = {Evolutionary Stability with Equilibrium Entrants},
  journal = {Journal of Economic Theory},
  year = {1992},
  volume = {57},
  pages = {306-332}
}

@ARTICLE{pTay:essttp,
  author = {Peter D. Taylor},
  title = {Evolutionarily Stable Strategies with Two Types of Player},
  journal = {Journal of Applied Probability},
  year = {1979},
  volume = {16},
  pages = {76-83}
}

@ARTICLE{2017arXiv170501454T,
       author = {Yu-Sung Tu and Wei-Torng Juang},
        title = {The Payoff Region of a Strategic Game and Its Extreme Points},
      journal = {arXiv e-prints},
         year = {2018}, 
archivePrefix = {arXiv},
       eprint = {1705.01454},
 primaryClass = {cs.GT}, 
          url = {https://arxiv.org/abs/1705.01454}
}

@BOOK{rCre:scegt,
  AUTHOR =    {Ross Cressman},
  TITLE =     {The Stability Concept of Evolutionary Game Theory: A Dynamic Approach},
  PUBLISHER = {Springer-Verlag Berlin Heidelberg},
  YEAR =      {1992},
  series =    {Lecture Notes in Biomathematics},
  VOLUME =    {94}
}

@BOOK{vDam:spne,
   AUTHOR = {van Damme, Eric},
    TITLE = {Stability and Perfection of {Nash} Equilibria},
     YEAR = {1991},
publisher = {Springer-Verlag Berlin Heidelberg}
}

@book{rDaw:sg,
  author = {Richard Dawkins},
  title  = {The Selfish Gene},
  publisher = {Oxford University Press},
  year   = {1976}
}

@BOOK{jMay:etg,
   AUTHOR = {Maynard Smith, John},
    TITLE = {Evolution and the Theory of Games},
     YEAR = {1982},
publisher = {Cambridge University Press}
}

@BOOK{wSan:pged,
  author = {William H. Sandholm},
  title = {Population Games and Evolutionary Dynamics},
  publisher = {The MIT Press},
  year = {2010}
}

@BOOK{jWei:egt,
  author = {J{\"{o}}rgen W. Weibull},
  title = {Evolutionary Game Theory},
  publisher = {The MIT Press},
  year = {1995}
}

@BOOK{svW:eneup,
  author = {von Widekind, Sven},
  title = {Evolution of Non-Expected Utility Preferences},
  publisher = {Springer-Verlag Berlin Heidelberg},
  year = {2008},
  series = {Lecture Notes in Economics and Mathematical Systems},
  VOLUME = {606}
}

@INCOLLECTION{rAum:apgcng,
   AUTHOR = {Robert J. Aumann},
   EDITOR = {Albert William Tucker and Robert Duncan Luce},
    TITLE = {Acceptable Points in General Cooperative $n$-Person Games},
BOOKTITLE = {Contributions to the Theory of Games, Volume IV},
    PAGES = {287-324},
PUBLISHER = {Princeton University Press},
     YEAR = {1959}
}

@INCOLLECTION{wGut-mYaa:erbssg,
   AUTHOR = {Werner G{\"{u}}th and Menahem E. Yaari},
   EDITOR = {Ulrich Witt},
    TITLE = {Explaining Reciprocal Behavior in Simple Strategic Games: An Evolutionary Approach},
BOOKTITLE = {Explaining Process and Change: Approaches to Evolutionary Economics},
    PAGES = {23-34},
PUBLISHER = {University of Michigan Press},
     YEAR = {1992}
}

@INCOLLECTION{aRob-lSam,
author = {Arthur J. Robson and Larry Samuelson},
editor = {Jess Benhabib and Alberto Bisin and Matthew O. Jackson},
title = {The Evolutionary Foundations of Preferences},
booktitle = {Handbook of Social Economics},
publisher = {North-Holland},
pages = {221-310},
volume = {1},
year = {2011}
}



\end{document}